\documentclass[fleqn,usenatbib]{mnras}

\usepackage{newtxtext,newtxmath}

\usepackage[T1]{fontenc}

\DeclareRobustCommand{\VAN}[3]{#2}
\let\VANthebibliography\thebibliography
\def\thebibliography{\DeclareRobustCommand{\VAN}[3]{##3}\VANthebibliography}

\usepackage{graphicx}	
\usepackage{amsmath}	
\usepackage{multicol}        
\usepackage{bm}		
\usepackage{pdflscape}	
\usepackage{graphicx}	
\usepackage{threeparttable,lscape}
\usepackage{natbib}
\usepackage{rotating}
\usepackage{color}
\usepackage{xcolor}
\usepackage{appendix}
\usepackage{multirow, array}
\usepackage{longtable}
\usepackage{ulem}

\newcommand{\Msun}{M$_\odot$}
\newcommand{\Rsun}{R$_\odot$}
\newcommand{\nodata}{\centering\arraybackslash --} 
\newcommand{\kms}{km s$^{-1}$}
\newcommand{\cofs}{$^{56}$Co}
\newcommand{\nifs}{$^{56}$Ni}
\newcommand{\sn}{SN~2020wnt}
\newcommand{\ha}{H$\alpha$}
\newcommand{\naid}{\ion{Na}{i}\,D\,}

\newcommand{\orcid}[1]{\href{https://orcid.org/#1}{\includegraphics[width=9pt]{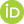}}}

\definecolor{yaleblue}{rgb}{0.1,0.3,0.9}
\definecolor{ultramarine}{rgb}{0, 0, 150}
\definecolor{bostonuniversityred}{rgb}{0.8, 0.0, 0.0}
\definecolor{lava}{rgb}{0.81, 0.06, 0.13}
\definecolor{forestgreen}{rgb}{0.0, 0.27, 0.13}
\hypersetup{colorlinks=true, linkcolor=lava, urlcolor=forestgreen, citecolor=yaleblue}

\title[SN~2020wnt: an unusual SLSN-I]{SN~2020wnt: a slow-evolving carbon-rich superluminous supernova with no \ion{O}{ii} lines and a bumpy light curve}

\author[Guti\'errez et al.]{
\parbox{\textwidth}{
\Large
C.~P.~Guti\'errez$^{\orcid{0000-0003-2375-2064},1,2,}$\thanks{E-mail: claudia.gutierrez@utu.fi},
A.~Pastorello,$^{\orcid{0000-0002-7259-4624},3}$
M.~Bersten,$^{4,5,6}$ 
S.~Benetti,$^{3}$
M.~Orellana,$^{7,8}$
A.~Fiore,$^{9,10,3}$
E.~Karamehmetoglu,$^{\orcid{0000-0001-6209-838X},11}$
T.~Kravtsov,$^{2}$
A.~Reguitti,$^{\orcid{0000-0003-4254-2724},12,13,3}$
T.~M.~Reynolds,$^{14,15}$
G.~Valerin,$^{16,3}$
P.~Mazzali,$^{17,18}$ 
M.~Sullivan,$^{\orcid{0000-0001-9053-4820},19}$ 
Y.-Z.~Cai,$^{20}$ 
N.~Elias-Rosa,$^{\orcid{0000-0002-1381-9125},3,21}$
M.~Fraser,$^{\orcid{0000-0003-2191-1674},22}$, 
E.~Y.~Hsiao,$^{\orcid{0000-0003-1039-2928},23}$
E.~Kankare,$^{\orcid{0000-0001-8257-3512},2,24}$,
R.~Kotak,$^{2}$
H.~Kuncarayakti,$^{2,1}$
Z.~Li,$^{25,26}$ 
S.~Mattila,$^{2}$
J.~Mo,$^{27}$ 
S.~Moran,$^{2}$
P.~Ochner,$^{3,16}$
M.~Shahbandeh,$^{\orcid{0000-0002-9301-5302},23}$
L.~Tomasella,$^{\orcid{0000-0002-3697-2616},3}$
X.~Wang,$^{20,27}$ 
S.~Yan,$^{20}$ 
J.~Zhang,$^{28,29}$  
T.~Zhang,$^{25,26}$ 
M.~D.~Stritzinger$^{\orcid{0000-0002-5571-1833},11}$\\
\textit{Affiliations are listed at end of paper}
}
}

\date{Accepted XXX. Received YYY; in original form ZZZ}

\pubyear{2022}

\begin{document}
\label{firstpage}
\pagerange{\pageref{firstpage}--\pageref{lastpage}}
\maketitle

\begin{abstract}
We present the analysis of \sn, an unusual hydrogen-poor super-luminous supernova (SLSN-I), at a redshift of 0.032. The light curves of \sn\ are characterised by an early bump lasting $\sim5$ days, followed by a bright main peak. The SN reaches a peak absolute magnitude of M$_{r}^{max}=-20.52\pm0.03$ mag at $\sim77.5$ days from explosion. This magnitude is at the lower end of the luminosity distribution of SLSNe-I, but the rise-time is one of the longest reported to date. Unlike other SLSNe-I, the spectra of \sn\ do not show \ion{O}{ii}, but strong lines of \ion{C}{ii} and \ion{Si}{ii} are detected. Spectroscopically, \sn\ resembles the Type Ic SN~2007gr, but its evolution is significantly slower. Comparing the bolometric light curve to hydrodynamical models, we find that \sn\ luminosity can be explained by radioactive powering. The progenitor of \sn\ is likely a massive and extended star with a pre-SN mass of 80 \Msun\ and a pre-SN radius of 15~\Rsun\ that experiences a very energetic explosion of $45\times10^{51}$ erg, producing 4 \Msun\ of \nifs. In this framework, the first peak results from a post-shock cooling phase for an extended progenitor, and the luminous main peak is due to a large nickel production. These characteristics are compatible with the pair-instability SN scenario. We note, however, that a significant contribution of interaction with circumstellar material cannot be ruled out.
\end{abstract}




\section{Introduction}

The rise of wide-field sky surveys in the last decade revealed the existence of very bright supernovae (SNe), now known as superluminous supernovae (SLSNe; \citealt{GalYam12}). SLSNe are around two orders of magnitude brighter than classical SNe \citep[$M_V<-19.5$ mag;][]{Quimby11,GalYam12,Angus19,GalYam19,Inserra19,Nicholl21}, and show a large diversity in both their light curves and spectra. 
Their host galaxies are generally found to be faint dwarf galaxies \citep{Neill11} with low metallicity ($<0.5$ Z$_\odot$) and low stellar masses \citep[e.g.][]{Stoll11,Lunnan14,Perley16,Chen17,Schulze18}. 

Initially, SLSNe were sub-classified into two classes:  hydrogen-poor (SLSN-I) and hydrogen-rich events (SLSNe-II; \citealt{GalYam12}). However, with the increase in the number of objects, especially those with better data sets, more detailed sub-classifications have been necessary. Within the SLSNe-I class, it is possible to identify two sub-groups based on their distinctive photometric properties: the slow-evolving SLSNe-I that show long rise times ($>50$ days) to the main peak, and the  fast-evolving SLSNe-I that have rise times shorter than $\sim30$ days \citep{Inserra17,Quimby18,Inserra19}.
Furthermore, \citet{Inserra18} found that slow-evolving SLSNe-I have small expansion velocities ($v\lesssim 10000$ \kms) and almost non-existent velocity gradients ($\Delta v/ \Delta t$ in units of \kms day$^{-1}$, over the time interval [+10, +30]), while the fast-evolving subgroup members have large velocities and large velocity gradients. However, the identification of SLSNe-I with intermediate (or transitional) properties (e.g. Gaia16apd; \citealt{Kangas17}; SN~2017gci; \citealt{Fiore21}), suggests a continuum distribution \citep{Nicholl15a,DeCia18}. Therefore, the bimodality or separation found by \citet{Inserra18} could be the consequence of the small sample considered (only 18 SNe). 

Studies of single objects with good photometric and spectroscopic coverage have revealed a large number of unusual properties that can give insights on the progenitor and explosion mechanisms. For instance, observations revealed pre-peak bump light curve morphologies \citep{Leloudas12, Nicholl15,Smith16,Anderson18a,Angus19}, and light curve undulations \citep{Nicholl15,Inserra17,Yan17,Fiore21}.
To explain the pre-peak bump, at least three mechanisms have been proposed. These include shock breakout within a dense
circumstellar material \citep[CSM;][]{Moriya12}, shock cooling of extended material \citep[e.g.][]{Nicholl15, Piro15,Smith16,Vreeswijk17}, and an enhanced magnetar-driven shock breakout \citep{Kasen16}. 
On the other hand, the light curve undulations have been interpreted as a signature of the interaction of the ejecta with CSM (\citealt{GalYam09a,Inserra17,Inserra19}; but see, \citealt{Moriya22}; \citealt{Kaplan20}).

Spectroscopically, the W-shaped \ion{O}{ii} features are a key characteristic of SLSNe-I  and have been recognised as such since their discovery \citep{Quimby11,Mazzali16}, although recently, it has been found that SLSN-I can be separated into two different sub-classes based on their pre-maximum spectra: events showing the W-shaped \ion{O}{ii} features, and events which do not show such W-shaped absorption \citep[e.g.][]{Konyves-Toth21}.
At early times, SLSN-I spectra are also characterised by the presence of \ion{C}{ii} \citep[e.g.][]{Dessart12b,Mazzali16,Dessart19,GalYam19} and \ion{Si}{ii}  \citep{Inserra13a}. Despite the limited number of late-time spectra available for SLSNe-I, they appear to resemble SNe~Ic associated with gamma-ray bursts \citep[e.g. SN~1998bw;][]{Nicholl16a, Jerkstrand17a}. The similarities between SLSNe-I and SNe~Ic suggest they are somewhat related \citep{Pastorello10}.    

Diverse explosion scenarios have been proposed to explain SLSNe \citep[see review of][and references therein]{Moriya18}. These include pair-instability mechanism \citep{Heger02,GalYam09a}, the interaction of the SN ejecta with CSM \citep[e.g.][]{Chevalier11,Chatzopoulos11,Ginzburg12,Dessart15,Sorokina16}, and the spin-down of a rapidly rotating, highly magnetic neutron star \citep{Kasen10, Woosley10,Bersten16}. Stars with initial masses larger than 140 \Msun~are predicted to undergo pair instability and explode completely \citep{Barkat67,Rakavy67}. The light curves of these pair-instability supernovae (PISNe) are expected to be very luminous, therefore, they have been proposed as a good alternative to explain the high luminosity, and in turn, the large amounts of synthesised \nifs\ in SLSNe. However, the light curves and spectra of some observed objects are not compatible with this scenario \citep[e.g.][]{Dessart13,Jerkstrand16,Mazzali19}.
Another scenario is the interaction of the SN ejecta with CSM produced by mass-loss of the progenitor star prior the explosion. This mechanism offers a proper explanation for a luminous and bumpy (fluctuations in brightness) light curves, but the absence of narrow emission lines in the SN spectra is currently a major issue \citep[but see,][]{Chevalier11}. The most accepted alternative of powering source of many SLSNe~I has been found in the magnetar scenario, which can explain most of the observed properties in SLSNe. However,  \citet{Soker17} found that the energy of the explosion in the magnetar model are more than what the neutrino-driven explosion mechanism  can supply, therefore, a jet feedback mechanism from jets launched at magnetar birth may be involved.

Given that several open questions remain regarding both the explosion mechanism and the progenitors of SLSNe, studying nearby SLSNe in detail allows us to discriminate among various scenarios. In this paper, we present \sn, one of the closest (z=0.032) SLSNe-I discovered to date. The excellent coverage from explosion to $\sim500$ days allows us to characterise its properties. Its light curves show an early bump, followed by a slow rise to the main peak. Some fluctuations in brightness are also observed at late time. On the other hand, unlike other SLSN objects, \sn\ spectra do not show signs of \ion{O}{ii}, but its evolution resembles that of the type Ic carbon-rich SN~2007gr \citep{Valenti08,Hunter09,Chen14}. This unusual similarity provides an excellent opportunity to understand the possible connection between H-poor SNe and SLSNe.

The paper is organised as follows. A description of the observations and data reduction is presented in Section~\ref{sec:obs}. The characterisation of \sn\ (host galaxy, photometric and spectral properties) is given in Section~\ref{sec:ana}. In Section~\ref{sec:comp}, comparisons with similar objects are presented, while in Section~\ref{sec:expl} the explosion progenitor properties are analysed through hydrodynamical modelling. Finally, in Section~\ref{sec:disc} and Section~\ref{sec:con}, we present the discussion and conclusions, respectively. Throughout this work, we will assume a flat $\Lambda$CDM universe, with a Hubble constant of $H_0=70$\,km\,s$^{-1}$\,Mpc$^{-1}$, and $\Omega_\mathrm{m}=0.3$.

\section{Observations of SN~2020wnt}
\label{sec:obs}

\subsection{Detection and classification}

\sn\ (a.k.a. ZTF20acjeflr and ATLAS20beko) was detected by the Zwicky Transient Facility \citep[ZTF;][]{Bellm19, Graham19} on 2020 October 14 (MJD$=59136.40$), at a magnitude m$_g=19.70\pm0.11$ mag. A couple of hours later (MJD$=59136.47$), a detection in the $r-$band confirmed the new object (m$_r=19.57\pm0.05$ mag). The discovery was reported to the Transient Name Server (TNS\footnote{\url{https://wis-tns.weizmann.ac.i}}) by the Automatic Learning for the Rapid Classification of Events (ALeRCE) broker \citep{Forster21} on MJD$=59136.79$. \sn\ was also detected by the Asteroid Terrestrial-impact Last Alert System (ATLAS; \citealt{Tonry18, Smith20}) on 2020 October 14 (MJD$=59136.50$; m$_c=19.70\pm0.11$ mag). The last non-detection obtained by ZTF was on 2020 October 12 (MJD$=59134.45$) with a detection limit of m$_r\sim20.70$ mag. A deeper non-detection in the $g$ band ($\sim21.00$ mag) occurred earlier the same night (MJD$=59134.39$). 
Using these constraints, we adopt the mid-point between the last non-detection and first detection as the explosion epoch (MJD$=59135.42\pm0.98$; 2020 October 13).
\sn\ was spectroscopically observed on 2020 November 15 (MJD$=59168.0$) by the UC Santa Cruz group and classified as a SN~I at a redshift of 0.032 \citep{Tinyanont20}.

\subsection{Photometry}
\label{slc}

\sn\ was observed photometrically for 72 weeks, from 2020 October 14 to 2022 February 27, using various facilities. Most of the observations were carried out by two wide-field imaging surveys, namely ATLAS and ZTF. From 2020 October 14 to 2021 November 4, photometry in the orange ($o$) filter (a red filter that covers a wavelength range of 5600 to 8200 \AA) and cyan ($c$) filter (wavelength range 4200 to 6500 \AA) was obtained by the twin 0.5 m ATLAS telescope system \citep{Tonry18}. ATLAS photometry (\citealt{Tonry18} and \citealt{Smith20}) was obtained through the ATLAS forced photometry server\footnote{\url{https://fallingstar-data.com/forcedphot/}}. ZTF obtained $g-$ and $r-$band images from 2020 October 14 to 2021 November 9. The ZTF photometry was obtained through the ZTF forced-photometry service \citep{Masci19}. The light curves were generated following the steps presented in the ZTF documentation\footnote{\url{https://irsa.ipac.caltech.edu/data/ZTF/docs/forcedphot.pdf}}.

Optical imaging was obtained with the Copernico 1.82 m telescope equipped with Asiago Faint Object Spectrograph and Camera (AFOSC) and the 67/91 Schmidt Telescope equipped with Moravian G4-16000LC at the Asiago Observatory (Italy); the 0.8-m Tsinghua University–NAOC (National Astronomical Observatories of China) Telescope (TNT) at Xinglong Observatory of NAOC \citep{Huang12}; the 2-m Liverpool Telescope (LT) using the IO:O imager, the Low-Resolution Spectrograph (LRS) at the 3.6-m Telescopio Nazionale Galileo (TNG), and the 2.56-m Nordic Optical Telescope (NOT) using the Alhambra Faint Object Spectrograph and Camera (ALFOSC) at the Roque de Los Muchachos Observatory (Spain). 10 epochs of near-infrared (NIR; $JHK$) photometry were obtained with NOTCam at NOT, while six epochs of UltraViolet (UV) and Optical observations were obtained with the UltraViolet/Optical Telescope (UVOT) onboard the Neil Gehrels \textit{Swift} Observatory spacecraft. All NOT observations were obtained through the NOT Unbiased Transient Survey 2 (NUTS2\footnote{\url{https://nuts.sn.ie}}) allocated time.

Data reduction and SN photometry measurements for Asiago, LT and NOT were performed using the \textsc{python/pyraf} SNOoPY pipeline \citep{Cappellaro14}, whereas the TNT and TNG images were reduced with \textsc{iraf} following standard procedures. The photometry for TNT and TNG was performed using the \textsc{python} package \textsc{photutils} \citep{Bradley19} of \textsc{astropy} \citep{Astropy18}. All $ugriz$ magnitudes were calibrated using observations of local Sloan and Pan-STARRS sequences \citep{Chambers16,Magnier20}. The $BV$ magnitudes were derived using Pan-STARRS and the transformations in \citet{Chonis08}, while the $JHK$ magnitudes were calibrated using 2MASS \citep{Skrutskie06}. UVOT reductions and the resulting photometry were performed by using the \textsc{heasoft} Software\footnote{\url{http://heasarc.gsfc.nasa.gov/ftools}} \citep{Swift14} and aperture photometry. 

Optical ($uBgVriz$), NIR and UVOT photometry are presented in Tables~A1, A2 and A3, respectively. The mean magnitudes from ATLAS are listed in Table~A4, while the $gr$ photometry from ZTF is in Table~A5.

\subsection{Spectroscopy}
\label{sspect}

26 optical spectra of \sn\ were obtained spanning phases between 32 and 293 days from explosion. These observations were acquired with seven different instruments: ALFOSC at the Nordic Optical Telescope (NOT), AFOSC at the Copernico 1.82-m Telescope (Mount Ekar); LRS at the Nazionale Galileo (TNG), Yunnan Faint Object Spectrograph and Camera (YFOSC) at the Lijiang 2.4-m Telescope (LJT), Beijing Faint Object Spectrograph and Camera (BFOSC) at the Xinglong 2.16-m Telescope (XLT), Optical System for Imaging and low-Intermediate-Resolution Integrated Spectroscopy (OSIRIS) at the 10.4-m Gran Telescopio Canarias (GTC), and Kast double spectrograph on the 3.0-m Shane telescope at the Lick Observatory\footnote{Public spectrum obtained from the TNS webpage: \url{https://www.wis-tns.org/object/2020wnt}}. All spectra  were  reduced using standard \textsc{iraf} routines (bias subtraction, flat-field correction, 1D extraction, and wavelength calibration).  The flux calibration was performed using spectra of standard stars obtained during the same night.
For the ALFOSC, AFOSC and OSIRIS spectra, the data were reduced using the \textsc{foscgui}\footnote{FOSCGUI is a graphical user interface aimed at extracting SN spectroscopy and photometry obtained with FOSC-like instruments. It was developed by E. Cappellaro. A package description can be found at \url{sngroup.oapd.inaf.it/foscgui.html.}} pipeline.

Additionally, a NIR spectrum was obtained at $\sim49$ days from explosion with the 0.7-5.3 Micron Medium-Resolution Spectrograph and Imager (SpeX instrument) on the 3.2-m NASA Infrared Telescope Facility. The spectrum was taken in cross-dispersed SXD mode with the 0.5 arcsec slit, and reduced using the \textsc{spextool} software package \citep{Cushing04} following the prescriptions described by \citet{Hsiao19}.
Details of the instruments used for the spectroscopic observations are reported in Table~A6. All spectra will be available through the WISeREP\footnote{\url{http://wiserep.weizmann.ac.il/home}} archive \citep{Yaron12}.

\section{Characterising SN~2020wnt}
\label{sec:ana}

\subsection{Host galaxy}
\label{sec:gal}

\begin{figure}
\centering
\includegraphics[width=0.85\columnwidth]{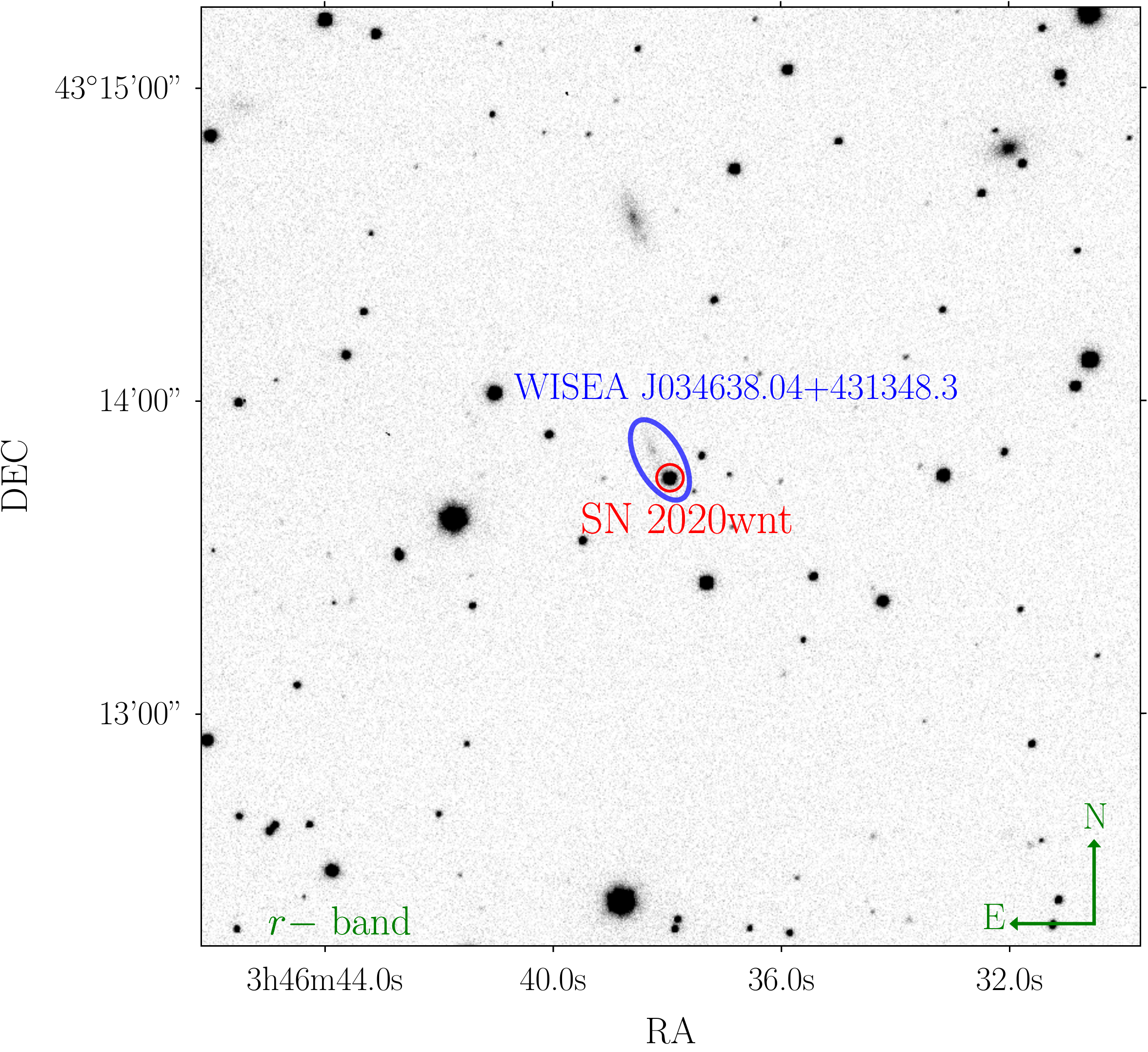}
\caption{NOT $r-$band image of \sn\ and its host galaxy, WISEA J034638.04+431348.3 The red circle marks the SN ($\textrm{RA}=03^{\rm h}46^{\rm m}37.\!\!^{\rm{s}}95$ $\textrm{Dec}=+43^{\circ}13'45.\!\!''30$ (J2000)), while the blue ellipse marks its host galaxy. The orientation of the image is indicated in the bottom-right
corner.
}
\label{fcsn20wnt}
\end{figure}

The host galaxy of \sn\ is WISEA J034638.04+431348.3, a faint galaxy with no published redshift or distance information. The redshift adopted in our analysis is derived from the narrow emission lines (H$\alpha$, \ion{[O}{iii]} $\lambda5007$) visible in the SN spectrum. These lines give us a mean redshift of 0.032. Given the lack of independent measurements of distance to this galaxy, we estimate the uncertainty in our measurements assuming a peculiar velocity of $\pm200$ km s$^{-1}$ \citep{Tully13}. With these values, we compute a distance of $d=140.4\pm3.0$ Mpc, which corresponds to a distance modulus of $\mu=35.74\pm0.05$ mag. Figure~\ref{fcsn20wnt} shows the NOT $r-$band image of \sn\ and its host galaxy.

The Galactic reddening is quite high with $E(B-V)=0.42$ mag \citep{Schlafly11}, while the host galaxy component is negligible. This is determined by the absence of narrow interstellar \naid lines ($\lambda\lambda5889$, 5895) at the rest wavelength of the host. As shown in section~\ref{sec:spec}, the spectra of \sn\ display a strong \naid line, but it corresponds to the Milky Way component. Therefore, we assume that the reddening in the  direction of \sn\ is totally dominated by the Milky Way.

To characterise the global properties of the \sn\ host galaxy, we got $griz$ photometry from the Pan-STARRS1 (PS1) public science archive\footnote{\url{https://catalogs.mast.stsci.edu/panstarrs/}}. To obtain estimates of the stellar mass ($M_*$) and star formation rate (SFR), we use a custom galaxy spectral energy distribution (SED) fitting code, following the procedure detailed in \citet{Sullivan10}.
The code is similar to \textsc{z-peg} \citep{LeBorgne02}, but uses the stellar population templates of \textsc{p\'{e}gase}.2 \citep{Fioc97}.  
The best-fitting templates correspond to $M_* = \log \left(M/M_{\odot}\right) = 8.22_{-0.12}^{+0.20}$ and $\log\left(\mathrm{SFR}\right) = -4.89_{-5.11}^{+3.09}$ \Msun yr$^{-1}$. The large uncertainties obtained for the SFR are mainly associated to the lack of photometry in bluer bands \citep{Childress13a}. Such low stellar mass and star formation obtained for the host galaxy of \sn\ are consistent with the expected range measured for other SLSN hosts \citep{Neill11,Chen13,Lunnan14,Leloudas15,Perley16,Schulze18}. 
In absence of a host spectrum, we could infer the metallicity through the mass -- metallicity relation \citep[e.g][]{Tremonti04}. Following the prescriptions of \citet{Kewley08}, we derive a value that corresponds to 12 + log(O/H)$=8.17\pm0.11$ dex in the O3N2 calibration and  and 12 + log(O/H)$=8.18\pm0.09$ dex in N2 calibration, respectively. These values suggest a low metallicity and are consistent with previous findings for SLSNe \citep[e.g.,][]{Chen17}.

\subsection{Light curves}
\label{sec:lc}

\begin{figure*}
\centering
\includegraphics[width=0.85\textwidth]{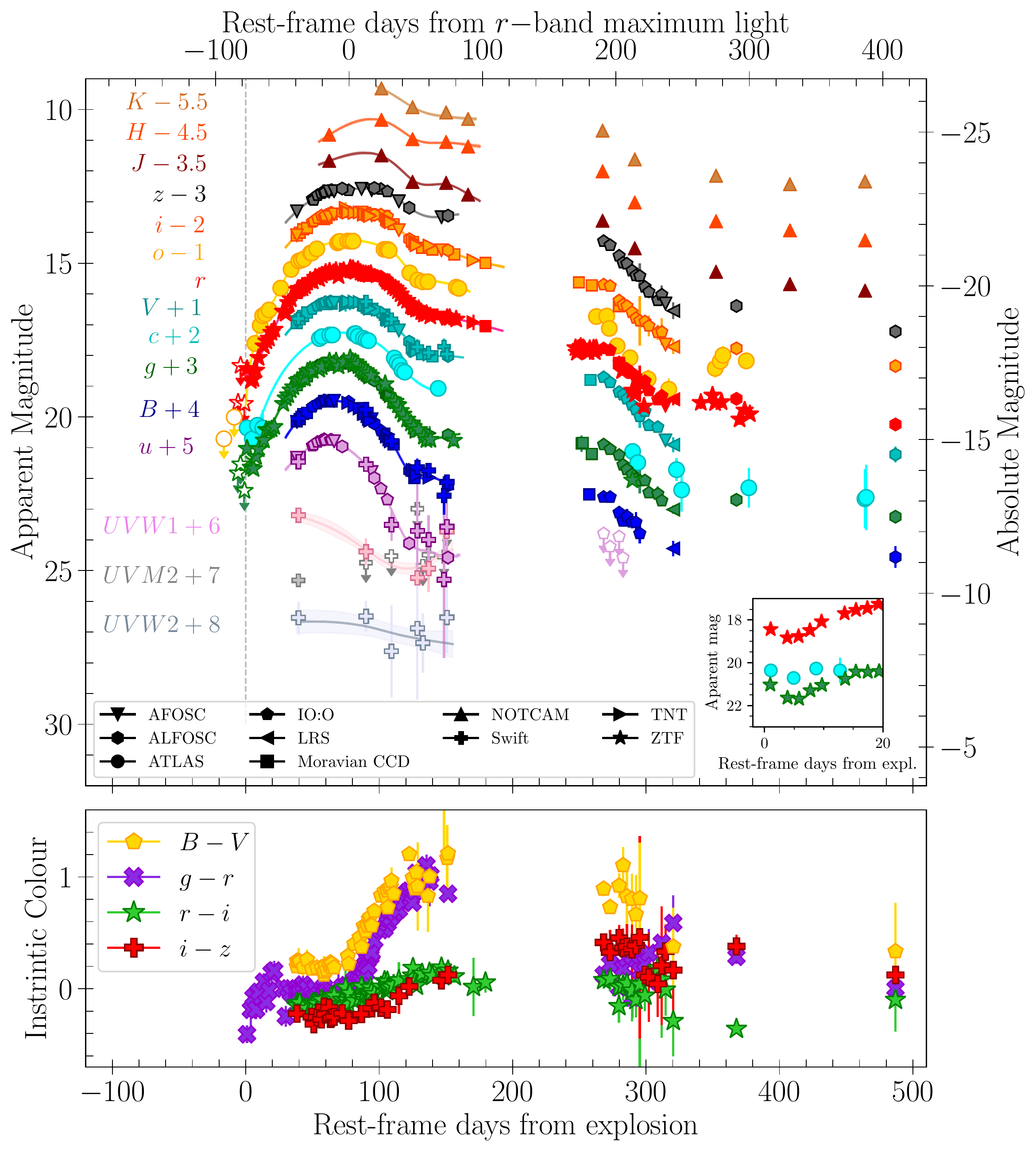}
\caption{\textbf{Top:} UV, optical and NIR light curves of SN~2020wnt. Upper limits are presented as open symbols. The explosion time is indicated as a vertical dashed line. The photometry is corrected for Milky Way extinction using the \citet{Cardelli89} extinction law. The solid lines show the Gaussian process (GP) interpolation before 180 days, and the shaded regions represent the errors from the GP. UV and $BVJHK$ photometry are in the Vega system, while $ugcroiz$ photometry is in the AB system. The inset plot shows the light curves in $gcr$ at very early phases. 
\textbf{Bottom:} Intrinsic colour curves of \sn.
}
\label{lcs}
\end{figure*}

Figure~\ref{lcs} (top panel) shows the rest-frame multiband light curves of \sn. The excellent photometric coverage during the first $\sim320$ days from explosion allows us to constrain exceptionally well the light curve shape, magnitudes at maximum, rise times, and decline rates in the different bands. To estimate the main parameters of the light curves, we use Gaussian processes (GPs). For this procedure, we use the \textsc{python} package \textsc{george} \citep{Ambikasaran16}, following the prescriptions of \citet{Gutierrez20}. 

As seen in the top panel of Figure~\ref{lcs}, \sn\ evolves quite slowly and shows several distinctive properties during its evolution. The dense sampling in three of the four filters with very early observations ($gcr$) allows us to detect an initial peak, which is brighter in the bluer filters with an absolute magnitude of M$_g=-17.71$ mag, M$_c=-17.38$ mag, M$_r=-17.30$ mag. 

Following this initial peak, the $gcr$ light curves show a decrease in brightness (between 0.5 in $r$ and 0.7 mag in $g$). At this point, the SN reaches a minimum value, with absolute magnitudes of  M$_g=-17.00$ mag, M$_c=-17.02$ mag, and M$_r=-16.8$ mag at $\sim4.9$ days. After this phase, a rise of $\sim3.3-3.7$ mag is observed in all bands. A high cadence follow-up in the $uBgVriz$ bands starts after about 38 days from the explosion. This permits to cover the SN maximum in all optical bands.

\begin{table}
\begin{center}
\caption{Light curve parameters of \sn.}
\setlength{\tabcolsep}{3pt}
\label{LCpar}
\begin{tabular}{ccccccccc}
\hline
\hline
     & Peak Abs. Mag   &   Rise time    &  $\Delta$M($_{\text{Peak}-130})$ & $\Delta$M($_{77.5-130})$ \\ 
Band & (mag)$^{\star}$ &(days)$^{\star}$&    (mag)$^{\dagger}$      &       (mag)              \\
(1)  &      (2)        &      (3)       &         (4)               &         (5)              \\
\hline                                                                                          
\hline                                                                                          
$u$  & $-20.00\pm0.02$ & $61.3\pm1.4$   &          3.58             &          3.25           \\
$B$  & $-20.27\pm0.01$ & $65.0\pm2.2$   &          2.49             &          2.45           \\
$g$  & $-20.52\pm0.03$ & $72.0\pm2.5$   &          2.27             &          2.25           \\
$c$  & $-20.50\pm0.03$ & $70.2\pm1.4$   &          1.67             &          1.62           \\
$V$  & $-20.50\pm0.02$ & $72.4\pm3.1$   &          1.69             &          1.65           \\
$r$  & $-20.52\pm0.03$ & $77.5\pm3.1$   &          1.32             &          1.32           \\
$o$  & $-20.49\pm0.01$ & $75.0\pm1.8$   &          1.29             &          1.29           \\
$i$  & $-20.40\pm0.02$ & $76.8\pm5.1$   &          1.09             &          1.08           \\
$z$  & $-20.16\pm0.01$ & $79.0\pm4.1$   &          0.81             &          0.81           \\
$J$  & $-20.84\pm0.10$ & $89.0\pm5.2$   &          1.01             &          0.94           \\
$H$  & $-20.93\pm0.08$ & $93.8\pm5.1$   &          0.72             &          0.55           \\
$K$  & $-20.92\pm0.03$ & $101.7\pm1.0^{\ddagger}$  &    0.71        &        \nodata          \\
\hline
\end{tabular}
\begin{list}{}{}
\item \textbf{Columns:} (1) Band; (2) Peak absolute magnitudes; (3) Rise time; (4) Change in magnitude from the peak to 130 d from explosion; (5) Change in magnitude from 77.5 to 130 d from explosion. 
\item $^{\star}$ Peak absolute magnitudes and rise times were obtained from Gaussian Process (GP) fits. Magnitudes are corrected by the Milky Way extinction.
\item $^{\dagger}$ Change in magnitude from the peak. Note that each band reaches the peak at different times.
\item $^{\ddagger}$ Due to the lack of data in $K$, the first point is assumed as the maximum and the uncertainty of the maximum time is the error from explosion epoch. 
\end{list}
\end{center}
\end{table}

From the GP fits, we find that \sn\ reaches the main peak in the optical bands between $\sim61$ and 79 days from explosion (all phases stated in this paper are in the rest frame). The rise times are different in each band, with a faster rise in $u$ and a more extended rise in the NIR bands. The absolute magnitude peaks are around $-20.00$ mag in $u$ and $-20.93$ mag in $H$. In $gcVro$ the absolute peak magnitudes are around $-20.50$ mag. These values place \sn\ at the bottom of the luminosity distribution of SLSNe-I \citep[e.g.][]{Angus19}. Table~\ref{LCpar} shows the absolute peak magnitudes and the rest-frame rise times obtained in all optical and NIR bands. 

One interesting characteristic observed in the light curves of \sn\ is the behaviour around peak. In the redder bands, the SN reaches the maximum a bit later than in the blue bands, and the luminosity stays quasi-constant for a longer time, displaying a kind of "plateau". After maximum, the decline in the blue bands is much faster than in the redder ones. From the main peak to 130 days, the SN dims by $\sim3.58$ mag in the $u$ band, $2.27$ mag in $g$ and  $1.32$ mag in $r$ (almost three times slower than $u$). As the $u$ peak occurs before than the peak in $g$ and $r$ (61.3, 72.0, and 77.5 days in $u$, $g$ and $r$, respectively), we can better compare these decreases by fixing a range of time. Measuring the change in magnitude from 77.5 days (the epoch of $r-$band peak) to 130 days, we see that the SN dims by $\sim3.26$ mag in $u$ and $2.25$ mag in $g$. The decline rates in these two ranges are also presented in Table~\ref{LCpar}.
Moreover, in the $UVW1$ and $UVW2$ bands, we see a flattening or upturn, which may be attributable to the known red leak\footnote{SWIFT $UVW1$ and $UVW2$ filters have extended red tails that reach into the optical. When the flux is optically dominated, the UV contribution can be minimal (\url{https://swift.gsfc.nasa.gov/analysis/uvot_digest/redleak.html})}. Meanwhile, for the $UVM2$ filter, which is not affected by the red leak, we only measure upper limits. Therefore, the evolution of the Swift UV bands indicates a decline in the UV flux and the emergence of an optically dominated  spectral energy distribution (SED) ${\sim}100$ days post explosion.

After $\sim130$ days, the drop in brightness slows down. Fitting a line to the observations obtained between 130 and 180 days (the last observation before the SN went behind the sun), we measure a slope of $1.01\pm0.03$ mag per 100 days in $r$, and $1.18\pm0.01$ per 100 days in $i$. The monitoring of \sn\ restarted at 247 days in $r$, and a couple of days later in $BgVcizJHK$. With the new data, we again fit a line to the data between 130 and 275 days and we found slower declines, with slopes of $0.90\pm0.01$ mag per 100 days in $r$ and $0.99\pm0.06$ mag per 100 days in $i$. These values are very close to those expected from the \cofs\ decay (0.98 mag per 100 days; \citealt{Woosley89}). Starting from 273 days, the SN is found to experience a drastic and sudden drop in brightness in all bands. Within $\sim35$ days, the magnitudes decrease by about $1.5$ mag. Fitting a line after 273 days in the $r-$band, we find a slope of $4.5\pm0.3$ mag per 100 days. From $\sim320$ days, fluctuations in brightness are observed both in the optical and the NIR bands. 

The intrinsic colour curves of \sn\ are presented in the bottom panel of Figure~\ref{lcs}. During the first $\sim35$ - 40 days, we can only infer $g-r$ colour information. In $\sim20$ days, \sn\ becomes redder, going from a $g-r=-0.41$ mag to $g-r=0.16$ mag. The $g-r$ colour shows an initial peak at 21.3 days. After this, the SN goes back to bluer colours, reaching a value of $g-r=-0.09$ mag at 42.5 days. From this epoch, the $g-r$ colour shows a quasi flat evolution up to $\sim70$ days. Later than 70 days, the SN becomes redder again, reaching its main peak at 135.4 days with a colour of $g-r=1.10$ mag. A gap in the observations prevented us from monitoring the SN evolution between 151 and 251 days, however, when the SN is recovered, we measure a colour of $g-r=0.10$ mag showing that \sn\ became bluer again, but after this, it gets redder one more time. 

Starting from $\sim38$ days, we also have the $B-V$, $r-i$ and $i-z$ colours. $B-V$ shows a similar behaviour than that observed in $g-r$, but with redder colours at all phases. Unlike $g-r$ and $B-V$, the evolution of  $r-i$ and $i-z$ shows little variation. Overall, they tend to be bluer up to $\sim150$ days. Of these two, the colour change is more significant in $i-z$, going from $-0.22$ to $0.12$ mag (at $\sim150$ days) in comparison to the evolution from $-0.13$ to $0.16$ mag in $r-i$. After the gap, the SN gets bluer in $B-V$, $r-i$ and $i-z$. This tendency is clear until 320 days. From there, the temporal coverage does not allow us to estimate a trend.

\subsection{Bolometric light curve}
\label{sec:bolo}

\begin{figure}
\centering
\includegraphics[width=\columnwidth]{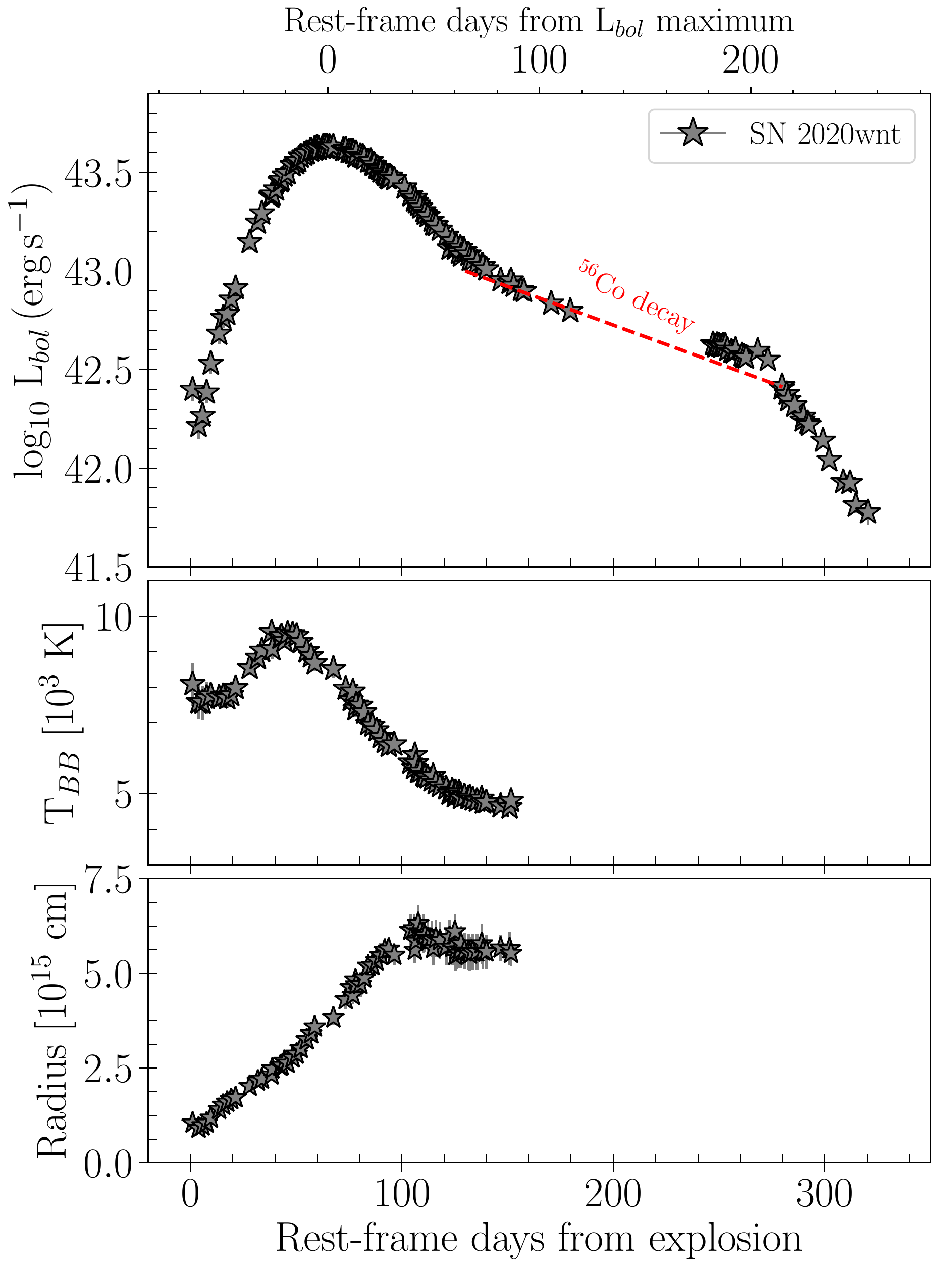}
\caption{\textbf{Top:} Bolometric light curve of \sn. The dashed red line shows the luminosity expected from \cofs\ decay (assuming full trapping). \textbf{Middle:} Temperature evolution of  \sn. \textbf{Bottom:} Evolution of blackbody radius of \sn. Error bars are comparable to the size of the symbols.
}
\label{bolo}
\end{figure}

To build the bolometric light curve of \sn\ using our reddening corrected photometry from UV to NIR bands, we employ the \textsc{superbol} code\footnote{\url{https://github.com/mnicholl/superbol/}} \citep{Nicholl18}. In order to have a similar coverage in the different bands at the same epochs, we interpolate and extrapolate the light curves assuming constant colours and using the $r-$band as a reference filter from explosion to 320 days after the explosion. We converted all magnitudes to flux and construct the SED at all  epochs. 
We computed multiple pseudo-bolometric light curves by performing trapezoidal integration just in the optical and NIR bands, and UV + optical + NIR. We also calculated a full bolometric light curve by fitting a blackbody to the SED. 
When comparing the bolometric light curve from UV to NIR with that obtained by extrapolating the SED constructed from the optical and NIR bands, we find that they are consistent. The bolometric light curve is presented in the top panel of Figure~\ref{bolo}.

Following the process described in Section~\ref{sec:lc}, we use a GP to estimate the main parameters in the bolometric light curve. 
We find a peak luminosity of $L_{bol} = 4.25 (\pm 0.30) \times 10^{43}$
erg s$^{-1}$ at 65 days. This maximum occurs earlier than most of the peaks obtained from the optical and NIR bands, except for the $uB$ filters. Between 140 and 270 days, the light curve declines at a rate of $0.77\pm0.02$ mag per 100 days. After that, we estimate a decline rate of $4.07\pm0.01$ mag per 100 days. 

From \textsc{superbol}, the blackbody temperature (T$_{BB}$) and radius (R$_{BB}$) are also obtained by fitting the SED of each epoch with a blackbody function. Figure~\ref{bolo} (middle and bottom panels) shows the temperature and radius evolution from explosion to $\sim150$ days. At early times the temperature is relatively low, with a value of T$_{BB}\approx8000$ K. The temperature increases and reaches a value of T$_{BB}\approx10000$ K at $\sim51$ days from explosion, and then it decreases. On the other hand, the radius shows a continuous increase up to 105 days, where it reaches its maximum value (R$_{BB}=6.2\times10^{15}$ cm). Fitting a line between explosion and the maximum value, we find a slope of $\sim6000$ \kms.
After this peak, the radius shows a slow and steady decline to $5.5\times10^{15}$ cm.

\subsection{Spectral evolution}
\label{sec:spec}

\begin{figure*}
\centering
\includegraphics[width=0.75\textwidth]{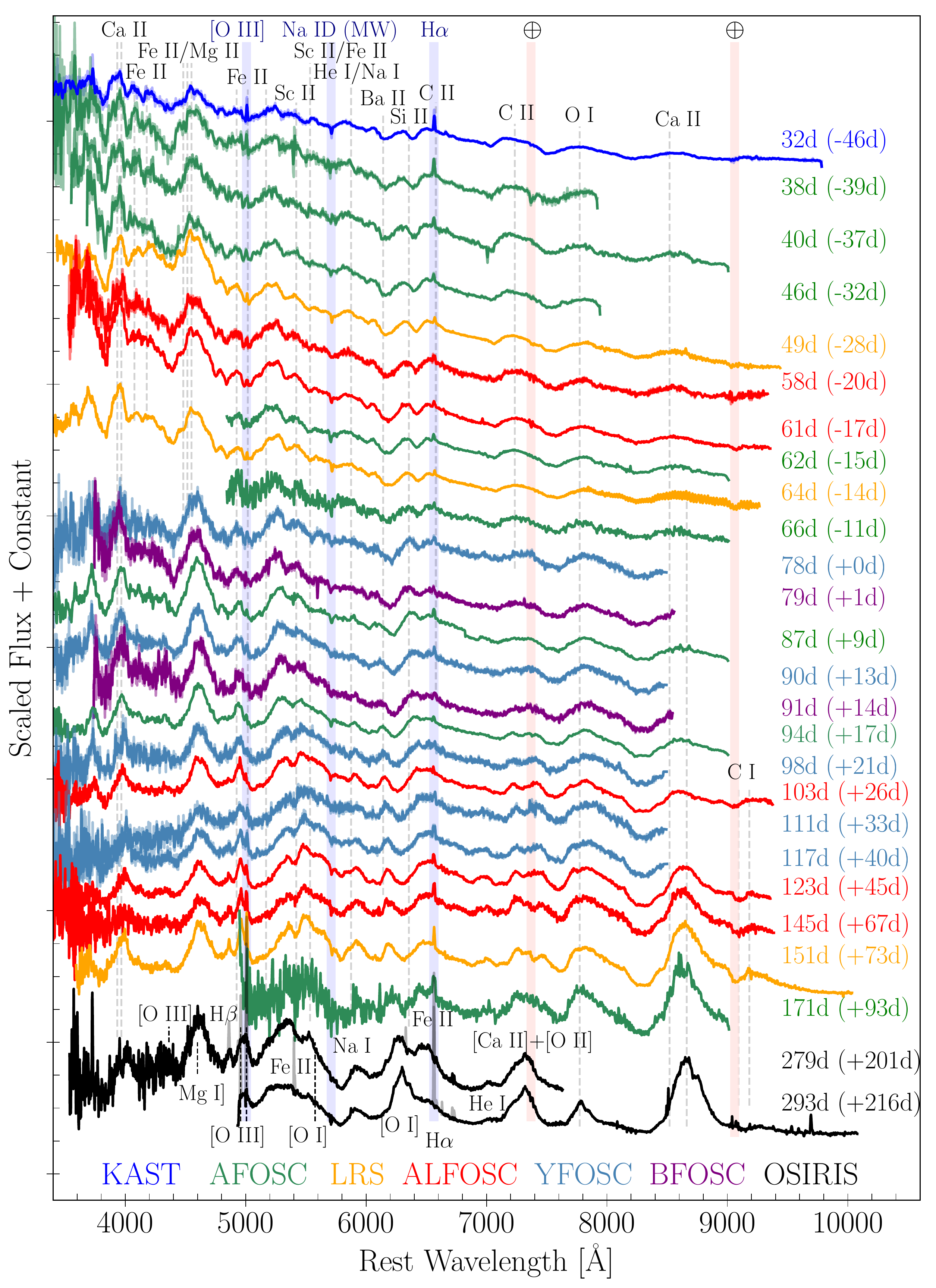}
\caption{Spectral sequence of \sn\ from 32 to 293 days from explosion in the rest-frame. The phases are labelled on the right. The numbers in parentheses are the phases with respect to the maximum light in the $r-$band. Each spectrum has been corrected for Milky Way (MW) reddening and shifted vertically by an arbitrary amount for presentation. The colour of the spectra represents the different instruments used to obtain the data. The vertical dashed lines indicate the rest position of the strongest lines, vertical blue lines indicate emission lines from the host galaxy (\ion{[O}{iii]} and \ha) and the narrow \naid interstellar feature from the MW, and the pink lines indicates the position of the telluric absorption ($\oplus$ symbol). In the last two spectra, narrow emission lines from the host galaxy (\ha , H$\beta$, \ion{[O}{iii]}, \ion{[N}{ii]}, and \ion{[S}{ii]}, visible in grey) were removed for presentation.
}
\label{spec}
\end{figure*}

Figure~\ref{spec} shows the spectral evolution of \sn\, covering the phases from 32 to 293 days after explosion. The slow evolution detected in the light curves is also visible in the spectra, where a blue continuum is observed for around 100 days. The first spectrum taken at 32 days ($-46$ d from the maximum light in the $r$ band) and used for the classification, is dominated by strong lines of \ion{O}{i} $\lambda7774$, \ion{Ca}{ii} (H\&K and NIR triplet), \ion{Si}{ii} $\lambda6355$, and \ion{C}{ii} $\lambda6580$, $\lambda7235$. \naid/\ion{He}{i} and the \ion{Fe}{ii} $\lambda4924$, 5018, 5169 lines are also clearly detected. From 32 to 78 days after explosion, there are limited changes in the spectra, with small variations in the relative line intensities. More precisely, \ion{C}{i} $\lambda6580$ and  $\lambda7235$ and \naid/\ion{He}{i} become weaker, while the \ion{Fe}{ii} $\lambda4924$, 5018, 5169 lines become stronger. During this period, the spectra do not show signs of \ion{O}{ii} lines. After 78 days, the \naid/\ion{He}{i} line vanishes while the flux in the bluer part of the spectra shows a significant decrease, mainly due to the line blanketing. After this epoch, a 'W'-shape profile is visible around 4800 \AA. This feature has been previously detected in several SNe~Ic (e.g. SN~2004aw; \citealt{Taubenberger06}; SN~2007gr; \citealt{Valenti08a,Hunter09,Chen14}) and SLSNe (e.g. SN~2015bn; \citealt{Nicholl16}). 

\begin{figure}
\centering
\includegraphics[width=\columnwidth]{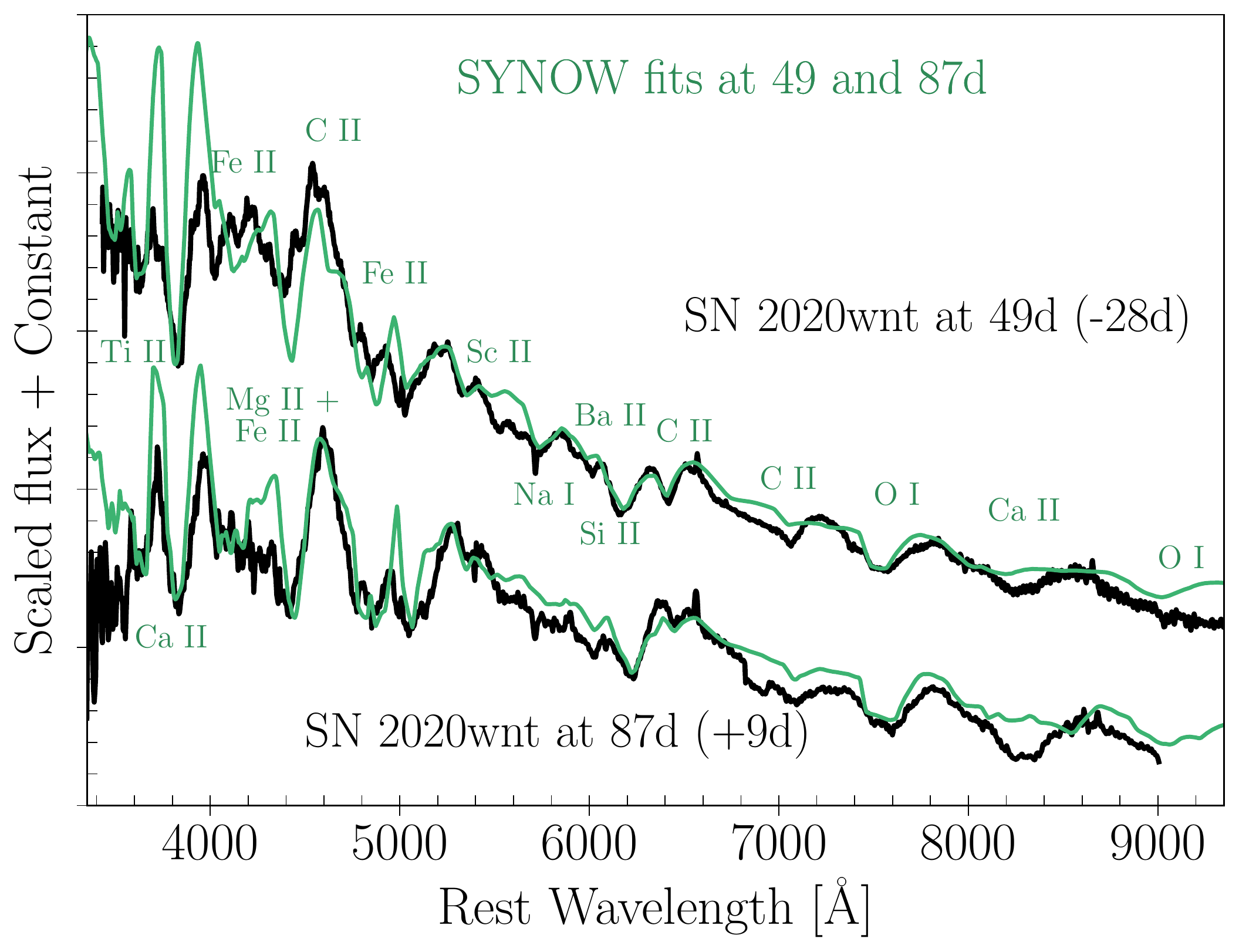}
\caption{Spectral comparison of \sn\ at 49 and 87 days from explosion ($-28$ and $+9$ from maximum, respectively) and the \textsc{synow} fits. The  \textsc{synow} synthetic spectra (green) are overplotted on the observed spectra (black).
}
\label{synow}
\end{figure}

To support our preliminary line identification in \sn, we employ the \textsc{synow} code \citep{Fisher00} and the best quality spectra at 49 days ($-28$ days from the $r-$band maximum) and 87 days ($+9$ days from the $r-$band maximum). For the spectrum before peak, we assume a blackbody temperature of $T_\text{bb}=10800$ K and a photospheric velocity of $v_\text{ph}=8000$ \kms, while for the spectrum after peak, we use a $T_\text{bb}=8000$ K and a photospheric velocity of $v_\text{ph}=6000$ \kms. To reproduce the observed features in both spectra, we include the lines of  \ion{Ca}{ii}, \ion{O}{i}, \ion{Si}{ii}, \ion{C}{ii}, \ion{Na}{i}, \ion{Fe}{ii}, \ion{Sc}{ii}, \ion{Ba}{ii}, \ion{Mg}{ii} and \ion{Ti}{ii}. As shown in Figure~\ref{synow}, the synthetic spectra at 49 and 87 days reproduce relatively well the observed features of \sn, allowing us to confirm the presence of \ion{C}{ii} and \ion{Si}{ii}. 

Returning to the spectroscopic evolution, we see that from day 87 (+9 days from maximum), the \ion{Ca}{ii} NIR triplet and \ion{O}{i} become stronger, and the region below 5500 \AA\ is almost entirely dominated by the iron-group lines. At 103 days, the \ion{C}{ii} $\lambda6580$ and $\lambda7235$ lines disappear while the continuum becomes redder. We detect a feature at $\sim9000$ \AA\ that is possibly due to \ion{C}{i} $\lambda9183$. As the temperature decreases, \ion{C}{i} lines start to be detected. 

After 123 days (+45 from the peak), \sn\ starts the transition to the nebular phase. This is indicated by the \ion{Ca}{ii} NIR triplet, which shows signs of an emission component. From 123 to 171 days, the spectral evolution is slow. The most significant difference is the strengthening of the emission component in the \ion{Ca}{ii} NIR triplet, as well as in the lines detected at $\sim4600$ \AA, $\sim6000$ \AA, and $\sim7300$ \AA.

\begin{figure}
\centering
\includegraphics[width=\columnwidth]{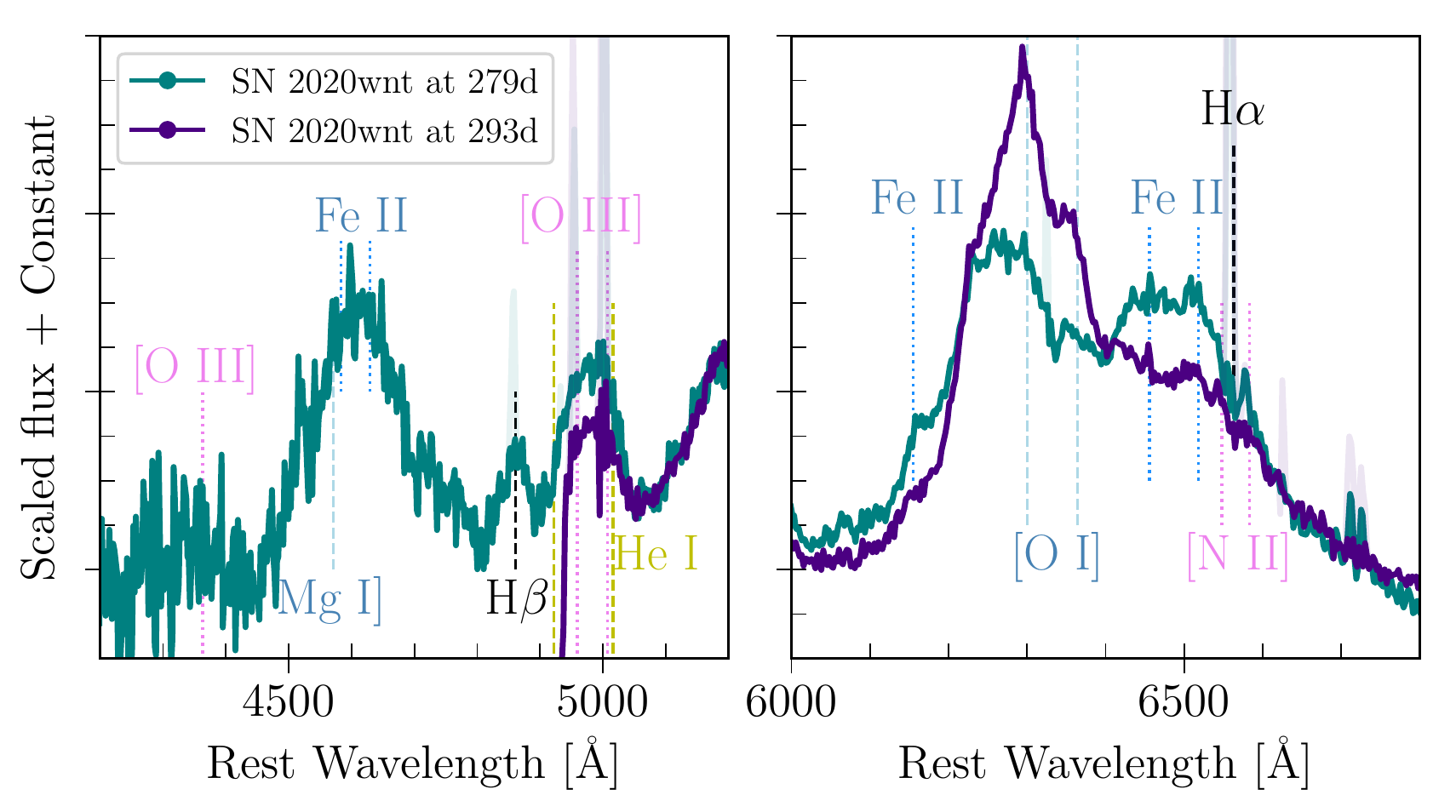}
\caption{Observed emission profiles of \sn\ at 279 and 293 days from explosion. Vertical lines indicate the rest position of the strongest lines. \textbf{Left panel:} Emission profiles between 4200 and 5200 \AA. \textbf{Right panel:} Emission profiles between 6000 and 6800 \AA.}
\label{nebH}
\end{figure}

The last two spectra of \sn\ were obtained at 279 and 293 days ($+201$ and $+216$ days from the peak), once the SN returned from behind the Sun. At 279 days, the spectrum shows emission lines of \ion{[Ca}{ii]} $\lambda\lambda7291$, 7324 with the possible contribution of \ion{[O}{ii]} $\lambda\lambda7320$, 7330), \ion{[O}{i]} $\lambda\lambda6300$, 6364, \ion{[O}{i]} $\lambda5577$, \ion{Na}{i}, \ion{Mg}{i]} $\lambda4571$ plus \ion{Fe}{ii}, a weak feature near $\sim7100$ \AA, possibly caused by \ion{He}{i} $\lambda7065$, and a broad emission at $\sim5000$, which could be identified as either the broad \ion{[O}{iii]} $\lambda4959$, $\lambda5007$ components, \ion{Fe}{ii}, \ion{He}{i} or \ion{[Fe}{ii]}. We explored all these identification scenarios and concluded that this feature is probably caused by \ion{[O}{iii]} $\lambda4959$, $\lambda5007$ with some contribution of \ion{He}{i} (see Figure~\ref{nebH}). Despite the bluer part of the spectrum is a bit noisy, we clearly see an emission line produced by \ion{Ca}{ii} H\&K. An excess around 4300 \AA\ is  noticed and could be consistent with \ion{[O}{iii]} $\lambda4363$. 

One remarkable characteristic in this phase is the identification of an emission line on the red side of \ion{[O}{i]} $\lambda\lambda6300$, 6364, which could be caused either by H$\alpha$,  \ion{Fe}{ii} $\lambda6456$, 6518 or \ion{[N}{ii]} $\lambda\lambda6548$, 6583. Fitting a Gaussian, we find that the line is centred at $\sim6480$ \AA. If this feature is caused entirely by either H$\alpha$ or \ion{[N}{ii]}, a significant blueshift ($>3000$ \kms) has to be taken into account, while for \ion{Fe}{ii}, the wavelength match is satisfactory. The possible detection of \ion{Fe}{ii} $\lambda6155$ could support this hypothesis. However, a broad line at the position of H$\beta$ is also found, which suggests that H$\alpha$ should contribute to the boxy feature. Therefore, this boxy emission line is probably a result of blended lines of \ion{Fe}{ii}, H$\alpha$, and probably \ion{[N}{ii]}. A zoom around the \ha\ and H$\beta$ regions is presented in Figure~\ref{nebH}.

The last spectrum, at 293 days, displays the same lines as the spectrum at 279 days (from 5000 \AA) plus emission features of the \ion{Ca}{ii} NIR triplet and \ion{O}{i} $\lambda7774$ (detected thanks to the larger wavelength coverage). With the identification of these features, we deduce that the strongest line at this phase is the \ion{Ca}{ii} NIR triplet. From 279 to 293 days, the spectra display two major changes, related to the decrease in the intensity of the lines near 5300, 6155 \AA\ (\ion{Fe}{ii}) and 6500 \AA\ (the boxy emission line). The consistent decrease in the strength of these lines supports the idea that the \ion{Fe}{ii} $\lambda6456$, 6518 lines should contribute to the line emission observed on the red side of \ion{[O}{i]} $\lambda\lambda6300$, 6364.

\begin{figure}
\centering
\includegraphics[width=\columnwidth]{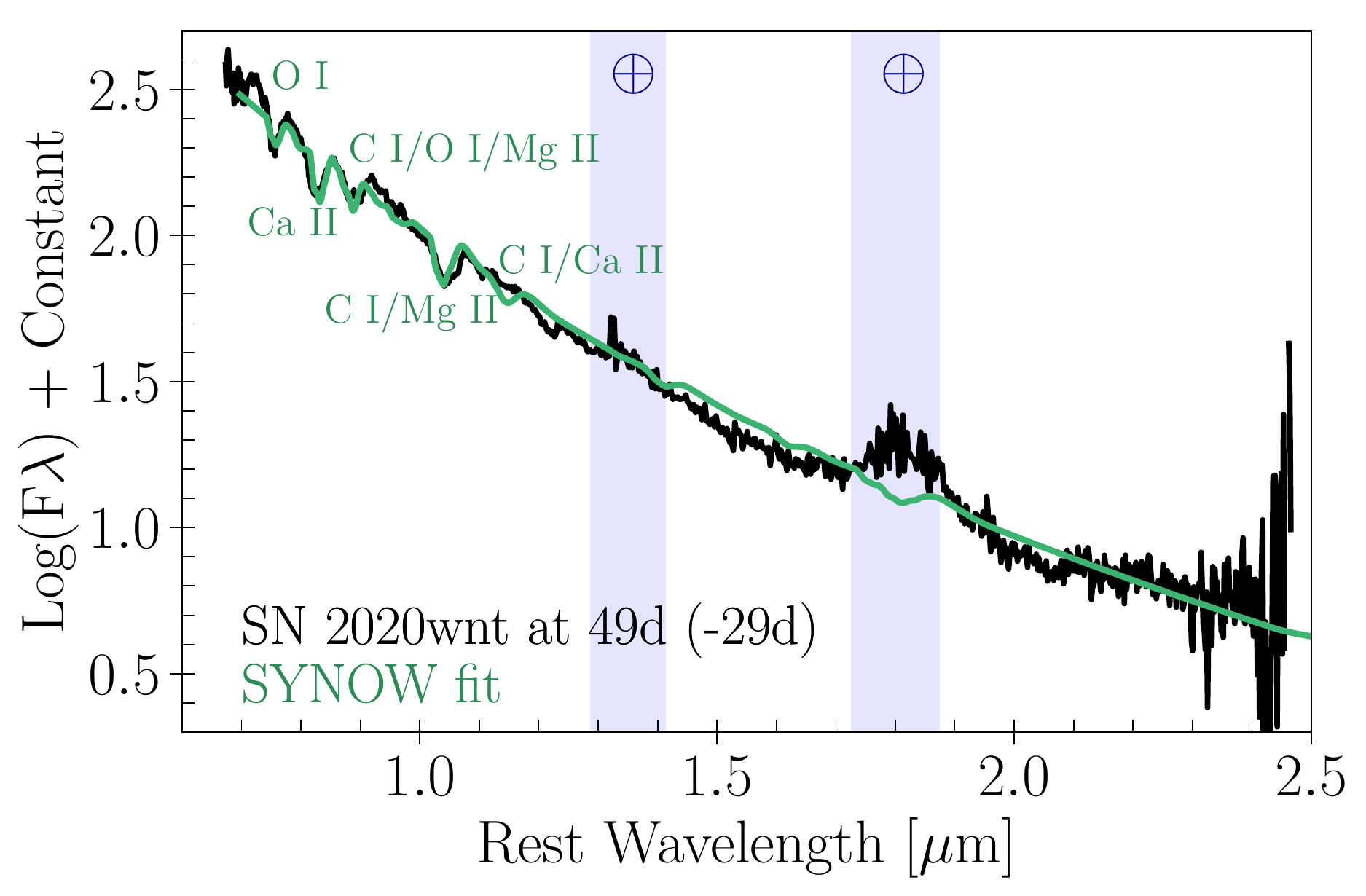}
\caption{NIR spectrum of SN~2020wnt (black) obtained at $\sim49$ days from explosion ($-29$ days from peak) with SpeX compared with the \textsc{synow} fit. Regions of strong telluric absorption are masked in blue.}
\label{nir}
\end{figure}

Figure~\ref{nir} shows a NIR spectrum obtained at $\sim49$ days from explosion ($-29$ days from peak), covering a wavelength range between 0.63 and 2.5 $\mu$m. At this phase, the spectrum displays a blue continuum with clear features of \ion{O}{I}, \ion{Ca}{II}, previously identified in the optical spectra, plus multiple features of \ion{C}{I}. We also detect an absorption line at $\sim1.05$ $\mu$m that could be either \ion{He}{i} 1.083 $\mu$m, \ion{C}{i} 1.069 $\mu$m, or even a mix of them. If this line is due to \ion{He}{i}, we would also see a clear line detection near 2.058 $\mu$m; however, this is not the case. The lack of this line suggests that the contribution of \ion{He}{i} to the 1.05 $\mu$m absorption is small. Similarly, the absorption features at 1.26 and 2.12 $\mu$m could be also produced by \ion{C}{i}. There are several carbon lines near those wavelengths (see \citealt{Millard99,Valenti08a,Hunter09}). 

To identify the ions that produce the lines in the NIR, we create a synthetic spectrum with \textsc{synow} by using the same parameters as mentioned before. We consider a blackbody temperature of $T_\text{bb}=10800$ K and a photospheric velocity of $v_\text{ph}=8000$ \kms. We reproduce the observed spectrum only including \ion{Ca}{ii}, \ion{O}{i}, \ion{C}{i} and \ion{Mg}{ii}, as shown in Figure~\ref{nir}, but beyond 1.8~$\mu$m none of the ions seems to produce  absorption lines. The identification of C lines in the NIR reinforces the idea that \sn\ has a carbon-rich progenitor.

\subsection{Expansion velocities}
\label{sec:vel}

\begin{figure}
\centering
\includegraphics[width=\columnwidth]{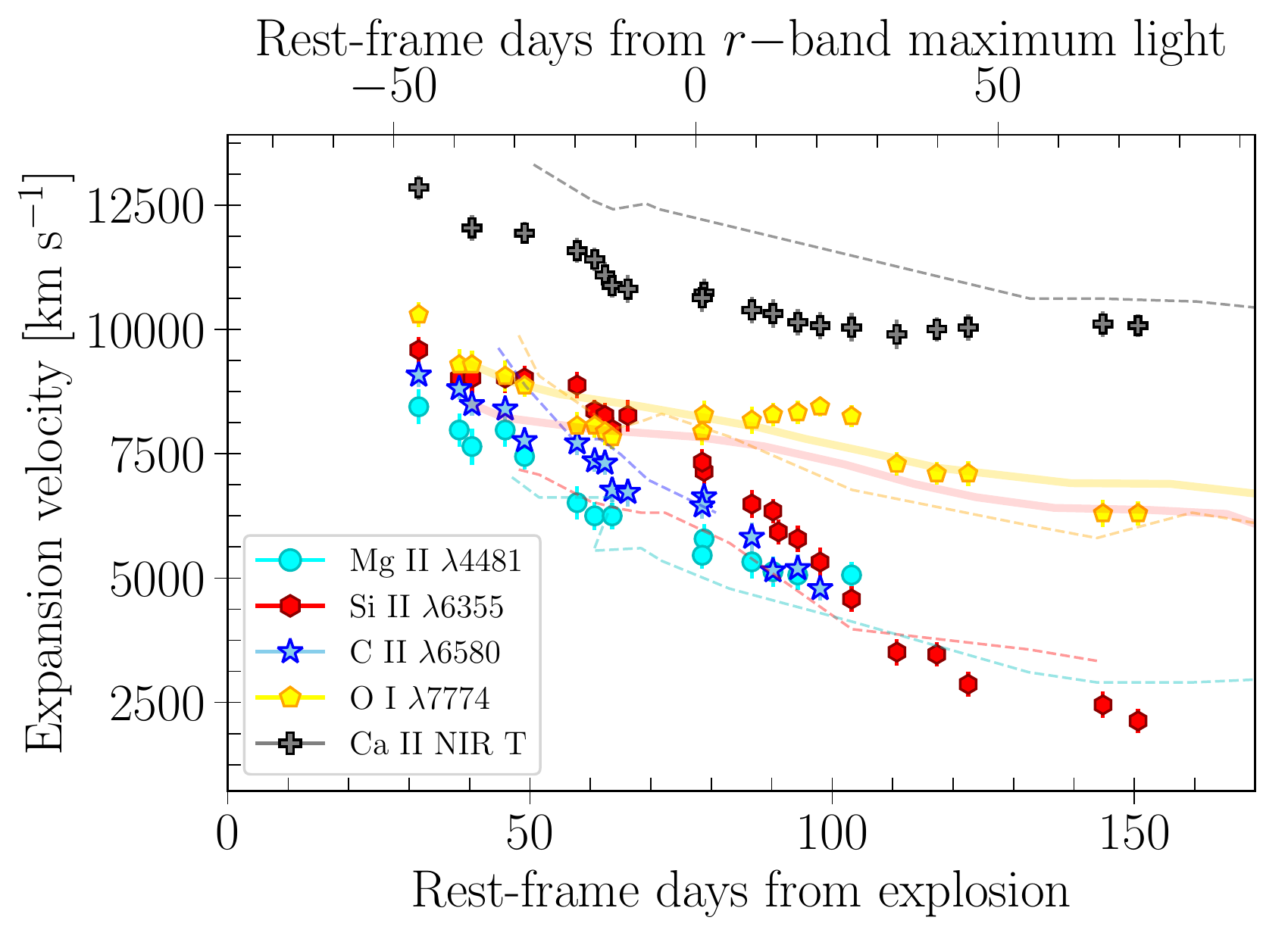}
\caption{Evolution of expansion velocities derived from the minimum flux of several absorption lines of \sn\ \citep[see][for details in the estimation of velocities and their uncertainties]{Gutierrez17a}. For comparison, we also include the \ion{Si}{ii} and \ion{O}{i} velocities of the SLSN 2015bn (solid lines), and the velocities of the carbon-rich type Ic SN~2007gr (dashed lines). For this object, we stretched the time by a factor of 4 ($t=4\times t_0$). This factor is obtained from the light curve analysis (see Section~\ref{sec:sncomp}).
}
\label{vel}
\end{figure}

We measure the expansion velocities of the ejecta for five spectral lines (\ion{Mg}{ii}, \ion{Si}{ii}, \ion{C}{ii}, \ion{O}{i} and \ion{Ca}{ii} NIR triplet) using the spectra covering the phases from 32 and 151 days from explosion. These velocities were obtained from the minimum flux of the absorption component and are presented in Figure~\ref{vel}. In this analysis, we do not include the iron lines (\ion{Fe}{ii} $\lambda4924$, 5018, 5169 \AA) due to complications in estimating their velocities as these lines seem to be blended with other ions. Figure~\ref{vel} shows that the evolution of the expansion velocities can be split into two groups: one characterised by an initial decrease that then flattens, and a second group that shows a monotonic decline with time. In the first group, we find \ion{Ca}{ii} NIR triplet and \ion{O}{i}, whereas \ion{Mg}{ii}, \ion{Si}{ii} and \ion{C}{ii} are found in the second group. In addition to this behaviour, we also see that the first group shows higher velocities, suggesting that the \ion{Ca}{ii} NIR triplet and \ion{O}{i} lines mostly form in the outer part of the ejecta, while the \ion{Mg}{ii}, \ion{Si}{ii} and \ion{C}{ii} lines form in the inner layers. The \ion{Ca}{ii} NIR triplet has the highest expansion velocities, decreasing from $\sim12800$ \kms\ at 32 days to $\sim10100$ \kms\ at 151 days. On the other hand, \ion{Si}{ii} decreases from 9600 \kms\ to just 2100 \kms. 

For comparison, in Figure~\ref{vel} we include the expansion velocities of the carbon-rich type Ic SN~2007gr \citep{Hunter09} corrected by a temporal factor of $\sim4$ in order to match the overall evolution of \sn\ (See Section~\ref{sec:sncomp}), and the \ion{Si}{ii} and \ion{O}{i} velocities of the SLSN 2015bn \citep{Nicholl16}. Overall, the velocities of \sn\ and SN~2007gr are comparable, except for that inferred from the \ion{Ca}{ii} NIR triplet, which has higher values in SN~2007gr.  When comparing \sn\ with SN~2015bn, we see that the \ion{O}{i} velocities are very similar, but the \ion{Si}{ii} velocities show a completely different behaviour. While the velocities of SN~2015bn shows a slow drop, in \sn\ we notice a more rapid decrease. The velocity values for all objects are low compared to those observed in typical SNe~Ibc \citep[e.g.][]{Prentice18} and SLSNe, respectively. Similarly, low velocities were also found for SN 2007gr \citep{Hunter09} and SN 2015bn \citep{Nicholl16}.

\section{Comparison with other SNe}
\label{sec:comp}

\sn\ is a hybrid object sharing the properties of both SLSN and SN~Ic classes. The light-curve morphology is comparable to that observed in several SLSNe (pre-peak bumps, long rise to the main peak, slow evolution); with absolute magnitude at peak within the luminosity distribution of SLSNe-I, although at the lower end \citep{DeCia18,Angus19}. Despite this, \sn\ is brighter ($\sim-20.5$ mag) compared to standard SNe~Ic (peak absolute magnitudes ranging between $-17$ and $-18$ mag; e.g. \citealt{Taddia18}), but lies in the luminosity range studied by \citet{Gomez22} for luminous SNe. On the other hand, the spectra are more similar to the type Ic class than to SLSNe. This similarity is founded on the absence of the \ion{O}{ii} lines (one of the features characterising SLSNe), and the strength of different lines, such as the \ion{Ca}{ii} NIR triplet, \ion{Si}{ii} and \ion{C}{ii} lines. A major difference between \sn\ and SNe~Ic is the slow evolution observed in the spectra, which is consistent with SLSNe-I. Given these hybrid properties, we compare \sn\ with both SLSNe and SNe Ic. These are well observed slow-evolving SLSNe, SLSNe with pre-peak bumps, SLSNe with \ion{C}{ii} lines in their spectra, and two very well sampled type Ic SNe. Details of the comparison sample are presented in Table~\ref{compprop}.

\begin{table}
\begin{center}
\caption{Detailed properties of the comparison sample.}
\label{compprop}
\setlength{\tabcolsep}{2pt}
\begin{tabular}{ccccccccc}
\hline
\hline
SN         & Redshift & E(B$-$V)$_{MW}$ &  Characteristics$^{\star}$   &  References$^{\dagger}$    \\
           &          &     (mag)     &                    &                \\ 
\hline                                                                      
\hline       
\multicolumn{5}{c}{\textbf{SLSNe}}\\
SN~2006oz  &  0.376   &     0.042     &     Bumpy          &     (1)        \\ 
SN~2007bi  &  0.127   &     0.028     &     Slow           &  (2), (3)      \\
PTF12dam   &  0.107   &     0.012     & Slow; \ion{C}{ii}  & (4), (5), (6)  \\ 
LSQ14an    &  0.1637  &     0.074     &     Slow           &     (7)        \\
LSQ14bdq   &  0.345   &     0.056     &     Bumpy          &     (8)        \\ 
DES14X3taz &  0.608   &     0.022     &     Bumpy          &     (9)        \\ 
SN~2015bn  &  0.1     &     0.022     &     Slow           &   (10), (11)   \\
DES15S2nr  &  0.22    &     0.030     &     Bumpy          &    (12)        \\ 
SN~2017gci &  0.0873  &     0.116     & Slow; \ion{C}{ii}  &    (13)        \\ 
DES17X1amf &  0.92    &     0.022     &     Bumpy          &    (12)        \\ 
SN~2018bsz &  0.0267  &     0.214     & Bumpy; \ion{C}{ii} &(14), (15), (16)\\ 
\hline                                                                      
\hline 
\multicolumn{5}{c}{\textbf{SNe~Ic}}\\
SN~2004aw  &  0.0163  & 0.022   &     \nodata        &    (17)        \\ 
SN~2007gr  &  0.0017  & 0.055   &   \ion{C}{ii}      &(18), (19), (20)\\ 
\hline
\end{tabular}
\begin{list}{}{}
\item $^\star$\textbf{Characteristics:} \textit{Slow:} Slow-evolving SLSNe; \textit{Bumpy:} SLSNe with pre-peak bumps; \textit{\ion{C}{ii}:} \ion{C}{ii} lines in the spectra.\\
\item $^{\dagger}$\textbf{References:} (1) \citet{Leloudas12}; (2) \citet{GalYam09a}; (3) \citet{Young10}; (4) \citet{Nicholl13}; (5) \citet{Chen15}; (6) \citet{Vreeswijk17}; (7) \citet{Inserra17}; (8) \citep{Nicholl15}; (9) \citet{Smith16}; (10) \citet{Nicholl16}; (11) \citet{Nicholl16a}; (12) \citet{Angus19}; (13) \citet{Fiore21}; (14) \citet{Anderson18a}; (15) \citet{Chen21}; (16) \citet{Pursiainen22}; (17) \citet{Taubenberger06}; (18) \citet{Valenti08a}; (19) \citet{Hunter09}; (20) \citet{Chen14}. 
\end{list}
\end{center}
\end{table}

\subsection{Light curve comparison}
\label{sec:lccom}

\begin{figure}
\centering
\includegraphics[width=\columnwidth]{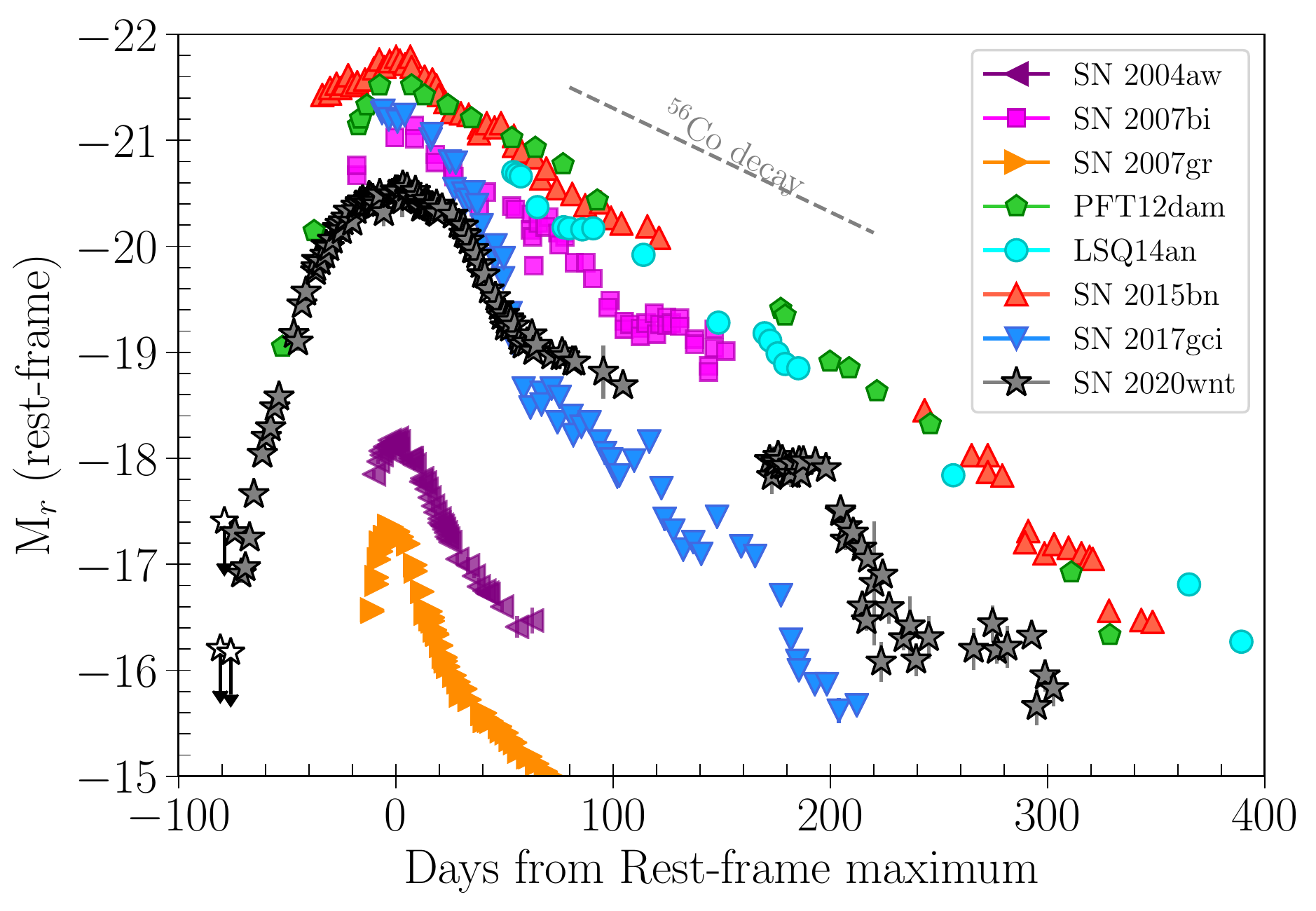}
\caption{Comparison of the $r-$band light curves of \sn\ with the slow-evolving SLSNe SN~2007bi, PTF12dam, LSQ14an SN~2015bn, and SN~2017gci, and the type Ic SN~2004aw and SN~2007gr (given as reference). All SLSNe have been K-corrected to rest-frame. Only corrections for Milky Way extinction have been applied.  
}
\label{compLC}
\end{figure}

Figure~\ref{compLC} shows the $r-$band absolute light curve of \sn\ compared to five well-sampled slow-evolving SLSNe-I and two normal SNe~Ic. From the comparison, the evolution of \sn\ is similar to that observed in SN~2017gci (this object has characteristics of both slow- and fast-evolving SLSNe-I; \citealt{Fiore21}). Excluding the type Ic SNe~2004aw and 2007gr, which are evidently faint, \sn\ is the faintest object in the (SLSN-I) sample. It is $>1$ mag fainter than the brightest object, SN~2015bn. Inspecting the evolution at the early phases, we notice incomplete information for the SLSN-I sample. Therefore, it is hard to know the rise-time duration and how these objects evolve at very early phases, i.e., if they show an initial peak (bump) or not. The only objects with an early detection were PTF12dam and SN~2015bn. For SN~2015bn, \citet{Nicholl16} estimated a rise time of 79 days, while for \sn, we estimate a rise time of $\sim77.5$ days in $r$. These rise times are among the longest presented to date.

Analysing the shape of the light curve, we see that \sn\ and SN~2017gci have a  similar behaviour. After maximum, the decline slope of both objects changes at $50-55$ days from peak. Later, a shoulder is observed. For \sn, after $\sim+50$ days from peak, the decline rate is comparable to that expected from the \cofs\ decay (see Section \ref{sec:lc}). This evolution is observed until $\sim+200$ days from peak. After this, the slope changes again, showing a very fast linear decline.

\begin{figure}
\centering
\includegraphics[width=\columnwidth]{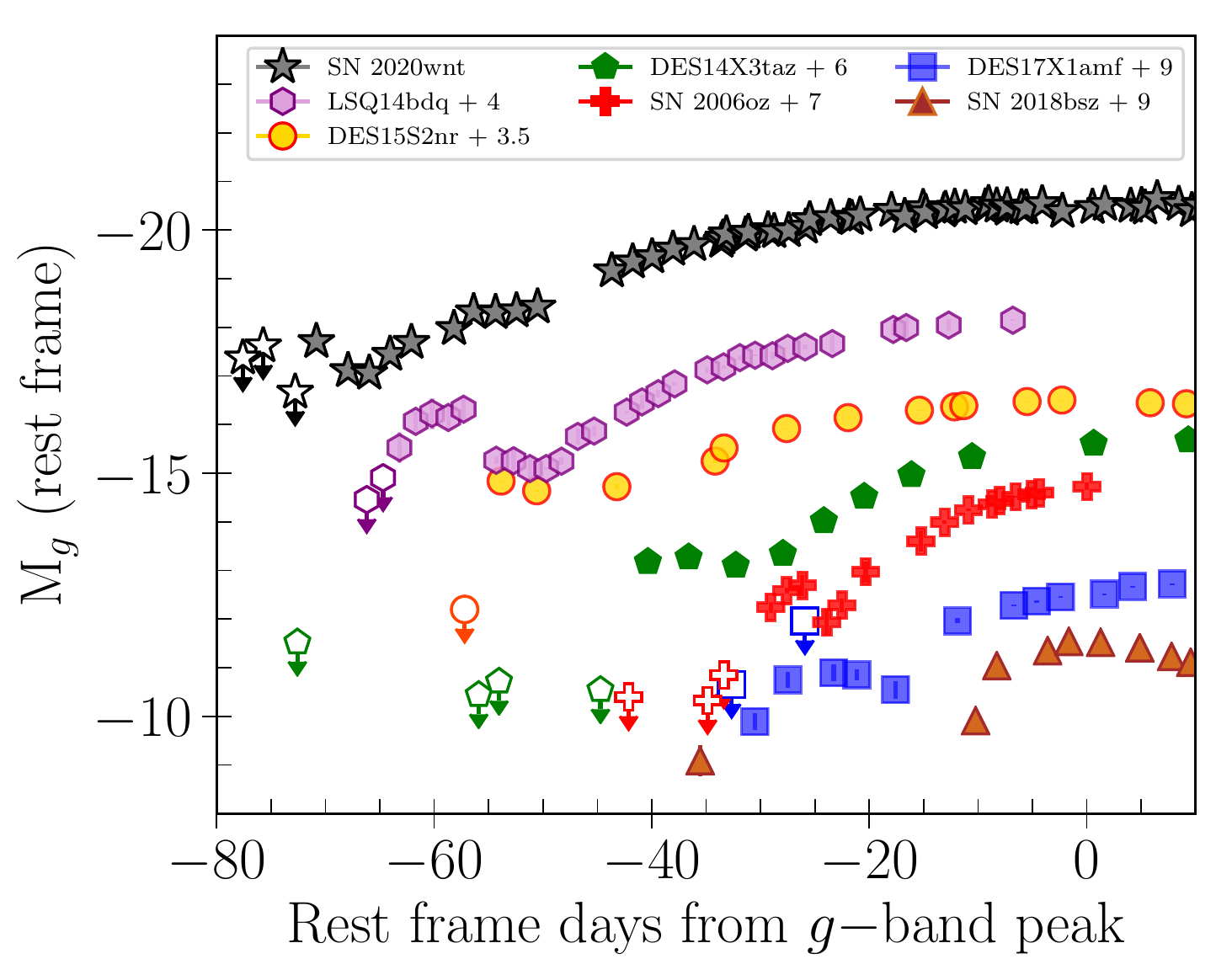}
\caption{Rest-frame, $g-$band light curve of SN~2020wnt compared with SLSNe showing pre-peak bumps: SN~2006oz, LSQ14bdq, DES14X3taz, DES15S2nr, DES17X1amf, and SN~2018bsz. }
\label{com_bumps}
\end{figure}

As mentioned in Section~\ref{sec:lc}, the $gcr-$band light curves of \sn\ have an initial bump with a relatively short duration. 
To analyse the pre-peak bump in \sn, we compare the $g-$band early light curve with a sample of SLSNe that show signs of an early bump. This comparison is presented in Figure~\ref{com_bumps}. To easily examine these objects, we arranged the light curves in terms of the rise-time to the main peak. \sn\ has the longest rise time ($\sim72$ days), while SN~2018bsz has the shortest one (around 10 days). An opposite behaviour is observed in the duration of the initial peak. Here, \sn\ shows the shortest bump with a duration $<5$ days, while the longest initial bump is observed for SN~2018bsz ($>25$ days; see \citealt{Anderson18a} for better constraints in other bands). 
In terms of luminosity, we find that LSQ14bdq is the most luminous object, DES15S2nr is the least luminous, followed by \sn\ and SN~2018bsz, which have similar absolute magnitudes at peak.

\subsection{Spectral comparison}
\label{sec:speccomp}

\begin{figure}
\centering
\includegraphics[width=\columnwidth]{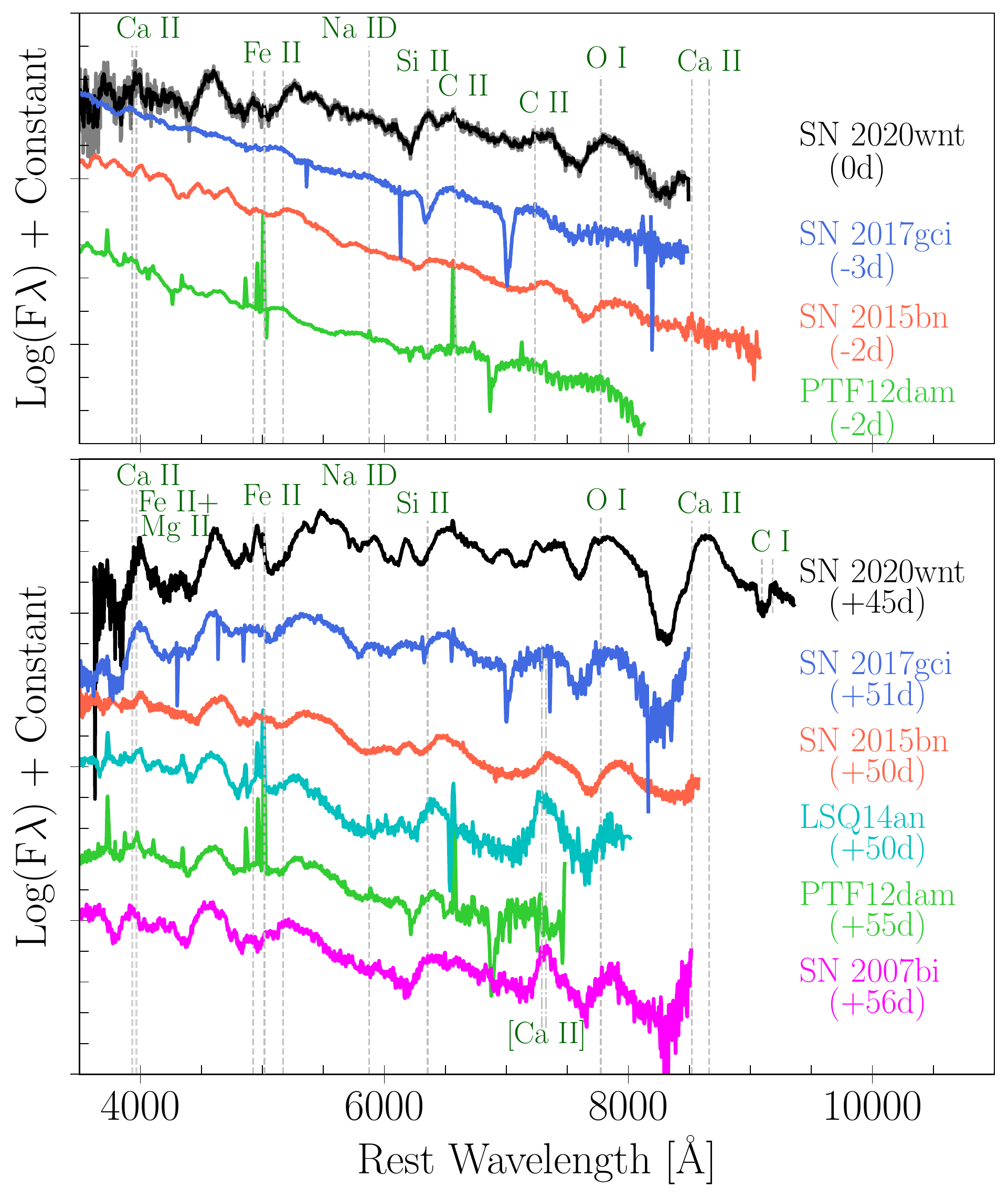}
\caption{\sn\ at around maximum light \textbf{(top) }and at $\sim+50$ days from $r-$band maximum \textbf{(bottom)} compared with well-sampled slow-evolving SLSNe:  SN~2007bi, PTF12dam, LSQ14an, SN~2015bn, and SN~2017gci. Each spectrum has been corrected for Milky Way reddening and shifted vertically by an arbitrary amount for presentation. Vertical lines indicate the rest position of the strongest lines. The phases and SN names are labeled on the right.
}
\label{compslow}
\end{figure}

The spectral comparison of \sn\ and slow-evolving SLSNe-I is presented in Figure~\ref{compslow}. To examine the similarities and differences between these objects, we select two reference epochs: around peak (top) and $\sim+50$ from maximum light (bottom). Firstly, we notice a large diversity. Around peak, \sn\ has a distinct spectrum, dominated by strong absorption lines. In particular, the W-shape profile due to Fe lines is a remarkable characteristic. Though the spectral coverage does not allow to see the full profile of the \ion{Ca}{ii} NIR triplet, based on the spectra before and after peak (Figure~\ref{spec}), this line is intense in \sn, but it seems to be absent in the other objects. Only a few common features are identified among \sn\ and the comparison sample: \ion{C}{ii} lines with SN~2017gci, and \ion{Si}{ii} and \ion{O}{i} lines with SN~2015bn.

At around $+50$ days from maximum light, the spectra are still quite heterogeneous, although \sn\ and SN~2017gci look more alike than before. This is seen in the bluer part, with the blended feature due to the \ion{Mg}{ii} and \ion{Fe}{ii} lines, and in the redder part, with the detection of strong features of \ion{O}{i} and \ion{Ca}{ii} NIR triplet. Both \sn\ and SN~2015bn still share similar profiles of \ion{Si}{ii} and \ion{O}{i}. \ion{Fe}{ii} and \ion{O}{i} are visible in all objects. In contrast to the photometric behaviour, the spectrum of \sn\ seems to evolve slower than that of SN~2007bi and LSQ14an. At the later phase, these objects show clear signs of nebular lines (e.g. \ion{[Ca}{ii]} + \ion{[O}{ii]}), while \sn\ still shows lines of the photospheric phase.

\subsection{Comparison with SN~2015bn and SN~2007gr}
\label{sec:sncomp}

\begin{figure}
\centering
\includegraphics[width=\columnwidth]{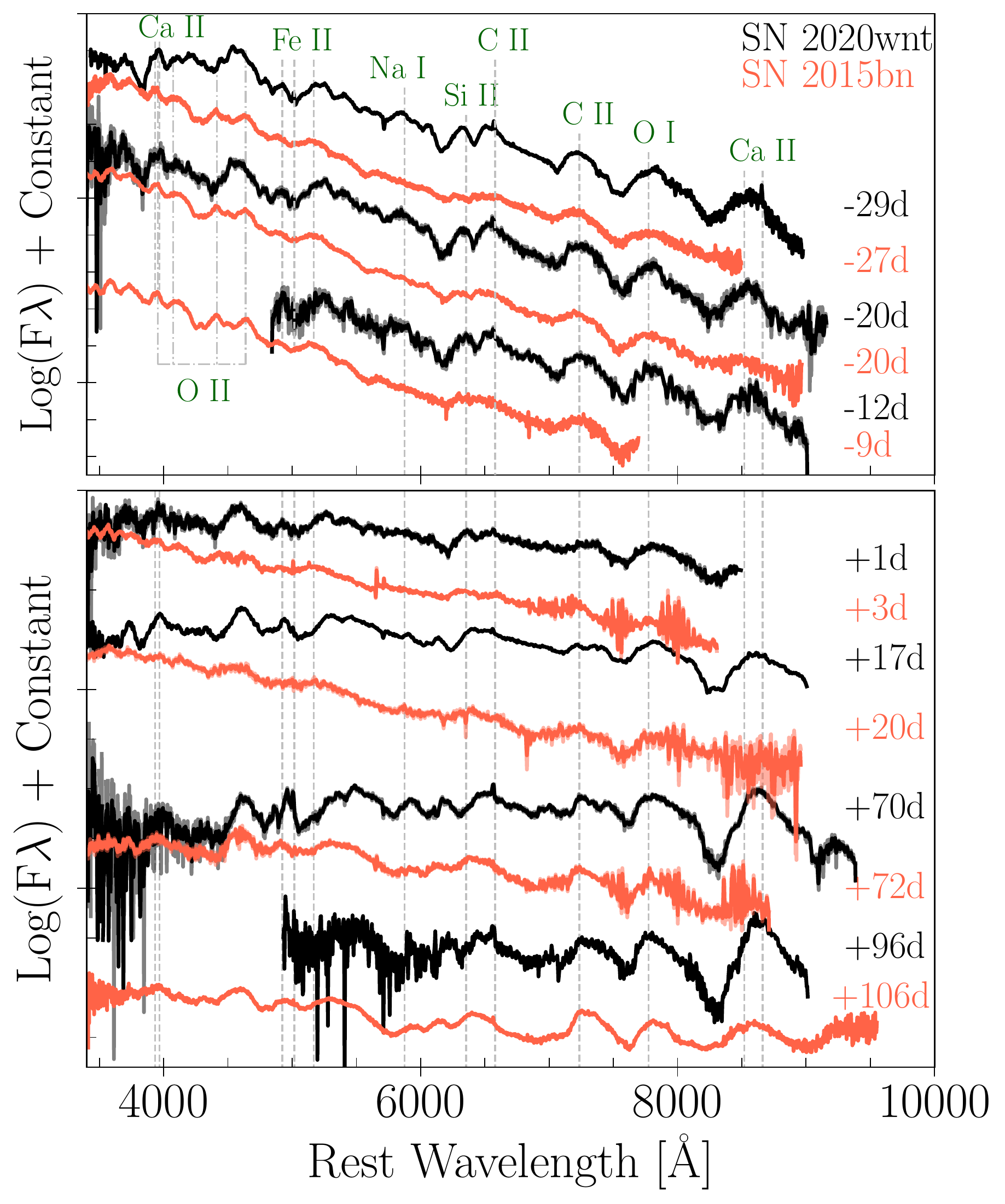}
\caption{Spectral comparison of \sn\ with the slow-evolving SN~2015bn at eight different epochs before (top panel) and after (bottom panel) the maximum light in the $r-$band. Each spectrum has been corrected for Milky Way reddening and shifted vertically by an arbitrary amount for presentation. Vertical lines indicate the rest position of the strongest lines. The phases are labeled on the right.
}
\label{comp15bn}
\end{figure}

We now compare \sn\ with the most extreme objects shown in Figure~\ref{compLC}: SN~2015bn, the brightest object in the comparison sample and one of the best observed SLSNe to date, and the carbon-rich type Ic SN~2007gr, the faintest SN in the plot, and the best-spectral match object found by \textsc{gelato} \citep{Harutyunyan08}. 

The spectral comparison between \sn\ and SN~2015bn at seven different epochs is presented in Figure~\ref{comp15bn}. Before the peak (top panel), the spectra of both objects are characterised by a blue continuum with lines of \ion{O}{i}, \ion{Si}{ii}, \ion{C}{ii} and \ion{Fe}{ii}. In \sn, these features are stronger at all phases. In contrast, the main differences lie in the \ion{O}{ii} and \ion{Ca}{ii} lines. The spectra of \sn\ do not show the W-shape \ion{O}{ii} features as observed in SN~2015bn, and other SLSN-I events. The presence/absence of these lines depends on the temperature (more details in Section~\ref{sec:disc}). Unlike SN~2015bn, \sn\ shows very strong \ion{Ca}{ii} absorptions. These \ion{Ca}{ii} absorption features are not observed in young SLSNe-I. After peak (bottom panel), in contrast with SN~2015bn, \sn\ has a redder continuum, more intense lines and the bluer part of the spectrum is dominated by Fe-group lines. Some signs of emission components are also detected, suggesting the beginning of the transition to the nebular phase. All these properties are delayed in SN~2015bn.     

\begin{figure}
\centering
\includegraphics[width=\columnwidth]{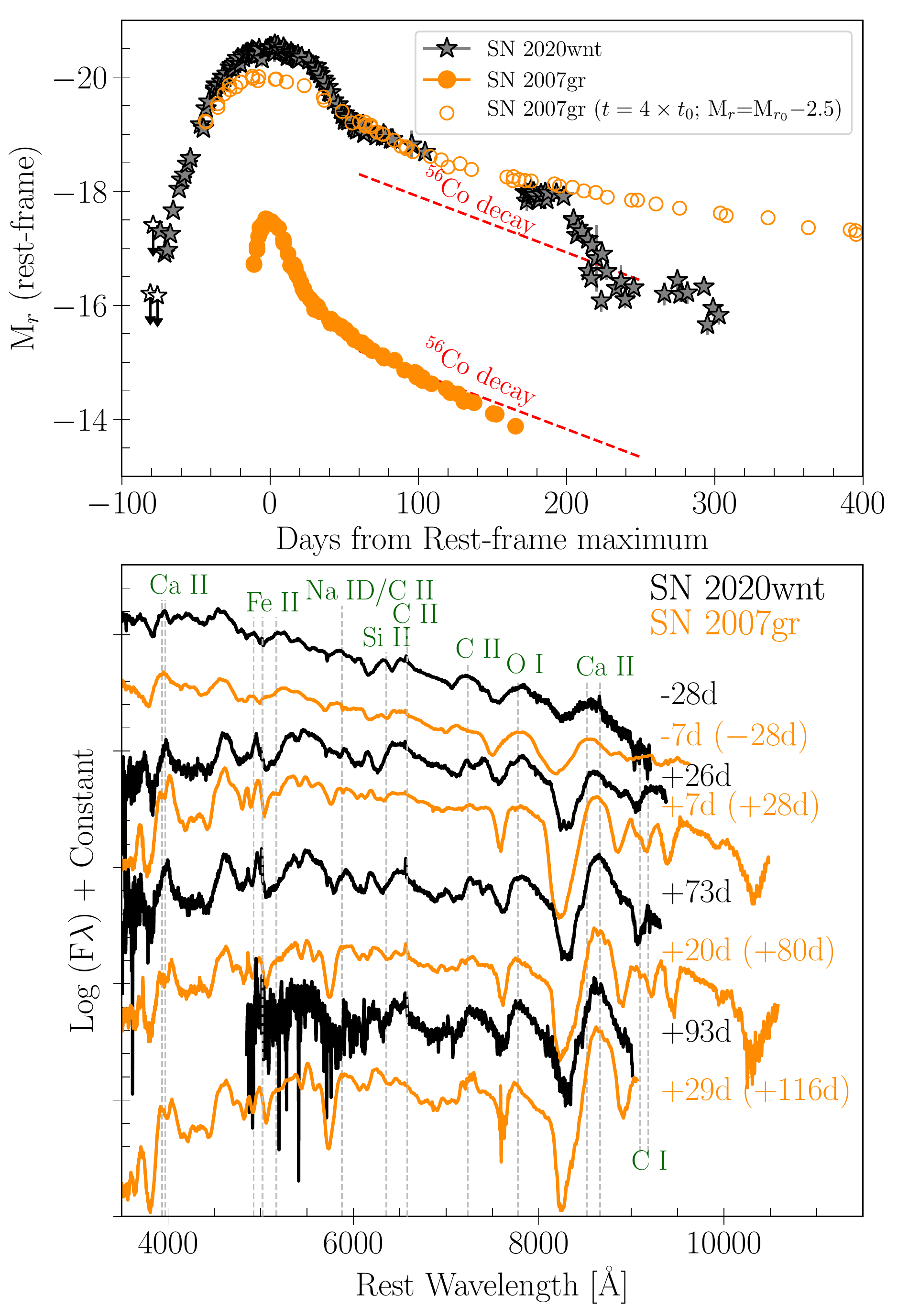}
\caption{Light curve and spectral comparison with the type Ic carbon-rich SN 2007gr. \textbf{Top:} Comparison of the $r-$band light curves. For the light curve of SN~2007gr, we applied a temporal correction ($t=4\times t_0$) to reproduce the same shape and width as \sn. A shift of 2.5 mag is applied to the magnitude in order to have a similar brightness in the transition and the tail (from $\sim+53$ days from peak). \textbf{Bottom:} Spectral comparison at four different epochs from the $r-$band maximum. Each spectrum has been corrected for Milky Way reddening and shifted vertically by an arbitrary amount for presentation. Vertical lines indicate the rest position of the strongest lines. The rest-frame phases are labeled on the right, while the epochs with the temporal correction (for SN~2007gr, $t=4\times t_0$) are in parenthesis. This correction gives a consistent epochs to \sn.
}
\label{comp07gr}
\end{figure}

In Figure~\ref{comp07gr}, we compare \sn\ and the type Ic carbon-rich SN~2007gr. Although the light curves (top panel) of these two objects are completely different at the first glance, they share a similar spectroscopic evolution (bottom panel) with some time lag ($t=4\times t_0$; where $t_0$ is the rest-frame maximum of SN~2007gr), which was initially found by the spectral matching from \textsc{gelato}. From the light curves in the $r-$band, we estimate that \sn\ is $\sim3$ mag brighter than SN~2007gr around the maximum light. \sn\ also has a pre-peak bump light curve morphology and evolves on a much longer timescale than its fainter counterpart. In order to have a broad light curve similar to that of \sn, we applied a temporal correction to SN~2007gr. We find that a factor of 4 can reproduce the light curve width of \sn, and it is, in turn, in agreement with that found by the spectral matching. This correction is included in the top panel of Figure~\ref{comp07gr} (open orange circles).

In the bottom panels of Figure~\ref{comp07gr}, the spectroscopic comparison between \sn\ and SN~2007gr is presented at four different epochs. Analysing the spectral features, one sees that the main similarity is the detection of the carbon lines, while the main difference is the strength of the \ion{Na}{i} line, which is more intense in SN~2007gr. From the spectral matching with \textsc{gelato}, we found a time lag of $t=4\times t_0$. Therefore, the \sn\ spectrum at $-27$ days from peak is compatible with that of SN~2007gr at $-7$ days from peak. Both SNe show strong features of \ion{Ca}{ii}, \ion{O}{i}, and the clear signs of \ion{Si}{ii}, \ion{C}{ii} and the W-shape \ion{Fe}{ii} lines at around 4800 -- 5200 \AA. After peak, both objects evolve maintaining the time lag identified in the spectra before peak, which is consistent with that obtained from the light curves. The similarity in the spectral evolution suggests that both objects may arise from a carbon-rich progenitors, however, the longer timescale and the brightness of the light curve may suggest  different explosion parameters.

\begin{figure}
\centering
\includegraphics[width=\columnwidth]{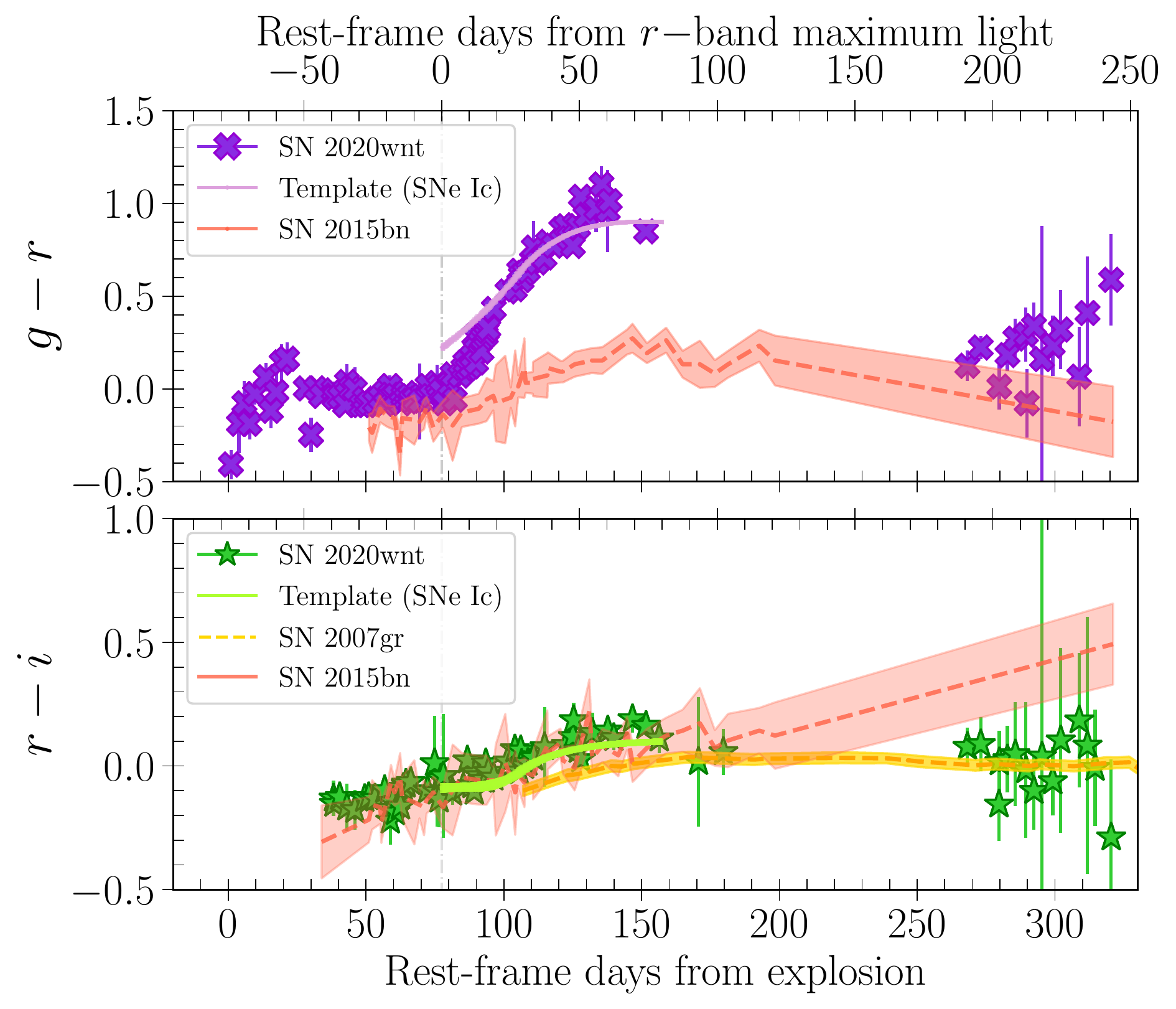}
\caption{$g-r$ and $r-i$ colour curves of \sn. For comparison, in solid lines (purple and green) we show the intrinsic SNe~Ic colour-curves templates ($g-r$ and $r-i$) from \citet{Stritzinger18b} multiplied by a temporal factor of 4 ($t=4\times t_0$). We also include the $r-i$ colour of SN~2007gr (multiplied by the temporal factor; dashed orange line) and the $g-r$ and $r-i$ colours of SN~2015bn (dashed light red line; \citet{Nicholl16}). The data used here for SN~2007gr were taken from \citet{Bianco14}.  
}
\label{com_colours}
\end{figure}

In Figure~\ref{com_colours}, we compare the $g-r$ and $r-i$ colours of \sn, SN~2015bn and SN~2007gr (applying the temporal correction previously mentioned). In $g-r$, we find that from the maximum light, the evolution of \sn\ and SN~2015bn are very different. While \sn\ evolves quickly to the red, reaching a peak $\sim50$ days later, SN~2015bn remains bluer a much longer time. At later phases, the colours are more alike. On the other hand, the $r-i$ colours are almost identical in the three objects up to $+150$ days from the maximum (SN~2007gr being the bluest). After this, the evolution diverges. For instance, \sn\ and SN~2007gr become a bit bluer, whereas SN~2015bn evolves to the red. In Figure~\ref{com_colours}, we also include the intrinsic colour templates of SNe~Ic from \citet{Stritzinger18b}. To these templates, we apply the same temporal corrections as that found for SN~2007gr. From this comparison, 1) using these templates to constrain the host extinction, we find that for \sn\ it is negligible; 2) the colour of \sn\ between the maximum light and $+40$ days from the maximum (temporal correction included), evolves similarly as SNe~Ic in both $g-r$ and $r-i$. This suggests that the colour of \sn\ is similar to that observed in standard SNe~Ic but on a longer timescale.

\begin{figure*}
\centering
\includegraphics[width=0.72\textwidth]{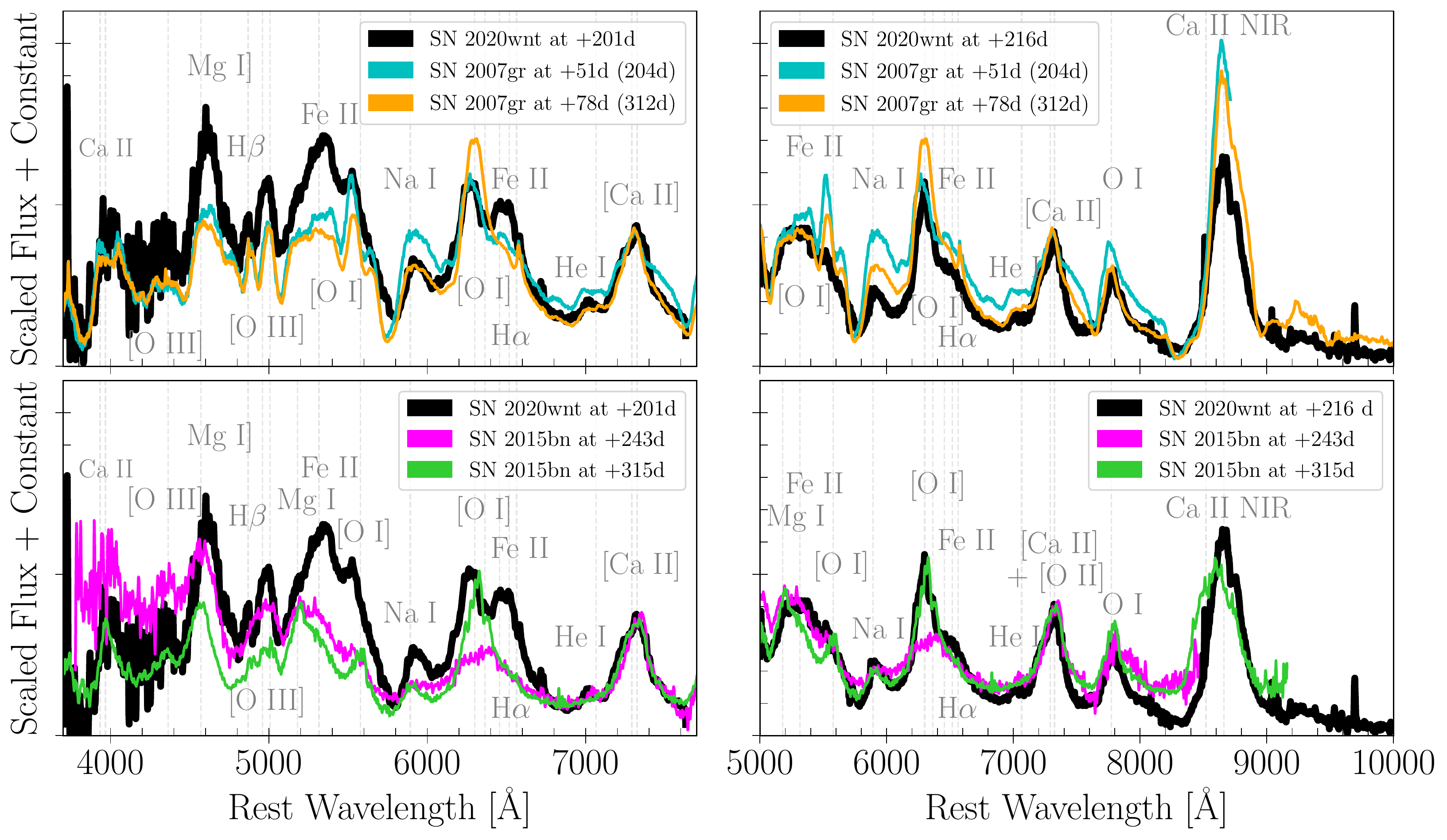}
\caption{Spectral comparison between \sn\ and SN~2007gr \textbf{(top)} and \sn\ and SN~2015bn \textbf{(bottom)}. Each spectrum has been corrected for Milky Way reddening. Spectra are scaled to match the \ion{[Ca}{ii]} line. Vertical lines indicate the rest wavelengths of the strongest lines. The phases are labeled in the legend. In \sn, the narrow emission lines from the host galaxy were removed for presentation. For SN~2007gr the phases in brackets correspond to the epochs with the temporal corrections ($t=4\times t_0$).
}
\label{compneb}
\end{figure*}

In Figure~\ref{compneb}, we compare the late phase spectra of \sn\ with those of SN~2007gr (top) and SN~2015bn (bottom) at similar epochs. As before, we find that the comparison with SN~2007gr has a time lag of $t=4\times t_0$, i.e., the spectrum of \sn\ at $+201$ days from peak is similar to the spectra of SN~2007gr after 50 days from peak. In the top panel of this Figure, we compare our spectra at $+201$ days (blue part, left panel) and $+216$ days (red part, right panel) with SN~2007gr at $+51$, and $+78$ days from peak (with the correction the epochs are $+204$, and $+312$). Analysing the blue part (left panel), we see that both objects have the similar lines, but they seem to be stronger in \sn. The shoulder on the red side of \ion{[O}{i]} $\lambda\lambda6300$, 6364 is also detected in SN~2007gr, though a bit fainter. From the red part (spectrum at $+216$ days, right panel), the lines observed in both objects fit very well. The larger discrepancy is in the intensity of the \ion{Ca}{ii} NIR triplet feature, which is stronger at all epochs in SN~2007gr. At this phase, the shoulder of the red side of the \ion{[O}{i]} line matches better than before. 

In the bottom panel of Figure~\ref{compneb}, we present the late spectra of \sn\ and SN~2015bn. For SN~2015bn we use two epochs ($+243$ and $+315$ days from peak) that have some similarities with our spectra. Examining the blue part of the spectra, we notice some differences between these two objects. For instance, in the three spectra of SN~2015bn, there are no signs of a feature on the red side of \ion{[O}{i]} $\lambda\lambda6300$, 6364, of the emission around 5300 \AA\ (attributed to \ion{Fe}{ii}) and the line at $\sim7100$ \AA\ (possibly \ion{He}{i}) detected in \sn. Now, the red part (right panel) is similar, although \sn\ has a stronger \ion{Ca}{ii} NIR triplet, but a weaker \ion{O}{i} than SN~2015bn at all epochs. Overall, the SN~2015bn spectrum that best matches \sn\ is that at $+315$ days. 
The similarity of the nebular spectra of \sn\ with those of SN~2007gr and SN~2015bn suggests that these objects are related, and they possibly have a common origin, including a similar chemical composition.

\section{Explosion and progenitor scenarios}
\label{sec:expl}

\subsection{Nebular properties}
\label{sec:neb}

To constrain the internal conditions and core structure of \sn, we investigate the nebular spectra in detail. Based on the findings of \cite{Jerkstrand14}, the \ion{O}{i} mass responsible for the line emission can be estimated from the oxygen luminosity, as follows: 

\begin{center}
M$_{O}=\frac{L_{6300,6364}/\beta_{6300,6364}}{9.7\times 10^{41}~ \mathrm{erg~s^{-1}}} \times exp(\frac{22720~K}{T})$ \Msun 
\end{center}

{\noindent}where L$_{6300,6364}$ is the line luminosity of \ion{[O}{i]} $\lambda\lambda6300$, 6364, $\beta_{6300,6364}$ is the is the Sobolev escape probability, and T is the temperature. Here, the temperature can be derived from the \ion{[O}{i]} $\lambda5577$ to \ion{[O}{i]} $\lambda\lambda6300$, 6364 ratio \citep{Houck96,Jerkstrand14}. Given that the \sn\ spectra are not fully nebular, the estimation of these parameters depends on how we define the line fluxes, and in turn, the temperature. To estimate the luminosities, a linear fit to the continuum is subtracted, and then, we fit a Gaussian to the line. As \ion{[O}{i]} $\lambda5577$ is hard to measure, we take the extreme values (minimum and maximum flux) obtained from the Gaussian fit. Thus, using the spectrum at $+216$ days from the maximum light, and assuming a $\beta_{5577}/\beta_{6300,6364}$ ratio equal to 1, we get a temperature between $\sim3800$ and $\sim4500$ K. For $\beta_{6300,6364}=1$ (i.e. assuming optically thin emission), we can obtain the minimum mass of oxygen required to produce the observed \ion{[O}{i]}. With these values, 
we derive a M$_{O}\approx 2-4$ \Msun. Using a similar approach, \citet{Nicholl16a} found M$_{O}\approx 9$ \Msun\ for SN 2015bn, while \citet{Mazzali10} found M$_{O}\approx 1$ \Msun\ by modelling the nebular spectra of SN~2007gr. Despite the uncertainties in the estimation of the \ion{O}{i} mass, the values found for \sn\ are intermediate between those derived for SN~2015bn and SN~2007gr.

Information on the core mass is usually inferred from the \ion{[Ca}{ii]} $\lambda\lambda7291,$\,7324/\ion{[O}{i]}$\lambda\lambda6300,$\,6364 ratio \citep{Fransson89,Elmhamdi04,Kuncarayakti15}, although we are aware that the ratio is sensitive to various parameters \citep[e.g.,][]{Li93,Jerkstrand17,Dessart21}. Nonetheless, we calculate this ratio in order to compare it with objects from the literature. Using the spectrum at $+216$ days, we compute a \ion{[Ca}{ii]}/\ion{[O}{i]} flux ratio of $\sim1.1$, which is within the range of values found for core-collapse SNe (lower than 1.43; e.g., \citealt{Kuncarayakti15,Terreran19,Gutierrez20a}), but larger than the value found for SN~2015bn (0.5; \citealt{Nicholl16a}). Although a \ion{[Ca}{ii]}/\ion{[O}{i]} ratio of $\sim1.1$, suggests a relatively low helium core mass, we note that the spectrum at $+216$ days is not completely nebular. Therefore, the ratio estimation can be affected by this issue, as seen in some core-collapse SNe with good nebular coverage \citep[e.g. SN~2017ivv;][]{Gutierrez20a}. Additionally, we also highlight that the \ion{[Ca}{ii]} feature could be contaminated by \ion{[O}{ii]}. In fact, if the nebular feature around $7300$ \AA\ is dominated by \ion{[O}{ii]} instead of \ion{[Ca}{ii]} \citep[e.g. SN~2007bi;][]{GalYam09a}, this ratio cannot be a good proxy of the progenitor core mass. In this case, the real O amount could be much higher, and more consistent with the very high value of the \nifs\ inferred in Section~\ref{sec:model}. 

In Section~\ref{sec:spec}, we pointed out that the strongest line observed in \sn\ at $+216$ days from peak (293 days from explosion) was the \ion{Ca}{ii} NIR triplet. From the nebular comparison, we found this feature is fainter in \sn\ than in SN 2007gr but stronger than in SN~2015bn. By modelling SLSN nebular spectra, \citet{Jerkstrand17a} found that an electron density of $n_e\gtrsim10^8$ cm$^{-3}$ is needed to reproduce a \ion{Ca}{ii} NIR/\ion{[Ca}{ii]} larger than 1. For SN~2015bn they derived a ratio of 1.7, which is unusually high for the nebular phase. Following this approach, we measure the \ion{Ca}{ii} NIR/\ion{[Ca}{ii]} ratio for \sn\ and we find a value of 3.3. This value is twice that derived for SN~2015bn. This suggests extraordinarily high electron densities for \sn. However, as mentioned before, the spectrum at +216 days is not fully nebular and the ratios we measure might correspond to some limits. 

In the spectrum of \sn\ at $+201$ days, we recognise two broad features around 4350 and 5000 \AA. These features, that  could be attributed to \ion{[O}{iii]} $\lambda4363$ and \ion{[O}{iii]} $\lambda\lambda4959$, 5007, have been detected in several core-collapse SNe at very late phases \citep[e.g.][]{Fesen99,Milisavljevic12} and in a few SLSNe during the nebular phase: PS1-14bj \citep{Lunnan16}, LSQ14an \citep{Inserra17}, SN~2015bn and SN~2010kd \citep{Kumar20}. As seen in Figures~\ref{spec}, \ref{nebH} and \ref{compneb}, \sn\ exhibits a relatively strong \ion{[O}{iii]} $\lambda\lambda4959$, 5007, but a weak \ion{[O}{iii]} $\lambda4363$, which suggest a high \ion{[O}{iii]} $\lambda\lambda4959$/ \ion{[O}{iii]} $\lambda4363$ flux ratio. This ratio can provide information about the temperature and electron density of the emitting region where they formed. Fitting a Gaussian to these lines, we measure a \ion{[O}{iii]} $\lambda\lambda4959$, 5007/ \ion{[O}{iii]} $\lambda4363$ flux ratio of $\sim3.5$. This value is much larger than those measured for PS1-14bj \citep{Lunnan16} and LSQ14an \citep{Inserra17}, and suggests electron densities greater than 10$^6$ cm$^{-3}$ \citep{Fesen99,Jerkstrand17a}.  

Summarising, from the line ratio analysis (e.g. \ion{Ca}{ii} NIR/\ion{[Ca}{ii]} and \ion{[O}{iii]} $\lambda\lambda4959$/, 5007 \ion{[O}{iii]} $\lambda4363$), we derive high electron densities ($n_e\sim10^6$ -- $10^8$ cm$^{-3}$). Density values of around 10$^8$ cm$^{-3}$ have been inferred before for the SNe~IIn SN~1995N \citep{Fransson02} and SN~2010jl \citep{Fransson14}, and more recently for the SLSN 2015bn \citep{Jerkstrand17a}.

\subsection{Light curve modelling}
\label{sec:model}

We explore several models, trying to explain the light curve morphology of \sn. Two important characteristics put some constraints in our modelling: 1) the long-rise time to the main peak and 2) the luminosity following the radioactive decay at 140 days from explosion. These two properties point out that this SN may belong to the rare class of SLSNe-I that are possibly powered by a large amount of nickel production, similar to SN 2007bi \citep{GalYam09a}. Here, we also analyse the possibility that the main peak can be powered by a magnetar.

\begin{figure}
\centering
\includegraphics[width=\columnwidth]{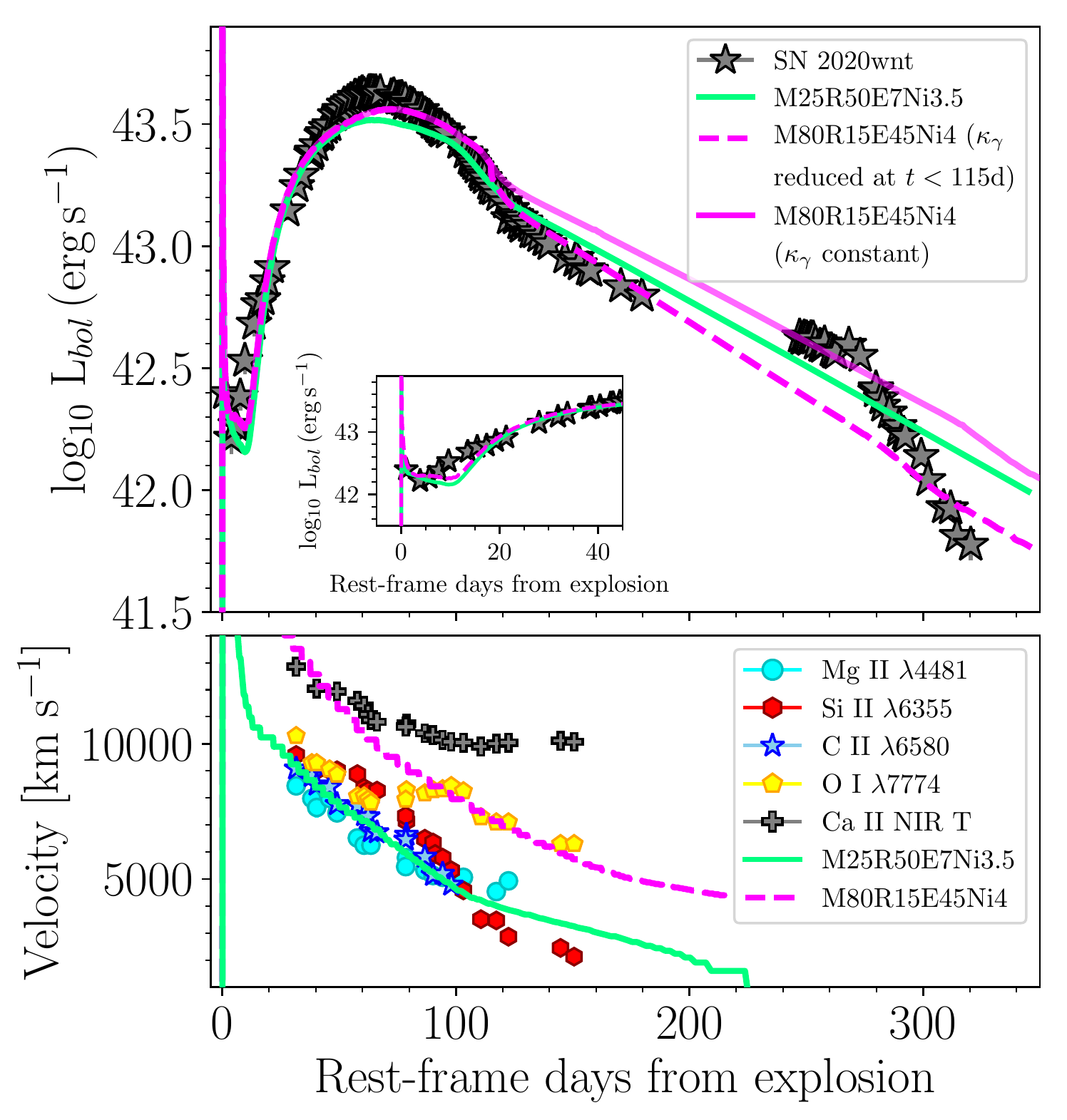}
\caption{\textbf{Top panel:} Comparison between our best hydrodynamical models and the SN observations (grey stars). For the more massive model (magenta), we use a progenitor with a pre-SN mass of 80 \Msun, $E_{\rm exp}=45\, \times 10^{51}$ erg, a \nifs\ mass of 4 \Msun\ and an initial radius of 15 \Rsun, while the less massive model (green solid line) has a a pre-SN mass of 25 \Msun\ and $E_{\rm exp}=7\, \times 10^{51}$ erg, a \nifs\ mass of 3.5 \Msun\ and an initial radius of 50 \Rsun. For the more massive model, we show the light curve with $\kappa_\gamma$ constant (solid line), and a light curve with $\kappa_\gamma$ reduction (dashed-line). The inset plot shows the light curves at very early phases. \textbf{Bottom panel:} Photospheric velocity of the models compared with the velocities of different species in \sn.
}
\label{model}
\end{figure}

To calculate the light curve and photospheric velocity, we use a one-dimensional local thermodynamical equilibrium (LTE) radiation hydrodynamical code presented in \citet{Bersten11}. The code follows the complete evolution of the light curve in a self-consistent way from the shock propagation to the nebular phase. The explosion initiates by injecting some energy near the progenitor core. This produces a powerful shock wave that propagates inside the progenitor until it arrives at the surface, where the photons begin to diffuse out. The code assumes a grey transport for gamma photons produced during the radioactive decay, but allows any distribution of this material inside the ejecta and assumes an opacity of $\kappa_\gamma=0.03$ cm$^2$ g$^{-1}$. Therefore, we are able to calculate the gamma-ray deposition in each part of the ejecta and estimate the gamma-ray escape as a function of time. The inclusion of a magnetar as an extra source to power the SN, including the relativistic effect, is accounted in the code \citep[see][for details]{Bersten16}. An extensive exploration of magnetar parameters for different types of progenitors was discussed in \citet{Orellana18}.   

An initial structure in hydrostatic equilibrium which simulates the condition of the star at the pre-SN stage is needed to start the hydrodynamical calculations. Here we assume both parametric and stellar evolutionary models. The initial emission and the rise time to the main peak of \sn\ may indicate that a more extended progenitor is required than the typical stripped (or compact) star assumed for H-free objects. Therefore, we test models typically used for H-rich SNe with extended and dense envelopes like red supergiant (RSG) or blue supergiant (BSG) stars, but manually modifying the chemical abundance to produce a H-free envelope. The RSG models are computed by stellar evolutionary calculation while the BSG progenitor is computed assuming a double polytropic model given that it not easy to generate these BSG progenitors with stellar evolution models. 

Assuming only a radioactive source, we cannot find a suitable solution for RSG (or stripped-envelope) progenitors, but models with a structure similar to those used for 87A-like objects or BSG progenitors (with our altered  chemical composition) seem to offer a good solution. We generate several initial configurations for different values of the pre-SN mass (between 15 and 100 \Msun) and radius (in the range of 15 to 80 \Rsun). We explore many values of the explosion energy and \nifs\ production for each configuration. Our best models are shown in the top panel of Figure~\ref{model} and correspond to two progenitors with a pre-SN radius of 50 and 15 \Rsun, pre-SN masses of 25 \Msun~(green) and 80 \Msun\ (magenta), and  explosion energies of $7\times10^{51}$ erg and $45\times10^{51}$ erg, respectively. In both cases, a large amount of \nifs\ (3.5 and 4 \Msun) is required to reproduce the main peak of the light curve. For the more massive model (80 \Msun), we slightly modified the gamma-ray opacity from $\kappa_\gamma=0.03$ to $\kappa_\gamma=0.013$ cm$^2$ g$^{-1}$ at around 115 days from the explosion to improve the fit from this epoch (i.e. allowing an easier leakage of gamma-ray photons; \citealp[see][for more details]{Gutierrez21}). However, we also include this model but with a $\kappa_\gamma$ constant (solid magenta line). 

As observed in several SNe~Ib/c \citep[e.g.][and references therein]{Sollerman00}, the gamma-ray leakage significantly affects the light curve at late phases.  More precisely, in several energetic SNe (e.g. SNe~Ic broad line), the nickel mass required to explain the peak luminosity generally overestimates the tail luminosity \citep[e.g.][]{Maeda03}. To partially solve this issue, it has been proposed that an enhancement in the gamma-ray escape may happen in latter epochs (after the main peak). This could be due to possible asymmetries in the ejecta as the presence of low-density (or clumps) zones \citep{Tominaga05, Folatelli06}. Such types of structures could be produced by jets \citep[e.g.][]{Soker22} or Rayleigh-Taylor instabilities. In these low-density regions, the gamma-rays could escape more efficiently, which can be simulated by reducing the kappa-gamma values. In addition, the grey transfer assumed for the gamma-rays can require a time varying factor, as shown in \citet{Wilk19} by solving the relativistic radioactive transfer equation for gamma-ray in SNe.

The photospheric velocities of the models are compared to the velocities of different species in the bottom panel of Figure~\ref{model}. The less massive model (25 \Msun) reproduces the observables exceptionally well, particularly for \ion{Mg}{ii}, \ion{Si}{ii} and \ion{C}{ii}; however, this is not the case for the 80 \Msun\ model. This model overestimates most of the velocities but has a good agreement with \ion{O}{i} from 100 to 160 days. 

Regarding the light curve, we found that the early emission can be attributed to the cooling phase for an extended progenitor of $\sim50$ and 15 \Rsun, respectively. However, none of the models can reproduce the luminosity at phases later than $\sim245$ days, since the hypotheses used in the code fail at these late times.
At these epochs, the luminosity declines slower than the \cofs\ decay (for about 30 days) and then suddenly drops. Between $\sim245$ and $\sim275$ days, the excess in luminosity is probably due to CSM interaction, which could be supported by the presence of H emission lines (H$\alpha$ and H$\beta$) in the late spectra (see Sections~\ref{sec:spec}). The sudden drop in luminosity may be explained by the end of this interaction phase or alternative by dust formation. When the SN reaches the minimum value in this decrease, it almost recovers the luminosity expected from radioactivity (80 \Msun\ model).  
Ignoring these late epochs, the 25 and 80 \Msun\ models well describe the overall light curve evolution. Considering the photospheric velocities, the 25 \Msun\ model better represents the observed properties of \sn. However, given the large amount of \nifs\ (3.5 \Msun), this model turns out to be unrealistic. Relaxing the velocity constraints and considering that reaching an $E\geq10^{52}$ requires a much higher and possibly nonphysical neutrino deposition fraction \citep{Janka12,Terreran17}, we find that the 80 \Msun\ model reproduces the light curve properties of \sn\ very well. Additionally, given that t$_{neb}\approx \sqrt{\kappa*M/v^2} \approx M*\sqrt{\kappa/E_{kin}}$ and t$_{peak}\approx \sqrt{\kappa*M/(v*c)}$, we have t$_{neb}^2\approx t_{peak}^2*c/v$. Since t$_{peak}$ is fixed and our more massive model has higher velocities than the less massive model, the former is more compatible with the expected values for t$_{neb}$.
Therefore, \sn\ is consistent with a massive progenitor (pre-SN mass of 80 \Msun), with a radius of 15 \Rsun, explosion energy of $45\times10^{51}$ erg, and 4 \Msun\ of \nifs. These characteristics are compatible with the PISNe scenario (e.g. \citealt{GalYam09a, Kozyreva14}, but see \citealt{Dessart13}).

As mentioned above, we also analyse the possibility that \sn\ is powered by a magnetar. For this, we apply a version of our one-dimensional LTE radiation hydrodynamics code that takes into account the power provided by a newborn magnetar that loses rotational energy. That energy is fully deposited in the inner zones of the ejecta. We use the $\dot{E}$ vacuum prescription that became popular after the work of \cite{Kasen10b} with the standard assumption of a magnetic dipole and a breaking index $n=3$. Modern and detailed treatments by \cite{Vurm21} indicate that this approach is roughly valid at early times from the explosion.
The parameters of this source are the surface magnetic field strength $B$ and the initial rotation period $P$.

\begin{figure}
\centering
\includegraphics[width=\columnwidth]{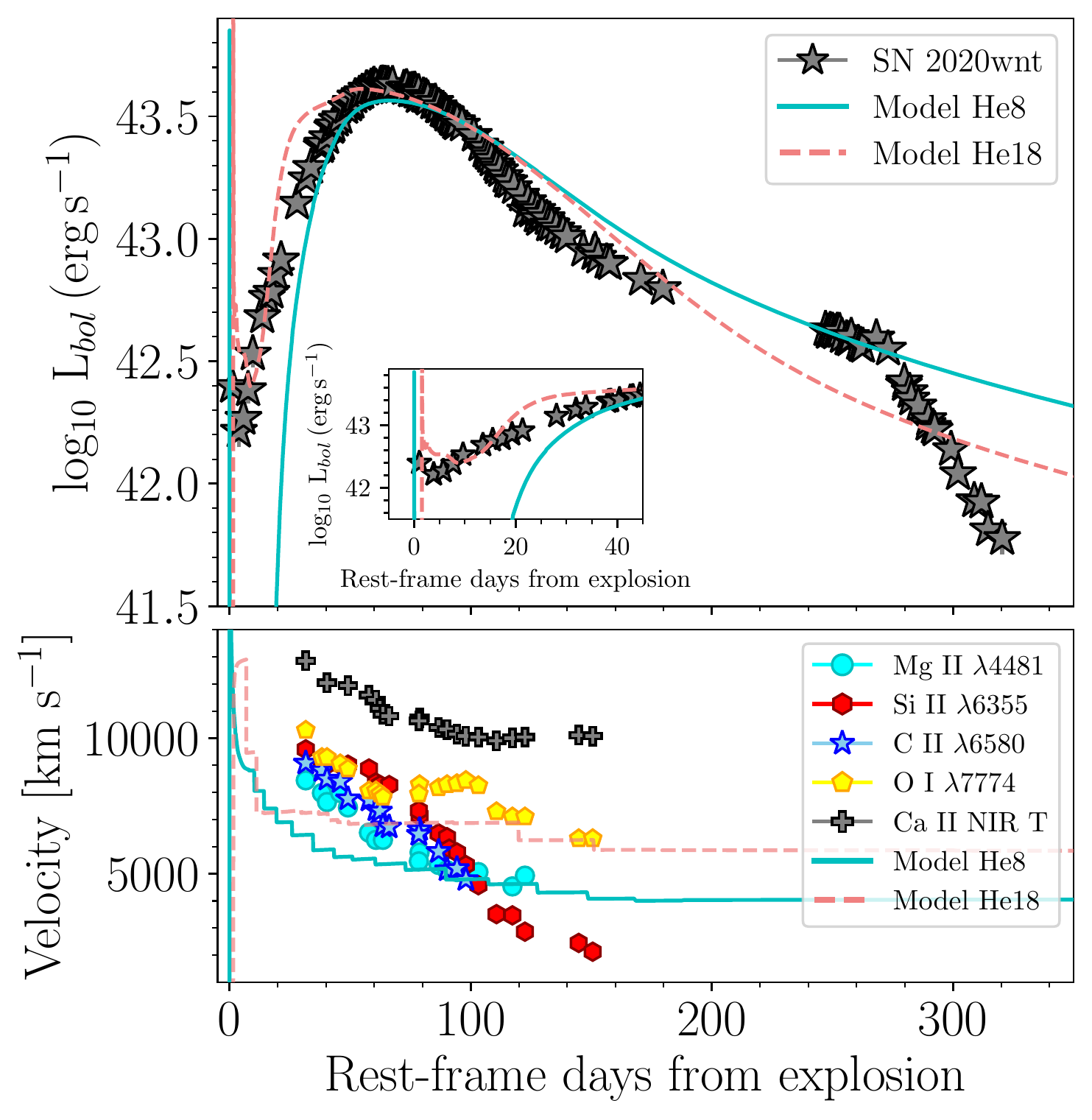}
\caption{\textbf{Top panel:} Comparison between the magnetar models and the SN observations (grey stars). For the compact configuration (cyan solid line), we use a progenitor with a pre-SN mass of 8 \Msun, $E_{\rm exp}=1\, \times 10^{51}$ erg and a magnetar with an initial period $P\sim 4.2$~ms and a magnetic field $B\sim 2\times10^{14}$~G. In a more extended configuration (pink dashed-line), we use a progenitor with a pre-SN mass of 18 \Msun, $E_{\rm exp}=1\, \times 10^{51}$ erg, $P\sim 2$~ms and a magnetic field $B\sim 3\times10^{14}$~G. The inset plot shows the light curves at very early phases. \textbf{Bottom panel:} Photospheric velocity of the magnetar models compared with the velocities of different species in \sn.
}
\label{magnetar}
\end{figure}

We test some of the H-poor progenitors, with a usual radioactive content of around 0.1 \Msun. For the compact configuration ($M=8$ \Msun\ with $R=1.3$ \Rsun\ and an explosion energy of $E_{\rm exp}=1\, \times 10^{51}$ erg ) the $L_\text{peak}$ of \sn\ can be reached with $P\sim 4.2$~ms and $B\sim 2\times10^{14}$~G, but the rise and post peak are not well reproduced (see Figure~\ref{magnetar}). We explore the same configuration with the mixing of the \nifs\ up to a fraction of 0.95 of the ejecta, but that makes no major changes. Then, we explore the possibility of a magnetar being applied to a more extended RSG structure. We use the results from \citet{Orellana18} to select a possible model for \sn. In Figure~\ref{magnetar}, we present the results for a progenitor with a pre-SN mass of 18 \Msun\ and a radius of 725 \Rsun. In this case the H is converted to He artificially, to have an inflated structure for a H-free star, though it is not the result of stellar evolution calculations. In order to obtain the time scale and the energy of the peak, the following magnetar parameters are required: $P\sim2$~ms and $B\sim3 \times10^{14}$~G, assuming an explosion energy of $E_{\rm exp}=1\, \times 10^{51}$ erg.  While this model provides an improvement with respect to the compact model, it produces a worse representation of the light curve than the \nifs\ powered models presented in Figure~\ref{model}. Additionally,  the velocities are not well reproduced. Therefore, this result disfavours a magnetar model to explain the overall light curve evolution of \sn. 

We remark that none of the light curve models presented here is  designed to fit the drastic drop in brightness observed from 273 days. Finally, although the \nifs\ powered models presented in Figure~\ref{model} provide a good explanation of \sn\ for both the light curve and the velocities, the progenitor models used were built parametrically (both the pre-SN density profile and the chemical composition). It would be desirable to investigate what kind of evolutionary path, if any, can generate this type of structure to provide a more solid physical framework for our modelling.

\subsection{CSM interaction}

Despite the spectra of \sn\ show remarkable similarities with SN~2007gr and non-interacting SNe Ic, the light curve fluctuations and the very high luminosity can be comfortably explained with the powering of ejecta-CSM interaction. Under this premise, we could consider that the actual \nifs-powered SN light curve would be below the observed one (similar to that observed for several SNe~Ic; see top panel of Figure~\ref{comp07gr}) dominated by ejecta-CSM interaction. We use \textsc{TigerFit}\footnote{\url{https://github.com/manolis07gr/TigerFit}}  \citep{Chatzopoulos16,Wheeler17} to explore this scenario by assuming a steady-state wind and a constant-density CSM shell, respectively. Although none of the fits reproduces the observed light curve of \sn, we found that a steady-state wind CSM can reach a better approximation. In this context, a potentially pervasive interaction phase lasts for $\sim273$ days and ends when the luminosity suddenly drops. Here, the CSM is probably H and He free. After the drop in brightness, the light curves continue to experience small fluctuations (see Figure~\ref{lcs}), most likely attributed to episodes of CSM interaction. We note that very massive stars (M$_{ZAMS}\sim100-140$ \Msun; \citealt{Heger03}) are expected to lose mass through pulsational pair-instability. The SN ejecta are then expected to interact with the CSM gathered through former mass-loss events. This would generate luminous light curves, which may also show large luminosity fluctuations. 
Even though this scenario could reproduce the observable properties of \sn, we do not have a model to test it.

\section{Discussion}
\label{sec:disc}

Similar pre-peak bump light curves have been detected in several SNe, which include SLSNe-I SN~2006oz, LSQ14bdq, DES14X3taz, DES15S2nr, DES17X1amf, SN~2018bsz and some signs of it in SN~2018hti \citep{Fiore22}, and the peculiar SNe~Ib/c SN~2005bf \citep{Anupama05, Tominaga05, Folatelli06, Maeda07}, PTF11mnb \citep{Taddia18b}, and SN~2019cad \citep{Gutierrez21}. These bumps have been suggested to be a common characteristic of the SLSN class \citep{Nicholl16a}, although more recent analyses with larger samples determined that such feature is not ubiquitous to all SLSNe-I \citep{Angus19}. 

Based on the analysis and comparison presented in previous sections, we found that \sn\ is an object that seems to connect SN~Ic and SLSN-I events. Although this connection was established before \citep[e.g.,][]{Pastorello10}, the excellent photometric and spectroscopic coverage of this SN provides important insights to better understand such relations, and the implications for the explosion and progenitor stars. We discuss below the consequences of these results.

\subsection{Unusual light curve evolution}
\label{lcevol}

As discussed in previous sections, the light curves of \sn\ show remarkable features such as a pre-peak bump, a tail resembling the \cofs\ decay followed by a sudden drop and minor luminosity fluctuations. Bumpy light curves have been observed before maximum in several SLSNe-I, as shown in Figure~\ref{com_bumps}, and have been suggested to be frequent in the SLSN class \citep{Nicholl16a}. However, more recent analyses with larger samples determined that such feature is not ubiquitous in SLSN-I \citep{Angus19}.  
Early bumps have been interpreted as resulting from the recombination wave in the ejecta \citep[][for SN~2006oz]{Leloudas12}, the shock breakout within a dense CSM \citep{Moriya12}, the shock cooling of extended material around the progenitor \citep{Nicholl15,Piro15,Smith16,Angus19} or an enhanced magnetar-driven shock breakout \citep{Kasen16}. In the case of \sn, we propose that such a feature is a consequence of a post-shock cooling phase in an extended progenitor (Section~\ref{sec:model}).

On the other hand, tails resembling the \cofs\ decay have also been previously observed in other SLSNe-I, however, it was suggested that they may be powered by magnetar energy injection rather than \cofs\ decay (\citealt{Inserra13a}, but see \citealt{GalYam12}). In the case of \sn, and as shown in Section~\ref{sec:model}, the \nifs\ models provide a better representation of the light curve of \sn\ than those obtained with the magnetar model, suggesting that the main source of power is radioactivity. Signs of CSM interaction (e.g. presence of H lines in the spectra, light curve fluctuations) may imply additional energy contribution from the CSM interaction.

Starting from 273 days from explosion, \sn\ shows a sudden drop in brightness. Although resembling breaks have been observed in a few SLSNe~I \citep[e.g.][]{Inserra17}, several interacting objects \citep[e.g.][]{Mattila08a,Pastorello08b,Ofek14,Tartaglia20}, and in some Super-Chandrasekhar SNe~Ia \citep[e.g.][]{Hsiao20}, the slope measured in \sn\ is unique. While for slow-evolving SLSNe~I, \citet{Inserra17} measured a decline that follows a power law of $t^{-5}$, interacting objects and Super-Chandrasekhar SNe~Ia usually show less steep declines \citep[e.g.][]{Fransson14,Moriya14}. On the contrary, for \sn, we find a much steeper slope, which follows a power law of $t^{-11}$.

It has been discussed that this break can either be caused by the breakout of the shock through the dense shell \citep{Ofek14,Fransson14}, or the result of a transition to a momentum-conserving phase, occurring when shock runs over a mass of CSM equivalent to the ejecta mass (\citealt{Ofek14}, but see \citealt{Moriya14}), or the result of CO formation \citep{Hsiao20}, or even the consequence of dust formation in a cool, dense shell \citep{Pastorello08b,Mattila08a,Smith08}. Despite the breakout of the shock through the dense shell being a plausible explanation for a type IIn event such as SN~2010jl, it is less reliable for \sn. While \sn\ shows some signatures of interaction, its spectra lack flat-topped profiles and intermediate-width features, which are expected from the interaction of the ejecta with dense CSM. However, as mentioned in Section~\ref{sec:expl}, ejecta-CSM interaction may play an important role in the sudden drop observed in the late optical light curve of \sn. Although CO and dust formation could be an option, it is not supported by the spectrum at 293 days from explosion (taken during the early part of the sudden drop). Here, we do not detect any emission line blueshifted, typically seen in SN spectra with dust forming in the ejecta. However, to confirm or reject this alternative, NIR observations are needed.

\subsection{Absence of \ion{O}{ii} lines and presence of \ion{C}{II} lines}
\label{sec:lines}

One of the most common features observed in the early phase spectra of SLSNe-I is the presence of \ion{O}{ii} lines around 4000 -- 4500 \AA\ (\citealt{Quimby11, Quimby18}; but see,  \citealt{Konyves-Toth21}). These lines appear at high temperatures (12000-15000 K; \citealt{Inserra19}) as a consequence of non-thermal excitation \citep{Mazzali16}. In the case of \sn, we noticed that these lines are not visible in the spectra at any epoch (see Figure~\ref{spec}). As shown in Section~\ref{sec:bolo} (and Figure~\ref{bolo}), the temperature of \sn\ evolves in a range of values lower than 10000 K. This temperature is indeed not high enough to ionise the oxygen. \citet{Quimby18} suggested that the lack of these features, at least in standard luminosity SNe~Ic, may be either the product of rapid cooling  or due to a lack of non-thermal sources of excitation. Given that the initial emission can be reproduced as the results of the cooling of an extended progenitor, this is not compatible with a rapid cooling phase. Therefore, the lack of a non-thermal source seems to be the most reliable explanation. This is also supported by our modelling, which does not require a very extended mixing of radioactive material.

\citet{GalYam19} found that, in addition to the O lines, the C lines are also  typical of SLSNe~I at around peak, and suggested that they result from the emission of an almost pure C/O envelope, without significant contamination of higher-mass elements from deeper layers. Inspecting the spectra of \sn, we detect strong and persistent lines of \ion{C}{ii}.
These lines are predicted by spectral models \citep{Dessart12b,Mazzali16,Dessart19}, and are identified in several SLSNe-I (e.g., PTF09cnd \citep{Quimby18}, PTF10aagc \citep{Quimby18}, PTF12dam, SN~2015bn, Gaia16apd \citep{Yan17}, iPFT16bad \citep{Yan17a}, SN~2017gci, SN~2018bsz, SN~2018hti \citep{Lin20,Fiore22}, although in most of the cases, the strength of the lines is moderate. However, \sn\ along with SN~2018bsz, have stronger lines than these other objects, and furthermore their \ion{C}{ii} lines agree with the models of \citet{Dessart19}, which typically overestimate the observed strength of these lines.

In the spectra of \sn, we also detect a strong \ion{Si}{ii}. A weak \ion{Si}{ii} feature has been detected in a few objects (e.g. PTF09cnd and SN~2015bn). From the modelling side, \ion{Si}{ii} is not predicted by the models presented in \citet{Mazzali16}, but it is reproduced by the magnetar model presented in \citet{Dessart12b}. They argue that the presence of the \ion{Si}{ii}, \ion{C}{ii}, and \ion{He}{i} lines is a result of the extra energy from a magnetar that heats the material and thermally excites the gas. Although the detection of these lines in \sn\ supports this scenario, our light curve modelling (Section~\ref{sec:model}) disfavours it.

\subsection{\sn: an extreme case of SN~2007gr?}

As discussed before (Section~\ref{sec:sncomp}), the spectral evolution of \sn\ resembles that of the carbon-rich type Ic SN~2007gr. In Section~\ref{sec:comp}, we showed that their light curves evolve differently. Indeed, \sn\ is over 3 magnitudes brighter than SN~2007gr, it evolves much more slowly (it is $\sim4$ times slower) and has the pre-peak bump that was not detected in its fainter counterpart. However, the remarkable similarity of the spectra suggests that they may have a progenitor with a similar composition. The question that arises from this comparison is why \sn\ and SN~2007gr show comparable spectroscopic behaviour but completely different photometric properties?

In our attempt to provide an answer, we analyse the environments of these objects. In Section~\ref{sec:gal} we mentioned that \sn\ is in a metal-poor environment (12 + log(O/H)$=8.175$ dex by using the mass-metallicity relation of \citealt{Kewley08}). More precisely, its host galaxy is faint, has a low stellar mass and very little star formation, similar to those observed for SLSNe-I. On the other hand, the host of SN~2007gr was identified as a nearby spiral galaxy (M$_{B}=-18.89$\footnote{http://leda.univ-lyon1.fr/}; \citealt{Makarov14}) that also hosted SN~1969L and SN~1961V \citep{Hunter09,Chen14}. According to \citet{Maund16}, SN~2007gr was located at near the centre of a dense young, massive star association \citep{Kuncarayakti13a}. \citet{Modjaz11} measured the metallicity near the position of SN~2007gr by using the O3N2 diagnostic method \citep{Pettini04} and found it was 12 + log(O/H)$=8.64$ dex, indicating a metal-rich environment. These results lead us to conclude that the host environments of \sn\ and SN~2007gr are different, and based on these estimations, also the metallicity of their progenitor stars.  

The nature of the progenitor star and explosion of SN~2007gr has been broadly discussed \citep[e.g.,][]{Crockett08, Chen14,Maund16}. \citet{Chen14} suggested that the progenitor of SN~2007gr was a low-mass Wolf–Rayet star resulting from an interacting binary. \citet{Maund16} support a Wolf–Rayet progenitor, but add that it was an initially massive star. From the nebular modelling, \citet{Mazzali10} found that SN~2007gr was the explosion of a low-mass CO core, probably the result of a star with an initial mass of 15 \Msun.  
For \sn, our light curve modelling suggests a massive progenitor and an energetic explosion with lot of \nifs\ produced. These parameters are, in all respects, more extreme than those found for SN~2007gr. 

\section{Conclusions}
\label{sec:con}

\sn\ is a slow-evolving carbon-rich SLSN-I. Its light curves show an early bump lasting $\sim5$ days followed by a slow rise to the main peak. The peak is reached at different times, occurring faster in the bluer bands. With an absolute peak magnitudes of around $\sim-20.5$ mag, \sn\ is in the low end of the luminosity distribution of SLSNe-I. After 130 days from explosion, the light curves show a linear decline in all bands, with slopes being around the expected decline rate of the \cofs\ decay. Later, from 273 days, a sudden drop in brightness is observed, implying a significant leakage of gamma-ray photons. Our last observations (after 350 days from explosion), show an increase in brightness, which may suggest interaction between the ejecta and the CSM. Indeed, minor light curve fluctuations support this scenario.

During the photospheric phase, the optical spectra show clear lines of \ion{C}{ii} and \ion{Si}{ii}, while the classical \ion{O}{ii} lines that typically characterise SLSNe-I are not detected. The lack of \ion{O}{ii} lines is probably related to the low temperatures of this object (below 10000 K). Late-time spectra display strong lines of \ion{[O}{i]}, \ion{[Ca}{ii]}, \ion{Ca}{ii}, \ion{Mg}{i]}, as well as, a broad emission of \ion{[O}{iii]} and Balmer lines. 

We modelled the light curve and the expansion velocities of \sn\ using a one-dimensional hydrodynamics code. Two scenarios were investigated, with the radioactive nickel and the magnetar as primary powering sources. In both cases, we found that an extended progenitor was required to reproduce the time scale of the peaks. However, the magnetar model produces a much worse fit to the data. Therefore, we consider the \nifs\ as the main power source. Specifically, we found that \sn\ can be explained by a progenitor with a pre-SN mass of 80 \Msun, a pre-SN radius of 15~\Rsun, an explosion energy of $45\times10^{51}$ erg, and ejecting 4 \Msun\ of \nifs.
In this scenario, the first peak results from a post-shock cooling phase for the extended progenitor, and the luminous main peak is due to a large \nifs\ production. The values of the parameters obtained are consistent with those expected for a PISN, which provide support for this scenario in the case of \sn. Although our model reproduces the almost complete evolution of the light curve reasonably well, it fails to explain the excess of flux at $\sim245$ days and the shape of the light curve after that. We propose that this behaviour is probably due to an additional contribution of ejecta-CSM interaction. The drop in brightness after 273 days from explosion could be attributed to either the end of earlier ejecta-CSM interaction, or the formation of molecules and dust in the SN ejecta or in a shocked cool dense shell. However, NIR observation are needed to confirm this suggestion.

We noticed remarkable spectral similarities between \sn\ and carbon-rich type Ic SN~2007gr, but on a longer time scale. This resemblance may suggest a connection between these two events, most probably associated with the structures of their progenitor stars. 
Although we have found a model that can explain the main photometric properties of \sn, we also discussed a possibility where the CSM interaction is the primary energy source already at early phases. However, we did not explore in detail this alternative scenario due to the lack of an available model with such extreme input parameters as those observed in \sn.

\section*{Acknowledgements}

We thank the anonymous referee for the comments and suggestions that have helped to improve the paper.
M.B. and M.O. acknowledge support from UNRN PI2020 40B885 and grant PICT-2020-SERIEA-01141 and PIP 112-202001-10034. 
A.R. acknowledges support from ANID BECAS/DOCTORADO NACIONAL 21202412. 
T.M.R. acknowledges the financial support of the Finnish Academy of Science and Letters. 
N.E.R. acknowledges partial support from MIUR, PRIN 2017 (grant 20179ZF5KS), from the Spanish MICINN grant PID2019-108709GB-I00 and FEDER funds, and from the program Unidad de Excelencia María de Maeztu CEX2020-001058-M. 
M.F. is supported by a Royal Society - Science Foundation Ireland University Research Fellowship. 
S.M. acknowledges support from the Magnus Ehrnrooth Foundation and the Vilho, Yrjö and Kalle Väisälä Foundation. 
M.S. was a visiting astronomer at the Infrared Telescope Facility, which is operated by the University of Hawaii under contract 80HQTR19D0030 with the National Aeronautics and Space Administration. 
X.W. is supported by the National Science Foundation of China (NSFC grants 12033003 and 11633002), the Scholar Program of Beijing Academy of Science and Technology (DZ:BS202002), and the Tencent Xplorer Prize. 
J.Z. is supported by the NSFC (grants 12173082, 11773067), by the Youth Innovation Promotion Association of the CAS (grant 2018081), and by the Ten Thousand Talents Program of Yunnan for Top-notch Young Talents. 
Funding for the LJT has been provided by Chinese Academy of Sciences and the People’s Government of Yunnan Province. The LJT is jointly operated and administrated by Yunnan Observatories and Center for Astronomical Mega-Science, CAS. 
This work is supported by the National Natural Science Foundation of China (NSFC grants 12033003, 11633002, 11325313, and 11761141001), the National Program on Key Research and Development Project (grant no. 2016YFA0400803). 
This work is partially supported by China Manned Spaced Project (CMS-CSST-2021-A12). 
This work is funded by China Postdoctoral Science Foundation (grant no. 2021M691821).\\
We are grateful to Jan Aaltonen, Ville Antila, Katja Matilainen and Sofia Suutarinen, who observed this target as part of the "NOT course 2021" organised by the Department of Physics and Astronomy at the University of Turku in October 2021.\\
Based on observations made with the Nordic Optical Telescope, owned in collaboration by the University of Turku and Aarhus University, and operated jointly by Aarhus University, the University of Turku and the University of Oslo, representing Denmark, Finland and Norway, the University of Iceland and Stockholm University at the Observatorio del Roque de los Muchachos, La Palma, Spain, of the Instituto de Astrofisica de Canarias.\\
Observations from the NOT were obtained through the NUTS2 collaboration which are supported in part by the Instrument Centre for Danish Astrophysics (IDA). The data presented here were obtained in part with ALFOSC, which is provided by the Instituto de Astrofisica de Andalucia (IAA) under a joint agreement with the University of Copenhagen and NOTSA. \\
Partially based on observations collected at Copernico and Schmidt telescopes (Asiago, Italy) of the INAF - Osservatorio Astronomico di Padova.\\
Based on observations made with the Gran Telescopio Canarias (GTC), installed in the Spanish Observatorio del Roque de los Muchachos of the Instituto de Astrofísica de Canarias, in the island of La Palma.\\
We acknowledge the support of the staff of the Lijiang 2.4m, Xinglong 2.16m and 80cm telescopes.\\
The Liverpool Telescope is operated on the island of La Palma by Liverpool John Moores University in the Spanish Observatorio del Roque de los Muchachos of the Instituto de Astrofisica de Canarias with financial support from the UK Science and Technology Facilities Council.\\
This work has made use of data from the Asteroid Terrestrial-impact Last Alert System (ATLAS) project. The Asteroid Terrestrial-impact Last Alert System (ATLAS) project is primarily funded to search for near earth asteroids through NASA grants NN12AR55G, 80NSSC18K0284, and 80NSSC18K1575; byproducts of the NEO search include images and catalogs from the survey area. This work was partially funded by Kepler/K2 grant J1944/80NSSC19K0112 and HST GO-15889, and STFC grants ST/T000198/1 and ST/S006109/1. The ATLAS science products have been made possible through the contributions of the University of Hawaii Institute for Astronomy, the Queen’s University Belfast, the Space Telescope Science Institute, the South African Astronomical Observatory, and The Millennium Institute of Astrophysics (MAS), Chile.\\
The ZTF forced-photometry service was funded under the Heising-Simons Foundation grant \#12540303 (PI: Graham).

\section*{Data Availability}

The data underlying this article are available in the appendix A (Tables A1 -- A5) and through the WISeREP (\url{https://wiserep.weizmann.ac.il/home}) archive \citep{Yaron12}.



\bibliographystyle{mnras}
\bibliography{Bibliography}

\begin{thebibliography}{}
\makeatletter
\relax
\def\mn@urlcharsother{\let\do\@makeother \do\$\do\&\do\#\do\^\do\_\do\%\do\~}
\def\mn@doi{\begingroup\mn@urlcharsother \@ifnextchar [ {\mn@doi@}
  {\mn@doi@[]}}
\def\mn@doi@[#1]#2{\def\@tempa{#1}\ifx\@tempa\@empty \href
  {http://dx.doi.org/#2} {doi:#2}\else \href {http://dx.doi.org/#2} {#1}\fi
  \endgroup}
\def\mn@eprint#1#2{\mn@eprint@#1:#2::\@nil}
\def\mn@eprint@arXiv#1{\href {http://arxiv.org/abs/#1} {{\tt arXiv:#1}}}
\def\mn@eprint@dblp#1{\href {http://dblp.uni-trier.de/rec/bibtex/#1.xml}
  {dblp:#1}}
\def\mn@eprint@#1:#2:#3:#4\@nil{\def\@tempa {#1}\def\@tempb {#2}\def\@tempc
  {#3}\ifx \@tempc \@empty \let \@tempc \@tempb \let \@tempb \@tempa \fi \ifx
  \@tempb \@empty \def\@tempb {arXiv}\fi \@ifundefined
  {mn@eprint@\@tempb}{\@tempb:\@tempc}{\expandafter \expandafter \csname
  mn@eprint@\@tempb\endcsname \expandafter{\@tempc}}}

\bibitem[\protect\citeauthoryear{{Ambikasaran} et~al.}{{Ambikasaran}
  et~al.}{2016}]{Ambikasaran16}
{Ambikasaran} S.,  et~al., 2016, IEEE Trans. Pattern Anal. Mach. Intell., 38

\bibitem[\protect\citeauthoryear{{Anderson} et~al.,}{{Anderson}
  et~al.}{2018}]{Anderson18a}
{Anderson} J.~P.,  et~al., 2018, \mn@doi [\aap] {10.1051/0004-6361/201833725},
  \href {https://ui.adsabs.harvard.edu/abs/2018A&A...620A..67A} {620, A67}

\bibitem[\protect\citeauthoryear{{Angus} et~al.,}{{Angus}
  et~al.}{2019}]{Angus19}
{Angus} C.~R.,  et~al., 2019, \mn@doi [\mnras] {10.1093/mnras/stz1321}, \href
  {https://ui.adsabs.harvard.edu/abs/2019MNRAS.487.2215A} {487, 2215}

\bibitem[\protect\citeauthoryear{{Anupama}, {Sahu}, {Deng}, {Nomoto},
  {Tominaga}, {Tanaka}, {Mazzali}  \& {Prabhu}}{{Anupama}
  et~al.}{2005}]{Anupama05}
{Anupama} G.~C.,  {Sahu} D.~K.,  {Deng} J.,  {Nomoto} K.,  {Tominaga} N.,
  {Tanaka} M.,  {Mazzali} P.~A.,   {Prabhu} T.~P.,  2005, \mn@doi [\apjl]
  {10.1086/497336}, \href
  {https://ui.adsabs.harvard.edu/abs/2005ApJ...631L.125A} {631, L125}

\bibitem[\protect\citeauthoryear{{Astropy Collaboration} et~al.,}{{Astropy
  Collaboration} et~al.}{2018}]{Astropy18}
{Astropy Collaboration} et~al., 2018, \mn@doi [\aj] {10.3847/1538-3881/aabc4f},
  \href {https://ui.adsabs.harvard.edu/abs/2018AJ....156..123A} {156, 123}

\bibitem[\protect\citeauthoryear{{Barkat}, {Rakavy}  \& {Sack}}{{Barkat}
  et~al.}{1967}]{Barkat67}
{Barkat} Z.,  {Rakavy} G.,   {Sack} N.,  1967, \mn@doi [\prl]
  {10.1103/PhysRevLett.18.379}, \href
  {https://ui.adsabs.harvard.edu/abs/1967PhRvL..18..379B} {18, 379}

\bibitem[\protect\citeauthoryear{{Bellm} et~al.,}{{Bellm}
  et~al.}{2019}]{Bellm19}
{Bellm} E.~C.,  et~al., 2019, \mn@doi [\pasp] {10.1088/1538-3873/aaecbe}, \href
  {https://ui.adsabs.harvard.edu/abs/2019PASP..131a8002B} {131, 018002}

\bibitem[\protect\citeauthoryear{{Bersten}, {Benvenuto}  \& {Hamuy}}{{Bersten}
  et~al.}{2011}]{Bersten11}
{Bersten} M.~C.,  {Benvenuto} O.,   {Hamuy} M.,  2011, \mn@doi [\apj]
  {10.1088/0004-637X/729/1/61}, \href
  {http://adsabs.harvard.edu/abs/2011ApJ...729...61B} {729, 61}

\bibitem[\protect\citeauthoryear{{Bersten}, {Benvenuto}, {Orellana}  \&
  {Nomoto}}{{Bersten} et~al.}{2016}]{Bersten16}
{Bersten} M.~C.,  {Benvenuto} O.~G.,  {Orellana} M.,   {Nomoto} K.,  2016,
  \mn@doi [\apjl] {10.3847/2041-8205/817/1/L8}, \href
  {http://adsabs.harvard.edu/abs/2016ApJ...817L...8B} {817, L8}

\bibitem[\protect\citeauthoryear{{Bianco} et~al.,}{{Bianco}
  et~al.}{2014}]{Bianco14}
{Bianco} F.~B.,  et~al., 2014, \mn@doi [\apjs] {10.1088/0067-0049/213/2/19},
  \href {https://ui.adsabs.harvard.edu/abs/2014ApJS..213...19B} {213, 19}

\bibitem[\protect\citeauthoryear{Bradley et~al.,}{Bradley
  et~al.}{2019}]{Bradley19}
Bradley L.,  et~al., 2019, astropy/photutils: v0.6,
  \mn@doi{10.5281/zenodo.2533376}, \url
  {https://doi.org/10.5281/zenodo.2533376}

\bibitem[\protect\citeauthoryear{{Cappellaro}}{{Cappellaro}}{2014}]{Cappellaro14}
{Cappellaro} E.,  2014, {SNOoPY: a package for SN photometry, {\url{
  http://sngroup.oapd.inaf.it/snoopy.html}}}

\bibitem[\protect\citeauthoryear{{Cardelli}, {Clayton}  \& {Mathis}}{{Cardelli}
  et~al.}{1989}]{Cardelli89}
{Cardelli} J.~A.,  {Clayton} G.~C.,   {Mathis} J.~S.,  1989, \mn@doi [\apj]
  {10.1086/167900}, \href {http://adsabs.harvard.edu/abs/1989ApJ...345..245C}
  {345, 245}

\bibitem[\protect\citeauthoryear{{Chambers} et~al.,}{{Chambers}
  et~al.}{2016}]{Chambers16}
{Chambers} K.~C.,  et~al., 2016, preprint, \href
  {http://adsabs.harvard.edu/abs/2016arXiv161205560C} {} (\mn@eprint {arXiv}
  {1612.05560})

\bibitem[\protect\citeauthoryear{{Chatzopoulos} et~al.}{{Chatzopoulos}
  et~al.}{2011}]{Chatzopoulos11}
{Chatzopoulos} E.,  et~al., 2011, ArXiv e-prints:1101.3581, \href
  {http://adsabs.harvard.edu/abs/2011arXiv1101.3581C} {}

\bibitem[\protect\citeauthoryear{{Chatzopoulos}, {Wheeler}, {Vinko}, {Nagy},
  {Wiggins}  \& {Even}}{{Chatzopoulos} et~al.}{2016}]{Chatzopoulos16}
{Chatzopoulos} E.,  {Wheeler} J.~C.,  {Vinko} J.,  {Nagy} A.~P.,  {Wiggins}
  B.~K.,   {Even} W.~P.,  2016, \mn@doi [\apj] {10.3847/0004-637X/828/2/94},
  \href {https://ui.adsabs.harvard.edu/abs/2016ApJ...828...94C} {828, 94}

\bibitem[\protect\citeauthoryear{{Chen} et~al.,}{{Chen} et~al.}{2013}]{Chen13}
{Chen} T.-W.,  et~al., 2013, \mn@doi [\apjl] {10.1088/2041-8205/763/2/L28},
  \href {https://ui.adsabs.harvard.edu/abs/2013ApJ...763L..28C} {763, L28}

\bibitem[\protect\citeauthoryear{{Chen} et~al.,}{{Chen} et~al.}{2014}]{Chen14}
{Chen} J.,  et~al., 2014, \mn@doi [\apj] {10.1088/0004-637X/790/2/120}, \href
  {https://ui.adsabs.harvard.edu/abs/2014ApJ...790..120C} {790, 120}

\bibitem[\protect\citeauthoryear{{Chen} et~al.,}{{Chen} et~al.}{2015}]{Chen15}
{Chen} T.~W.,  et~al., 2015, \mn@doi [\mnras] {10.1093/mnras/stv1360}, \href
  {https://ui.adsabs.harvard.edu/abs/2015MNRAS.452.1567C} {452, 1567}

\bibitem[\protect\citeauthoryear{{Chen}, {Smartt}, {Yates}, {Nicholl},
  {Kr{\"u}hler}, {Schady}, {Dennefeld}  \& {Inserra}}{{Chen}
  et~al.}{2017}]{Chen17}
{Chen} T.-W.,  {Smartt} S.~J.,  {Yates} R.~M.,  {Nicholl} M.,  {Kr{\"u}hler}
  T.,  {Schady} P.,  {Dennefeld} M.,   {Inserra} C.,  2017, \mn@doi [\mnras]
  {10.1093/mnras/stx1428}, \href
  {https://ui.adsabs.harvard.edu/abs/2017MNRAS.470.3566C} {470, 3566}

\bibitem[\protect\citeauthoryear{{Chen} et~al.,}{{Chen} et~al.}{2021}]{Chen21}
{Chen} T.~W.,  et~al., 2021, arXiv e-prints, \href
  {https://ui.adsabs.harvard.edu/abs/2021arXiv210907942C} {p. arXiv:2109.07942}

\bibitem[\protect\citeauthoryear{{Chevalier} \& {Irwin}}{{Chevalier} \&
  {Irwin}}{2011}]{Chevalier11}
{Chevalier} R.~A.,  {Irwin} C.~M.,  2011, \mn@doi [\apjl]
  {10.1088/2041-8205/729/1/L6}, \href
  {https://ui.adsabs.harvard.edu/abs/2011ApJ...729L...6C} {729, L6}

\bibitem[\protect\citeauthoryear{{Childress} et~al.,}{{Childress}
  et~al.}{2013}]{Childress13a}
{Childress} M.,  et~al., 2013, \mn@doi [\apj] {10.1088/0004-637X/770/2/107},
  \href {https://ui.adsabs.harvard.edu/abs/2013ApJ...770..107C} {770, 107}

\bibitem[\protect\citeauthoryear{{Chonis} \& {Gaskell}}{{Chonis} \&
  {Gaskell}}{2008}]{Chonis08}
{Chonis} T.~S.,  {Gaskell} C.~M.,  2008, \mn@doi [\aj]
  {10.1088/0004-6256/135/1/264}, \href
  {https://ui.adsabs.harvard.edu/\#abs/2008AJ....135..264C} {135, 264}

\bibitem[\protect\citeauthoryear{{Crockett} et~al.,}{{Crockett}
  et~al.}{2008}]{Crockett08}
{Crockett} R.~M.,  et~al., 2008, \mn@doi [\apjl] {10.1086/527299}, \href
  {https://ui.adsabs.harvard.edu/abs/2008ApJ...672L..99C} {672, L99}

\bibitem[\protect\citeauthoryear{{Cushing}, {Vacca}  \& {Rayner}}{{Cushing}
  et~al.}{2004}]{Cushing04}
{Cushing} M.~C.,  {Vacca} W.~D.,   {Rayner} J.~T.,  2004, \mn@doi [\pasp]
  {10.1086/382907}, \href
  {https://ui.adsabs.harvard.edu/abs/2004PASP..116..362C} {116, 362}

\bibitem[\protect\citeauthoryear{{De Cia} et~al.,}{{De Cia}
  et~al.}{2018}]{DeCia18}
{De Cia} A.,  et~al., 2018, \mn@doi [\apj] {10.3847/1538-4357/aab9b6}, \href
  {https://ui.adsabs.harvard.edu/abs/2018ApJ...860..100D} {860, 100}

\bibitem[\protect\citeauthoryear{{Dessart}}{{Dessart}}{2019}]{Dessart19}
{Dessart} L.,  2019, \mn@doi [\aap] {10.1051/0004-6361/201834535}, \href
  {https://ui.adsabs.harvard.edu/abs/2019A&A...621A.141D} {621, A141}

\bibitem[\protect\citeauthoryear{{Dessart}, {Hillier}, {Waldman}, {Livne}  \&
  {Blondin}}{{Dessart} et~al.}{2012}]{Dessart12b}
{Dessart} L.,  {Hillier} D.~J.,  {Waldman} R.,  {Livne} E.,   {Blondin} S.,
  2012, \mn@doi [\mnras] {10.1111/j.1745-3933.2012.01329.x}, \href
  {http://adsabs.harvard.edu/abs/2012MNRAS.426L..76D} {426, L76}

\bibitem[\protect\citeauthoryear{{Dessart}, {Waldman}, {Livne}, {Hillier}  \&
  {Blondin}}{{Dessart} et~al.}{2013}]{Dessart13}
{Dessart} L.,  {Waldman} R.,  {Livne} E.,  {Hillier} D.~J.,   {Blondin} S.,
  2013, \mn@doi [\mnras] {10.1093/mnras/sts269}, \href
  {http://adsabs.harvard.edu/abs/2013MNRAS.428.3227D} {428, 3227}

\bibitem[\protect\citeauthoryear{{Dessart}, {Audit}  \& {Hillier}}{{Dessart}
  et~al.}{2015}]{Dessart15}
{Dessart} L.,  {Audit} E.,   {Hillier} D.~J.,  2015, \mn@doi [\mnras]
  {10.1093/mnras/stv609}, \href
  {https://ui.adsabs.harvard.edu/abs/2015MNRAS.449.4304D} {449, 4304}

\bibitem[\protect\citeauthoryear{{Dessart}, {Hillier}, {Sukhbold}, {Woosley}
  \& {Janka}}{{Dessart} et~al.}{2021}]{Dessart21}
{Dessart} L.,  {Hillier} D.~J.,  {Sukhbold} T.,  {Woosley} S.~E.,   {Janka}
  H.~T.,  2021, \mn@doi [\aap] {10.1051/0004-6361/202140839}, \href
  {https://ui.adsabs.harvard.edu/abs/2021A&A...652A..64D} {652, A64}

\bibitem[\protect\citeauthoryear{{Elmhamdi}, {Danziger}, {Cappellaro}, {Della
  Valle}, {Gouiffes}, {Phillips}  \& {Turatto}}{{Elmhamdi}
  et~al.}{2004}]{Elmhamdi04}
{Elmhamdi} A.,  {Danziger} I.~J.,  {Cappellaro} E.,  {Della Valle} M.,
  {Gouiffes} C.,  {Phillips} M.~M.,   {Turatto} M.,  2004, \mn@doi [\aap]
  {10.1051/0004-6361:20041318}, \href
  {https://ui.adsabs.harvard.edu/abs/2004A&A...426..963E} {426, 963}

\bibitem[\protect\citeauthoryear{{Fesen} et~al.,}{{Fesen}
  et~al.}{1999}]{Fesen99}
{Fesen} R.~A.,  et~al., 1999, \mn@doi [\aj] {10.1086/300751}, \href
  {http://adsabs.harvard.edu/abs/1999AJ....117..725F} {117, 725}

\bibitem[\protect\citeauthoryear{{Fioc} \& {Rocca-Volmerange}}{{Fioc} \&
  {Rocca-Volmerange}}{1997}]{Fioc97}
{Fioc} M.,  {Rocca-Volmerange} B.,  1997, \aap, \href
  {http://adsabs.harvard.edu/abs/1997A%26A...326..950F} {326, 950}

\bibitem[\protect\citeauthoryear{{Fiore} et~al.,}{{Fiore}
  et~al.}{2021}]{Fiore21}
{Fiore} A.,  et~al., 2021, \mn@doi [\mnras] {10.1093/mnras/staa4035}, \href
  {https://ui.adsabs.harvard.edu/abs/2021MNRAS.502.2120F} {502, 2120}

\bibitem[\protect\citeauthoryear{{Fiore} et~al.,}{{Fiore}
  et~al.}{2022}]{Fiore22}
{Fiore} A.,  et~al., 2022, \mn@doi [\mnras] {10.1093/mnras/stac744}, \href
  {https://ui.adsabs.harvard.edu/abs/2022MNRAS.512.4484F} {512, 4484}

\bibitem[\protect\citeauthoryear{{Fisher}}{{Fisher}}{2000}]{Fisher00}
{Fisher} A.~K.,  2000, PhD thesis, THE UNIVERSITY OF OKLAHOMA

\bibitem[\protect\citeauthoryear{{Folatelli} et~al.}{{Folatelli}
  et~al.}{2006}]{Folatelli06}
{Folatelli} G.,  et~al., 2006, \mn@doi [\apj] {10.1086/500531}, \href
  {http://adsabs.harvard.edu/abs/2006ApJ...641.1039F} {641, 1039}

\bibitem[\protect\citeauthoryear{{F{\"o}rster} et~al.,}{{F{\"o}rster}
  et~al.}{2021}]{Forster21}
{F{\"o}rster} F.,  et~al., 2021, \mn@doi [\aj] {10.3847/1538-3881/abe9bc},
  \href {https://ui.adsabs.harvard.edu/abs/2021AJ....161..242F} {161, 242}

\bibitem[\protect\citeauthoryear{{Fransson} \& {Chevalier}}{{Fransson} \&
  {Chevalier}}{1989}]{Fransson89}
{Fransson} C.,  {Chevalier} R.~A.,  1989, \mn@doi [\apj] {10.1086/167707},
  \href {https://ui.adsabs.harvard.edu/abs/1989ApJ...343..323F} {343, 323}

\bibitem[\protect\citeauthoryear{{Fransson} et~al.,}{{Fransson}
  et~al.}{2002}]{Fransson02}
{Fransson} C.,  et~al., 2002, \mn@doi [\apj] {10.1086/340295}, \href
  {https://ui.adsabs.harvard.edu/abs/2002ApJ...572..350F} {572, 350}

\bibitem[\protect\citeauthoryear{{Fransson} et~al.,}{{Fransson}
  et~al.}{2014}]{Fransson14}
{Fransson} C.,  et~al., 2014, \mn@doi [\apj] {10.1088/0004-637X/797/2/118},
  \href {https://ui.adsabs.harvard.edu/abs/2014ApJ...797..118F} {797, 118}

\bibitem[\protect\citeauthoryear{{Gal-Yam}}{{Gal-Yam}}{2012}]{GalYam12}
{Gal-Yam} A.,  2012, \mn@doi [Science] {10.1126/science.1203601}, \href
  {https://ui.adsabs.harvard.edu/abs/2012Sci...337..927G} {337, 927}

\bibitem[\protect\citeauthoryear{{Gal-Yam}}{{Gal-Yam}}{2019}]{GalYam19}
{Gal-Yam} A.,  2019, \mn@doi [\apj] {10.3847/1538-4357/ab2f79}, \href
  {https://ui.adsabs.harvard.edu/abs/2019ApJ...882..102G} {882, 102}

\bibitem[\protect\citeauthoryear{{Gal-Yam} et~al.,}{{Gal-Yam}
  et~al.}{2009}]{GalYam09a}
{Gal-Yam} A.,  et~al., 2009, \mn@doi [\nat] {10.1038/nature08579}, \href
  {https://ui.adsabs.harvard.edu/abs/2009Natur.462..624G} {462, 624}

\bibitem[\protect\citeauthoryear{{Ginzburg} \& {Balberg}}{{Ginzburg} \&
  {Balberg}}{2012}]{Ginzburg12}
{Ginzburg} S.,  {Balberg} S.,  2012, \mn@doi [\apj]
  {10.1088/0004-637X/757/2/178}, \href
  {https://ui.adsabs.harvard.edu/abs/2012ApJ...757..178G} {757, 178}

\bibitem[\protect\citeauthoryear{{Gomez}, {Berger}, {Nicholl}, {Blanchard}  \&
  {Hosseinzadeh}}{{Gomez} et~al.}{2022}]{Gomez22}
{Gomez} S.,  {Berger} E.,  {Nicholl} M.,  {Blanchard} P.~K.,   {Hosseinzadeh}
  G.,  2022, arXiv e-prints, \href
  {https://ui.adsabs.harvard.edu/abs/2022arXiv220408486G} {p. arXiv:2204.08486}

\bibitem[\protect\citeauthoryear{{Graham} et~al.,}{{Graham}
  et~al.}{2019}]{Graham19}
{Graham} M.~J.,  et~al., 2019, \mn@doi [\pasp] {10.1088/1538-3873/ab006c},
  \href {https://ui.adsabs.harvard.edu/abs/2019PASP..131g8001G} {131, 078001}

\bibitem[\protect\citeauthoryear{{Guti{\'e}rrez} et~al.,}{{Guti{\'e}rrez}
  et~al.}{2017}]{Gutierrez17a}
{Guti{\'e}rrez} C.~P.,  et~al., 2017, \mn@doi [\apj]
  {10.3847/1538-4357/aa8f52}, \href
  {http://adsabs.harvard.edu/abs/2017ApJ...850...89G} {850, 89}

\bibitem[\protect\citeauthoryear{{Guti{\'e}rrez} et~al.,}{{Guti{\'e}rrez}
  et~al.}{2020a}]{Gutierrez20}
{Guti{\'e}rrez} C.~P.,  et~al., 2020a, \mn@doi [\mnras]
  {10.1093/mnras/staa1452}, \href
  {https://ui.adsabs.harvard.edu/abs/2020MNRAS.496...95G} {496, 95}

\bibitem[\protect\citeauthoryear{{Guti{\'e}rrez} et~al.,}{{Guti{\'e}rrez}
  et~al.}{2020b}]{Gutierrez20a}
{Guti{\'e}rrez} C.~P.,  et~al., 2020b, \mn@doi [\mnras]
  {10.1093/mnras/staa2763}, \href
  {https://ui.adsabs.harvard.edu/abs/2020MNRAS.499..974G} {499, 974}

\bibitem[\protect\citeauthoryear{{Guti{\'e}rrez} et~al.,}{{Guti{\'e}rrez}
  et~al.}{2021}]{Gutierrez21}
{Guti{\'e}rrez} C.~P.,  et~al., 2021, \mn@doi [\mnras]
  {10.1093/mnras/stab1009}, \href
  {https://ui.adsabs.harvard.edu/abs/2021MNRAS.504.4907G} {504, 4907}

\bibitem[\protect\citeauthoryear{{Harutyunyan} et~al.}{{Harutyunyan}
  et~al.}{2008}]{Harutyunyan08}
{Harutyunyan} A.~H.,  et~al., 2008, \mn@doi [\aap]
  {10.1051/0004-6361:20078859}, \href
  {http://adsabs.harvard.edu/abs/2008A%26A...488..383H} {488, 383}

\bibitem[\protect\citeauthoryear{{Heger} \& {Woosley}}{{Heger} \&
  {Woosley}}{2002}]{Heger02}
{Heger} A.,  {Woosley} S.~E.,  2002, \mn@doi [\apj] {10.1086/338487}, \href
  {https://ui.adsabs.harvard.edu/abs/2002ApJ...567..532H} {567, 532}

\bibitem[\protect\citeauthoryear{{Heger}, {Fryer}, {Woosley}, {Langer}  \&
  {Hartmann}}{{Heger} et~al.}{2003}]{Heger03}
{Heger} A.,  {Fryer} C.~L.,  {Woosley} S.~E.,  {Langer} N.,   {Hartmann} D.~H.,
   2003, \mn@doi [\apj] {10.1086/375341}, \href
  {http://adsabs.harvard.edu/abs/2003ApJ...591..288H} {591, 288}

\bibitem[\protect\citeauthoryear{{Houck} \& {Fransson}}{{Houck} \&
  {Fransson}}{1996}]{Houck96}
{Houck} J.~C.,  {Fransson} C.,  1996, \mn@doi [\apj] {10.1086/176699}, \href
  {https://ui.adsabs.harvard.edu/abs/1996ApJ...456..811H} {456, 811}

\bibitem[\protect\citeauthoryear{{Hsiao} et~al.,}{{Hsiao}
  et~al.}{2019}]{Hsiao19}
{Hsiao} E.~Y.,  et~al., 2019, \mn@doi [\pasp] {10.1088/1538-3873/aae961}, \href
  {https://ui.adsabs.harvard.edu/abs/2019PASP..131a4002H} {131, 014002}

\bibitem[\protect\citeauthoryear{{Hsiao} et~al.,}{{Hsiao}
  et~al.}{2020}]{Hsiao20}
{Hsiao} E.~Y.,  et~al., 2020, \mn@doi [\apj] {10.3847/1538-4357/abaf4c}, \href
  {https://ui.adsabs.harvard.edu/abs/2020ApJ...900..140H} {900, 140}

\bibitem[\protect\citeauthoryear{{Huang}, {Li}, {Wang}, {Shang}, {Zhang}, {Hu},
  {Qiu}  \& {Jiang}}{{Huang} et~al.}{2012}]{Huang12}
{Huang} F.,  {Li} J.-Z.,  {Wang} X.-F.,  {Shang} R.-C.,  {Zhang} T.-M.,  {Hu}
  J.-Y.,  {Qiu} Y.-L.,   {Jiang} X.-J.,  2012, \mn@doi [Research in Astronomy
  and Astrophysics] {10.1088/1674-4527/12/11/012}, \href
  {https://ui.adsabs.harvard.edu/abs/2012RAA....12.1585H} {12, 1585}

\bibitem[\protect\citeauthoryear{{Hunter} et~al.}{{Hunter}
  et~al.}{2009}]{Hunter09}
{Hunter} D.~J.,  et~al., 2009, \mn@doi [\aap] {10.1051/0004-6361/200912896},
  \href {http://adsabs.harvard.edu/abs/2009A\%26A...508..371H} {508, 371}

\bibitem[\protect\citeauthoryear{{Inserra}}{{Inserra}}{2019}]{Inserra19}
{Inserra} C.,  2019, \mn@doi [Nature Astronomy] {10.1038/s41550-019-0854-4},
  \href {https://ui.adsabs.harvard.edu/abs/2019NatAs...3..697I} {3, 697}

\bibitem[\protect\citeauthoryear{{Inserra} et~al.}{{Inserra}
  et~al.}{2013}]{Inserra13a}
{Inserra} C.,  et~al., 2013, \mn@doi [\apj] {10.1088/0004-637X/770/2/128},
  \href {http://adsabs.harvard.edu/abs/2013ApJ...770..128I} {770, 128}

\bibitem[\protect\citeauthoryear{{Inserra} et~al.,}{{Inserra}
  et~al.}{2017}]{Inserra17}
{Inserra} C.,  et~al., 2017, \mn@doi [\mnras] {10.1093/mnras/stx834}, \href
  {https://ui.adsabs.harvard.edu/abs/2017MNRAS.468.4642I} {468, 4642}

\bibitem[\protect\citeauthoryear{{Inserra} et~al.,}{{Inserra}
  et~al.}{2018}]{Inserra18}
{Inserra} C.,  et~al., 2018, \mn@doi [\mnras] {10.1093/mnras/stx3179}, \href
  {https://ui.adsabs.harvard.edu/\#abs/2018MNRAS.475.1046I} {475, 1046}

\bibitem[\protect\citeauthoryear{{Janka}}{{Janka}}{2012}]{Janka12}
{Janka} H.-T.,  2012, \mn@doi [Annual Review of Nuclear and Particle Science]
  {10.1146/annurev-nucl-102711-094901}, \href
  {https://ui.adsabs.harvard.edu/abs/2012ARNPS..62..407J} {62, 407}

\bibitem[\protect\citeauthoryear{{Jerkstrand}}{{Jerkstrand}}{2017}]{Jerkstrand17}
{Jerkstrand} A.,  2017, {Spectra of Supernovae in the Nebular Phase}.
p.~795, \mn@doi{10.1007/978-3-319-21846-5_29}

\bibitem[\protect\citeauthoryear{{Jerkstrand}, {Smartt}, {Fraser}, {Fransson},
  {Sollerman}, {Taddia}  \& {Kotak}}{{Jerkstrand} et~al.}{2014}]{Jerkstrand14}
{Jerkstrand} A.,  {Smartt} S.~J.,  {Fraser} M.,  {Fransson} C.,  {Sollerman}
  J.,  {Taddia} F.,   {Kotak} R.,  2014, \mn@doi [\mnras]
  {10.1093/mnras/stu221}, \href
  {https://ui.adsabs.harvard.edu/abs/2014MNRAS.439.3694J} {439, 3694}

\bibitem[\protect\citeauthoryear{{Jerkstrand}, {Smartt}  \&
  {Heger}}{{Jerkstrand} et~al.}{2016}]{Jerkstrand16}
{Jerkstrand} A.,  {Smartt} S.~J.,   {Heger} A.,  2016, \mn@doi [\mnras]
  {10.1093/mnras/stv2369}, \href
  {https://ui.adsabs.harvard.edu/abs/2016MNRAS.455.3207J} {455, 3207}

\bibitem[\protect\citeauthoryear{{Jerkstrand} et~al.,}{{Jerkstrand}
  et~al.}{2017}]{Jerkstrand17a}
{Jerkstrand} A.,  et~al., 2017, \mn@doi [\apj] {10.3847/1538-4357/835/1/13},
  \href {https://ui.adsabs.harvard.edu/abs/2017ApJ...835...13J} {835, 13}

\bibitem[\protect\citeauthoryear{{Kangas} et~al.,}{{Kangas}
  et~al.}{2017}]{Kangas17}
{Kangas} T.,  et~al., 2017, \mn@doi [\mnras] {10.1093/mnras/stx833}, \href
  {https://ui.adsabs.harvard.edu/abs/2017MNRAS.469.1246K} {469, 1246}

\bibitem[\protect\citeauthoryear{{Kaplan} \& {Soker}}{{Kaplan} \&
  {Soker}}{2020}]{Kaplan20}
{Kaplan} N.,  {Soker} N.,  2020, \mn@doi [\mnras] {10.1093/mnras/staa020},
  \href {https://ui.adsabs.harvard.edu/abs/2020MNRAS.492.3013K} {492, 3013}

\bibitem[\protect\citeauthoryear{{Kasen}}{{Kasen}}{2010}]{Kasen10}
{Kasen} D.,  2010, \mn@doi [\apj] {10.1088/0004-637X/708/2/1025}, \href
  {http://adsabs.harvard.edu/abs/2010ApJ...708.1025K} {708, 1025}

\bibitem[\protect\citeauthoryear{{Kasen} \& {Bildsten}}{{Kasen} \&
  {Bildsten}}{2010}]{Kasen10b}
{Kasen} D.,  {Bildsten} L.,  2010, \mn@doi [\apj]
  {10.1088/0004-637X/717/1/245}, \href
  {http://adsabs.harvard.edu/abs/2010ApJ...717..245K} {717, 245}

\bibitem[\protect\citeauthoryear{{Kasen}, {Metzger}  \& {Bildsten}}{{Kasen}
  et~al.}{2016}]{Kasen16}
{Kasen} D.,  {Metzger} B.~D.,   {Bildsten} L.,  2016, \mn@doi [\apj]
  {10.3847/0004-637X/821/1/36}, \href
  {https://ui.adsabs.harvard.edu/abs/2016ApJ...821...36K} {821, 36}

\bibitem[\protect\citeauthoryear{{Kewley} \& {Ellison}}{{Kewley} \&
  {Ellison}}{2008}]{Kewley08}
{Kewley} L.~J.,  {Ellison} S.~L.,  2008, \mn@doi [\apj] {10.1086/587500}, \href
  {http://adsabs.harvard.edu/abs/2008ApJ...681.1183K} {681, 1183}

\bibitem[\protect\citeauthoryear{{K{\"o}nyves-T{\'o}th} \&
  {Vink{\'o}}}{{K{\"o}nyves-T{\'o}th} \& {Vink{\'o}}}{2021}]{Konyves-Toth21}
{K{\"o}nyves-T{\'o}th} R.,  {Vink{\'o}} J.,  2021, \mn@doi [\apj]
  {10.3847/1538-4357/abd6c8}, \href
  {https://ui.adsabs.harvard.edu/abs/2021ApJ...909...24K} {909, 24}

\bibitem[\protect\citeauthoryear{{Kozyreva}, {Yoon}  \& {Langer}}{{Kozyreva}
  et~al.}{2014}]{Kozyreva14}
{Kozyreva} A.,  {Yoon} S.~C.,   {Langer} N.,  2014, \mn@doi [\aap]
  {10.1051/0004-6361/201423641}, \href
  {https://ui.adsabs.harvard.edu/abs/2014A&A...566A.146K} {566, A146}

\bibitem[\protect\citeauthoryear{{Kumar} et~al.,}{{Kumar}
  et~al.}{2020}]{Kumar20}
{Kumar} A.,  et~al., 2020, \mn@doi [\apj] {10.3847/1538-4357/ab737b}, \href
  {https://ui.adsabs.harvard.edu/abs/2020ApJ...892...28K} {892, 28}

\bibitem[\protect\citeauthoryear{{Kuncarayakti} et~al.,}{{Kuncarayakti}
  et~al.}{2013}]{Kuncarayakti13a}
{Kuncarayakti} H.,  et~al., 2013, \mn@doi [\aj] {10.1088/0004-6256/146/2/30},
  \href {https://ui.adsabs.harvard.edu/abs/2013AJ....146...30K} {146, 30}

\bibitem[\protect\citeauthoryear{{Kuncarayakti} et~al.,}{{Kuncarayakti}
  et~al.}{2015}]{Kuncarayakti15}
{Kuncarayakti} H.,  et~al., 2015, \mn@doi [\aap] {10.1051/0004-6361/201425604},
  \href {https://ui.adsabs.harvard.edu/abs/2015A&A...579A..95K} {579, A95}

\bibitem[\protect\citeauthoryear{{Le Borgne} \& {Rocca-Volmerange}}{{Le Borgne}
  \& {Rocca-Volmerange}}{2002}]{LeBorgne02}
{Le Borgne} D.,  {Rocca-Volmerange} B.,  2002, \mn@doi [\aap]
  {10.1051/0004-6361:20020259}, \href
  {http://adsabs.harvard.edu/abs/2002A%26A...386..446L} {386, 446}

\bibitem[\protect\citeauthoryear{{Leloudas} et~al.,}{{Leloudas}
  et~al.}{2012}]{Leloudas12}
{Leloudas} G.,  et~al., 2012, \mn@doi [\aap] {10.1051/0004-6361/201118498},
  \href {https://ui.adsabs.harvard.edu/abs/2012A&A...541A.129L} {541, A129}

\bibitem[\protect\citeauthoryear{{Leloudas} et~al.,}{{Leloudas}
  et~al.}{2015}]{Leloudas15}
{Leloudas} G.,  et~al., 2015, \mn@doi [\mnras] {10.1093/mnras/stv320}, \href
  {https://ui.adsabs.harvard.edu/abs/2015MNRAS.449..917L} {449, 917}

\bibitem[\protect\citeauthoryear{{Li} \& {McCray}}{{Li} \&
  {McCray}}{1993}]{Li93}
{Li} H.,  {McCray} R.,  1993, \mn@doi [\apj] {10.1086/172401}, \href
  {https://ui.adsabs.harvard.edu/abs/1993ApJ...405..730L} {405, 730}

\bibitem[\protect\citeauthoryear{{Lin} et~al.,}{{Lin} et~al.}{2020}]{Lin20}
{Lin} W.~L.,  et~al., 2020, \mn@doi [\mnras] {10.1093/mnras/staa1918}, \href
  {https://ui.adsabs.harvard.edu/abs/2020MNRAS.497..318L} {497, 318}

\bibitem[\protect\citeauthoryear{{Lunnan} et~al.,}{{Lunnan}
  et~al.}{2014}]{Lunnan14}
{Lunnan} R.,  et~al., 2014, \mn@doi [\apj] {10.1088/0004-637X/787/2/138}, \href
  {https://ui.adsabs.harvard.edu/abs/2014ApJ...787..138L} {787, 138}

\bibitem[\protect\citeauthoryear{{Lunnan} et~al.,}{{Lunnan}
  et~al.}{2016}]{Lunnan16}
{Lunnan} R.,  et~al., 2016, \mn@doi [\apj] {10.3847/0004-637X/831/2/144}, \href
  {https://ui.adsabs.harvard.edu/abs/2016ApJ...831..144L} {831, 144}

\bibitem[\protect\citeauthoryear{{Maeda}, {Mazzali}, {Deng}, {Nomoto},
  {Yoshii}, {Tomita}  \& {Kobayashi}}{{Maeda} et~al.}{2003}]{Maeda03}
{Maeda} K.,  {Mazzali} P.~A.,  {Deng} J.,  {Nomoto} K.,  {Yoshii} Y.,  {Tomita}
  H.,   {Kobayashi} Y.,  2003, \mn@doi [\apj] {10.1086/376591}, \href
  {https://ui.adsabs.harvard.edu/abs/2003ApJ...593..931M} {593, 931}

\bibitem[\protect\citeauthoryear{{Maeda} et~al.,}{{Maeda}
  et~al.}{2007}]{Maeda07}
{Maeda} K.,  et~al., 2007, \mn@doi [\apj] {10.1086/520054}, \href
  {https://ui.adsabs.harvard.edu/abs/2007ApJ...666.1069M} {666, 1069}

\bibitem[\protect\citeauthoryear{{Magnier} et~al.,}{{Magnier}
  et~al.}{2020}]{Magnier20}
{Magnier} E.~A.,  et~al., 2020, \mn@doi [\apjs] {10.3847/1538-4365/abb82a},
  \href {https://ui.adsabs.harvard.edu/abs/2020ApJS..251....6M} {251, 6}

\bibitem[\protect\citeauthoryear{{Makarov}, {Prugniel}, {Terekhova}, {Courtois}
   \& {Vauglin}}{{Makarov} et~al.}{2014}]{Makarov14}
{Makarov} D.,  {Prugniel} P.,  {Terekhova} N.,  {Courtois} H.,   {Vauglin} I.,
  2014, \mn@doi [\aap] {10.1051/0004-6361/201423496}, \href
  {http://adsabs.harvard.edu/abs/2014A%26A...570A..13M} {570, A13}

\bibitem[\protect\citeauthoryear{{Masci} et~al.,}{{Masci}
  et~al.}{2019}]{Masci19}
{Masci} F.~J.,  et~al., 2019, \mn@doi [\pasp] {10.1088/1538-3873/aae8ac}, \href
  {https://ui.adsabs.harvard.edu/abs/2019PASP..131a8003M} {131, 018003}

\bibitem[\protect\citeauthoryear{{Mattila} et~al.,}{{Mattila}
  et~al.}{2008}]{Mattila08a}
{Mattila} S.,  et~al., 2008, \mn@doi [\mnras]
  {10.1111/j.1365-2966.2008.13516.x}, \href
  {https://ui.adsabs.harvard.edu/abs/2008MNRAS.389..141M} {389, 141}

\bibitem[\protect\citeauthoryear{{Maund} \& {Ramirez-Ruiz}}{{Maund} \&
  {Ramirez-Ruiz}}{2016}]{Maund16}
{Maund} J.~R.,  {Ramirez-Ruiz} E.,  2016, \mn@doi [\mnras]
  {10.1093/mnras/stv2760}, \href
  {https://ui.adsabs.harvard.edu/abs/2016MNRAS.456.3175M} {456, 3175}

\bibitem[\protect\citeauthoryear{{Mazzali}, {Maurer}, {Valenti}, {Kotak}  \&
  {Hunter}}{{Mazzali} et~al.}{2010}]{Mazzali10}
{Mazzali} P.~A.,  {Maurer} I.,  {Valenti} S.,  {Kotak} R.,   {Hunter} D.,
  2010, \mn@doi [\mnras] {10.1111/j.1365-2966.2010.17133.x}, \href
  {https://ui.adsabs.harvard.edu/abs/2010MNRAS.408...87M} {408, 87}

\bibitem[\protect\citeauthoryear{{Mazzali}, {Sullivan}, {Pian}, {Greiner}  \&
  {Kann}}{{Mazzali} et~al.}{2016}]{Mazzali16}
{Mazzali} P.~A.,  {Sullivan} M.,  {Pian} E.,  {Greiner} J.,   {Kann} D.~A.,
  2016, \mn@doi [\mnras] {10.1093/mnras/stw512}, \href
  {https://ui.adsabs.harvard.edu/abs/2016MNRAS.458.3455M} {458, 3455}

\bibitem[\protect\citeauthoryear{{Mazzali}, {Moriya}, {Tanaka}  \&
  {Woosley}}{{Mazzali} et~al.}{2019}]{Mazzali19}
{Mazzali} P.~A.,  {Moriya} T.~J.,  {Tanaka} M.,   {Woosley} S.~E.,  2019,
  \mn@doi [\mnras] {10.1093/mnras/stz177}, \href
  {https://ui.adsabs.harvard.edu/abs/2019MNRAS.484.3451M} {484, 3451}

\bibitem[\protect\citeauthoryear{{Milisavljevic}, {Fesen}, {Chevalier},
  {Kirshner}, {Challis}  \& {Turatto}}{{Milisavljevic}
  et~al.}{2012}]{Milisavljevic12}
{Milisavljevic} D.,  {Fesen} R.~A.,  {Chevalier} R.~A.,  {Kirshner} R.~P.,
  {Challis} P.,   {Turatto} M.,  2012, \mn@doi [\apj]
  {10.1088/0004-637X/751/1/25}, \href
  {http://adsabs.harvard.edu/abs/2012ApJ...751...25M} {751, 25}

\bibitem[\protect\citeauthoryear{{Millard} et~al.,}{{Millard}
  et~al.}{1999}]{Millard99}
{Millard} J.,  et~al., 1999, \mn@doi [\apj] {10.1086/308108}, \href
  {https://ui.adsabs.harvard.edu/abs/1999ApJ...527..746M} {527, 746}

\bibitem[\protect\citeauthoryear{{Modjaz}, {Kewley}, {Bloom}, {Filippenko},
  {Perley}  \& {Silverman}}{{Modjaz} et~al.}{2011}]{Modjaz11}
{Modjaz} M.,  {Kewley} L.,  {Bloom} J.~S.,  {Filippenko} A.~V.,  {Perley} D.,
  {Silverman} J.~M.,  2011, \mn@doi [\apjl] {10.1088/2041-8205/731/1/L4}, \href
  {http://adsabs.harvard.edu/abs/2011ApJ...731L...4M} {731, L4+}

\bibitem[\protect\citeauthoryear{{Moriya}}{{Moriya}}{2014}]{Moriya14}
{Moriya} T.~J.,  2014, arXiv e-prints, \href
  {https://ui.adsabs.harvard.edu/abs/2014arXiv1402.2519M} {p. arXiv:1402.2519}

\bibitem[\protect\citeauthoryear{{Moriya} \& {Maeda}}{{Moriya} \&
  {Maeda}}{2012}]{Moriya12}
{Moriya} T.~J.,  {Maeda} K.,  2012, \mn@doi [\apjl]
  {10.1088/2041-8205/756/1/L22}, \href
  {https://ui.adsabs.harvard.edu/abs/2012ApJ...756L..22M} {756, L22}

\bibitem[\protect\citeauthoryear{{Moriya}, {Sorokina}  \& {Chevalier}}{{Moriya}
  et~al.}{2018}]{Moriya18}
{Moriya} T.~J.,  {Sorokina} E.~I.,   {Chevalier} R.~A.,  2018, \mn@doi [\ssr]
  {10.1007/s11214-018-0493-6}, \href
  {https://ui.adsabs.harvard.edu/abs/2018SSRv..214...59M} {214, 59}

\bibitem[\protect\citeauthoryear{{Moriya}, {Murase}, {Kashiyama}  \&
  {Blinnikov}}{{Moriya} et~al.}{2022}]{Moriya22}
{Moriya} T.~J.,  {Murase} K.,  {Kashiyama} K.,   {Blinnikov} S.~I.,  2022,
  arXiv e-prints, \href {https://ui.adsabs.harvard.edu/abs/2022arXiv220203082M}
  {p. arXiv:2202.03082}

\bibitem[\protect\citeauthoryear{{Nasa High Energy Astrophysics Science Archive
  Research Center (Heasarc)}}{{Nasa High Energy Astrophysics Science Archive
  Research Center (Heasarc)}}{2014}]{Swift14}
{Nasa High Energy Astrophysics Science Archive Research Center (Heasarc)} 2014,
  {HEAsoft: Unified Release of FTOOLS and XANADU} (\mn@eprint {ascl}
  {1408.004})

\bibitem[\protect\citeauthoryear{{Neill} et~al.}{{Neill}
  et~al.}{2011}]{Neill11}
{Neill} J.~D.,  et~al., 2011, \mn@doi [\apj] {10.1088/0004-637X/727/1/15},
  \href {http://adsabs.harvard.edu/abs/2011ApJ...727...15N} {727, 15}

\bibitem[\protect\citeauthoryear{{Nicholl}}{{Nicholl}}{2018}]{Nicholl18}
{Nicholl} M.,  2018, \mn@doi [Research Notes of the American Astronomical
  Society] {10.3847/2515-5172/aaf799}, \href
  {https://ui.adsabs.harvard.edu/abs/2018RNAAS...2..230N} {2, 230}

\bibitem[\protect\citeauthoryear{{Nicholl}}{{Nicholl}}{2021}]{Nicholl21}
{Nicholl} M.,  2021, \mn@doi [Astronomy and Geophysics]
  {10.1093/astrogeo/atab092}, \href
  {https://ui.adsabs.harvard.edu/abs/2021A&G....62.5.34N} {62, 5.34}

\bibitem[\protect\citeauthoryear{{Nicholl} et~al.}{{Nicholl}
  et~al.}{2013}]{Nicholl13}
{Nicholl} M.,  et~al., 2013, \mn@doi [\nat] {10.1038/nature12569}, \href
  {http://adsabs.harvard.edu/abs/2013Natur.502..346N} {502, 346}

\bibitem[\protect\citeauthoryear{{Nicholl} et~al.,}{{Nicholl}
  et~al.}{2015a}]{Nicholl15a}
{Nicholl} M.,  et~al., 2015a, \mn@doi [\mnras] {10.1093/mnras/stv1522}, \href
  {https://ui.adsabs.harvard.edu/abs/2015MNRAS.452.3869N} {452, 3869}

\bibitem[\protect\citeauthoryear{{Nicholl} et~al.,}{{Nicholl}
  et~al.}{2015b}]{Nicholl15}
{Nicholl} M.,  et~al., 2015b, \mn@doi [\apjl] {10.1088/2041-8205/807/1/L18},
  \href {https://ui.adsabs.harvard.edu/abs/2015ApJ...807L..18N} {807, L18}

\bibitem[\protect\citeauthoryear{{Nicholl} et~al.,}{{Nicholl}
  et~al.}{2016a}]{Nicholl16}
{Nicholl} M.,  et~al., 2016a, \mn@doi [\apj] {10.3847/0004-637X/826/1/39},
  \href {https://ui.adsabs.harvard.edu/abs/2016ApJ...826...39N} {826, 39}

\bibitem[\protect\citeauthoryear{{Nicholl} et~al.,}{{Nicholl}
  et~al.}{2016b}]{Nicholl16a}
{Nicholl} M.,  et~al., 2016b, \mn@doi [\apjl] {10.3847/2041-8205/828/2/L18},
  \href {https://ui.adsabs.harvard.edu/abs/2016ApJ...828L..18N} {828, L18}

\bibitem[\protect\citeauthoryear{{Ofek} et~al.,}{{Ofek} et~al.}{2014}]{Ofek14}
{Ofek} E.~O.,  et~al., 2014, \mn@doi [\apj] {10.1088/0004-637X/781/1/42}, \href
  {https://ui.adsabs.harvard.edu/abs/2014ApJ...781...42O} {781, 42}

\bibitem[\protect\citeauthoryear{{Orellana}, {Bersten}  \& {Moriya}}{{Orellana}
  et~al.}{2018}]{Orellana18}
{Orellana} M.,  {Bersten} M.~C.,   {Moriya} T.~J.,  2018, \mn@doi [\aap]
  {10.1051/0004-6361/201832661}, \href
  {https://ui.adsabs.harvard.edu/abs/2018A&A...619A.145O} {619, A145}

\bibitem[\protect\citeauthoryear{{Pastorello} et~al.}{{Pastorello}
  et~al.}{2008}]{Pastorello08b}
{Pastorello} A.,  et~al., 2008, \mn@doi [\mnras]
  {10.1111/j.1365-2966.2008.13602.x}, 389, 113

\bibitem[\protect\citeauthoryear{{Pastorello} et~al.}{{Pastorello}
  et~al.}{2010}]{Pastorello10}
{Pastorello} A.,  et~al., 2010, \mn@doi [\apjl] {10.1088/2041-8205/724/1/L16},
  \href {http://adsabs.harvard.edu/abs/2010ApJ...724L..16P} {724, L16}

\bibitem[\protect\citeauthoryear{{Perley} et~al.,}{{Perley}
  et~al.}{2016}]{Perley16}
{Perley} D.~A.,  et~al., 2016, \mn@doi [\apj] {10.3847/0004-637X/830/1/13},
  \href {https://ui.adsabs.harvard.edu/abs/2016ApJ...830...13P} {830, 13}

\bibitem[\protect\citeauthoryear{{Pettini} \& {Pagel}}{{Pettini} \&
  {Pagel}}{2004}]{Pettini04}
{Pettini} M.,  {Pagel} B.~E.~J.,  2004, \mn@doi [\mnras]
  {10.1111/j.1365-2966.2004.07591.x}, \href
  {http://adsabs.harvard.edu/abs/2004MNRAS.348L..59P} {348, L59}

\bibitem[\protect\citeauthoryear{{Piro}}{{Piro}}{2015}]{Piro15}
{Piro} A.~L.,  2015, \mn@doi [\apj] {10.1088/2041-8205/808/2/L51}, \href
  {https://ui.adsabs.harvard.edu/abs/2015ApJ...808L..51P} {808, L51}

\bibitem[\protect\citeauthoryear{{Prentice} et~al.,}{{Prentice}
  et~al.}{2019}]{Prentice18}
{Prentice} S.~J.,  et~al., 2019, \mn@doi [\mnras] {10.1093/mnras/sty3399},
  \href {https://ui.adsabs.harvard.edu/abs/2019MNRAS.485.1559P} {485, 1559}

\bibitem[\protect\citeauthoryear{{Pursiainen} et~al.,}{{Pursiainen}
  et~al.}{2022}]{Pursiainen22}
{Pursiainen} M.,  et~al., 2022, arXiv e-prints, \href
  {https://ui.adsabs.harvard.edu/abs/2022arXiv220201635P} {p. arXiv:2202.01635}

\bibitem[\protect\citeauthoryear{{Quimby} et~al.,}{{Quimby}
  et~al.}{2011}]{Quimby11}
{Quimby} R.~M.,  et~al., 2011, \mn@doi [\nat] {10.1038/nature10095}, \href
  {https://ui.adsabs.harvard.edu/abs/2011Natur.474..487Q} {474, 487}

\bibitem[\protect\citeauthoryear{{Quimby} et~al.,}{{Quimby}
  et~al.}{2018}]{Quimby18}
{Quimby} R.~M.,  et~al., 2018, \mn@doi [\apj] {10.3847/1538-4357/aaac2f}, \href
  {https://ui.adsabs.harvard.edu/abs/2018ApJ...855....2Q} {855, 2}

\bibitem[\protect\citeauthoryear{{Rakavy} \& {Shaviv}}{{Rakavy} \&
  {Shaviv}}{1967}]{Rakavy67}
{Rakavy} G.,  {Shaviv} G.,  1967, \mn@doi [\apj] {10.1086/149204}, \href
  {https://ui.adsabs.harvard.edu/abs/1967ApJ...148..803R} {148, 803}

\bibitem[\protect\citeauthoryear{{Schlafly} \& {Finkbeiner}}{{Schlafly} \&
  {Finkbeiner}}{2011}]{Schlafly11}
{Schlafly} E.~F.,  {Finkbeiner} D.~P.,  2011, \mn@doi [\apj]
  {10.1088/0004-637X/737/2/103}, \href
  {http://adsabs.harvard.edu/abs/2011ApJ...737..103S} {737, 103}

\bibitem[\protect\citeauthoryear{{Schulze} et~al.,}{{Schulze}
  et~al.}{2018}]{Schulze18}
{Schulze} S.,  et~al., 2018, \mn@doi [\mnras] {10.1093/mnras/stx2352}, \href
  {https://ui.adsabs.harvard.edu/abs/2018MNRAS.473.1258S} {473, 1258}

\bibitem[\protect\citeauthoryear{{Skrutskie} et~al.,}{{Skrutskie}
  et~al.}{2006}]{Skrutskie06}
{Skrutskie} M.~F.,  et~al., 2006, \mn@doi [\aj] {10.1086/498708}, \href
  {https://ui.adsabs.harvard.edu/abs/2006AJ....131.1163S} {131, 1163}

\bibitem[\protect\citeauthoryear{{Smith}, {Foley}  \& {Filippenko}}{{Smith}
  et~al.}{2008}]{Smith08}
{Smith} N.,  {Foley} R.~J.,   {Filippenko} A.~V.,  2008, \mn@doi [\apj]
  {10.1086/587860}, \href
  {https://ui.adsabs.harvard.edu/abs/2008ApJ...680..568S} {680, 568}

\bibitem[\protect\citeauthoryear{{Smith} et~al.,}{{Smith}
  et~al.}{2016}]{Smith16}
{Smith} M.,  et~al., 2016, \mn@doi [\apjl] {10.3847/2041-8205/818/1/L8}, \href
  {https://ui.adsabs.harvard.edu/abs/2016ApJ...818L...8S} {818, L8}

\bibitem[\protect\citeauthoryear{{Smith} et~al.,}{{Smith}
  et~al.}{2020}]{Smith20}
{Smith} K.~W.,  et~al., 2020, arXiv e-prints, \href
  {https://ui.adsabs.harvard.edu/abs/2020arXiv200309052S} {p. arXiv:2003.09052}

\bibitem[\protect\citeauthoryear{{Soker}}{{Soker}}{2022}]{Soker22}
{Soker} N.,  2022, arXiv e-prints, \href
  {https://ui.adsabs.harvard.edu/abs/2022arXiv220509560S} {p. arXiv:2205.09560}

\bibitem[\protect\citeauthoryear{{Soker} \& {Gilkis}}{{Soker} \&
  {Gilkis}}{2017}]{Soker17}
{Soker} N.,  {Gilkis} A.,  2017, \mn@doi [\apj] {10.3847/1538-4357/aa9c83},
  \href {https://ui.adsabs.harvard.edu/abs/2017ApJ...851...95S} {851, 95}

\bibitem[\protect\citeauthoryear{{Sollerman}, {Kozma}, {Fransson},
  {Leibundgut}, {Lundqvist}, {Ryde}  \& {Woudt}}{{Sollerman}
  et~al.}{2000}]{Sollerman00}
{Sollerman} J.,  {Kozma} C.,  {Fransson} C.,  {Leibundgut} B.,  {Lundqvist} P.,
   {Ryde} F.,   {Woudt} P.,  2000, \mn@doi [\apjl] {10.1086/312763}, \href
  {https://ui.adsabs.harvard.edu/abs/2000ApJ...537L.127S} {537, L127}

\bibitem[\protect\citeauthoryear{{Sorokina}, {Blinnikov}, {Nomoto}, {Quimby}
  \& {Tolstov}}{{Sorokina} et~al.}{2016}]{Sorokina16}
{Sorokina} E.,  {Blinnikov} S.,  {Nomoto} K.,  {Quimby} R.,   {Tolstov} A.,
  2016, \mn@doi [\apj] {10.3847/0004-637X/829/1/17}, \href
  {https://ui.adsabs.harvard.edu/abs/2016ApJ...829...17S} {829, 17}

\bibitem[\protect\citeauthoryear{{Stoll}, {Prieto}, {Stanek}, {Pogge},
  {Szczygiel}, {Pojmanski}, {Antognini}  \& {Yan}}{{Stoll}
  et~al.}{2011}]{Stoll11}
{Stoll} R.,  {Prieto} J.~L.,  {Stanek} K.~Z.,  {Pogge} R.~W.,  {Szczygiel}
  D.~M.,  {Pojmanski} G.,  {Antognini} J.,   {Yan} H.,  2011, \mn@doi [\apj]
  {10.1088/0004-637X/730/1/34}, \href
  {http://adsabs.harvard.edu/abs/2011ApJ...730...34S} {730, 34}

\bibitem[\protect\citeauthoryear{{Stritzinger} et~al.,}{{Stritzinger}
  et~al.}{2018}]{Stritzinger18b}
{Stritzinger} M.~D.,  et~al., 2018, \mn@doi [\aap]
  {10.1051/0004-6361/201730843}, \href
  {https://ui.adsabs.harvard.edu/abs/2018A&A...609A.135S} {609, A135}

\bibitem[\protect\citeauthoryear{{Sullivan} et~al.}{{Sullivan}
  et~al.}{2010}]{Sullivan10}
{Sullivan} M.,  et~al., 2010, \mn@doi [\mnras]
  {10.1111/j.1365-2966.2010.16731.x}, \href
  {http://adsabs.harvard.edu/abs/2010MNRAS.406..782S} {406, 782}

\bibitem[\protect\citeauthoryear{{Taddia} et~al.,}{{Taddia}
  et~al.}{2018a}]{Taddia18b}
{Taddia} F.,  et~al., 2018a, \mn@doi [\aap] {10.1051/0004-6361/201629874},
  \href {https://ui.adsabs.harvard.edu/abs/2018A&A...609A.106T} {609, A106}

\bibitem[\protect\citeauthoryear{{Taddia} et~al.,}{{Taddia}
  et~al.}{2018b}]{Taddia18}
{Taddia} F.,  et~al., 2018b, \mn@doi [\aap] {10.1051/0004-6361/201730844},
  \href {https://ui.adsabs.harvard.edu/\#abs/2018A&A...609A.136T} {609, A136}

\bibitem[\protect\citeauthoryear{{Tartaglia} et~al.,}{{Tartaglia}
  et~al.}{2020}]{Tartaglia20}
{Tartaglia} L.,  et~al., 2020, \mn@doi [\aap] {10.1051/0004-6361/201936553},
  \href {https://ui.adsabs.harvard.edu/abs/2020A&A...635A..39T} {635, A39}

\bibitem[\protect\citeauthoryear{{Taubenberger} et~al.}{{Taubenberger}
  et~al.}{2006}]{Taubenberger06}
{Taubenberger} S.,  et~al., 2006, \mn@doi [\mnras]
  {10.1111/j.1365-2966.2006.10776.x}, 371, 1459

\bibitem[\protect\citeauthoryear{{Terreran} et~al.,}{{Terreran}
  et~al.}{2017}]{Terreran17}
{Terreran} G.,  et~al., 2017, \mn@doi [Nature Astronomy]
  {10.1038/s41550-017-0228-8}, \href
  {https://ui.adsabs.harvard.edu/\#abs/2017NatAs...1..713T} {1, 713}

\bibitem[\protect\citeauthoryear{{Terreran} et~al.,}{{Terreran}
  et~al.}{2019}]{Terreran19}
{Terreran} G.,  et~al., 2019, \mn@doi [\apj] {10.3847/1538-4357/ab3e37}, \href
  {https://ui.adsabs.harvard.edu/abs/2019ApJ...883..147T} {883, 147}

\bibitem[\protect\citeauthoryear{{Tinyanont}, {Dimitriadis}  \&
  {Foley}}{{Tinyanont} et~al.}{2020}]{Tinyanont20}
{Tinyanont} S.,  {Dimitriadis} G.,   {Foley} R.~J.,  2020, Transient Name
  Server Classification Report, \href
  {https://ui.adsabs.harvard.edu/abs/2020TNSCR3473....1T} {2020-3473, 1}

\bibitem[\protect\citeauthoryear{{Tominaga} et~al.,}{{Tominaga}
  et~al.}{2005}]{Tominaga05}
{Tominaga} N.,  et~al., 2005, \mn@doi [\apjl] {10.1086/498570}, \href
  {https://ui.adsabs.harvard.edu/abs/2005ApJ...633L..97T} {633, L97}

\bibitem[\protect\citeauthoryear{{Tonry} et~al.,}{{Tonry}
  et~al.}{2018}]{Tonry18}
{Tonry} J.~L.,  et~al., 2018, \mn@doi [Publications of the Astronomical Society
  of the Pacific] {10.1088/1538-3873/aabadf}, \href
  {https://ui.adsabs.harvard.edu/\#abs/2018PASP..130f4505T} {130, 064505}

\bibitem[\protect\citeauthoryear{{Tremonti} et~al.}{{Tremonti}
  et~al.}{2004}]{Tremonti04}
{Tremonti} C.~A.,  et~al., 2004, \mn@doi [\apj] {10.1086/423264}, 613, 898

\bibitem[\protect\citeauthoryear{{Tully} et~al.,}{{Tully}
  et~al.}{2013}]{Tully13}
{Tully} R.~B.,  et~al., 2013, \mn@doi [\aj] {10.1088/0004-6256/146/4/86}, \href
  {https://ui.adsabs.harvard.edu/abs/2013AJ....146...86T} {146, 86}

\bibitem[\protect\citeauthoryear{{Valenti} et~al.,}{{Valenti}
  et~al.}{2008a}]{Valenti08}
{Valenti} S.,  et~al., 2008a, \mn@doi [\mnras]
  {10.1111/j.1365-2966.2007.12647.x}, \href
  {https://ui.adsabs.harvard.edu/abs/2008MNRAS.383.1485V} {383, 1485}

\bibitem[\protect\citeauthoryear{{Valenti} et~al.,}{{Valenti}
  et~al.}{2008b}]{Valenti08a}
{Valenti} S.,  et~al., 2008b, \mn@doi [\apjl] {10.1086/527672}, \href
  {https://ui.adsabs.harvard.edu/abs/2008ApJ...673L.155V} {673, L155}

\bibitem[\protect\citeauthoryear{{Vreeswijk} et~al.,}{{Vreeswijk}
  et~al.}{2017}]{Vreeswijk17}
{Vreeswijk} P.~M.,  et~al., 2017, \mn@doi [\apj] {10.3847/1538-4357/835/1/58},
  \href {https://ui.adsabs.harvard.edu/abs/2017ApJ...835...58V} {835, 58}

\bibitem[\protect\citeauthoryear{{Vurm} \& {Metzger}}{{Vurm} \&
  {Metzger}}{2021}]{Vurm21}
{Vurm} I.,  {Metzger} B.~D.,  2021, \mn@doi [\apj] {10.3847/1538-4357/ac0826},
  \href {https://ui.adsabs.harvard.edu/abs/2021ApJ...917...77V} {917, 77}

\bibitem[\protect\citeauthoryear{{Wheeler}, {Chatzopoulos}, {Vink{\'o}}  \&
  {Tuminello}}{{Wheeler} et~al.}{2017}]{Wheeler17}
{Wheeler} J.~C.,  {Chatzopoulos} E.,  {Vink{\'o}} J.,   {Tuminello} R.,  2017,
  \mn@doi [\apjl] {10.3847/2041-8213/aa9d84}, \href
  {https://ui.adsabs.harvard.edu/abs/2017ApJ...851L..14W} {851, L14}

\bibitem[\protect\citeauthoryear{{Wilk}, {Hillier}  \& {Dessart}}{{Wilk}
  et~al.}{2019}]{Wilk19}
{Wilk} K.~D.,  {Hillier} D.~J.,   {Dessart} L.,  2019, \mn@doi [\mnras]
  {10.1093/mnras/stz1367}, \href
  {https://ui.adsabs.harvard.edu/abs/2019MNRAS.487.1218W} {487, 1218}

\bibitem[\protect\citeauthoryear{{Woosley}}{{Woosley}}{2010}]{Woosley10}
{Woosley} S.~E.,  2010, \mn@doi [\apjl] {10.1088/2041-8205/719/2/L204}, \href
  {https://ui.adsabs.harvard.edu/abs/2010ApJ...719L.204W} {719, L204}

\bibitem[\protect\citeauthoryear{{Woosley}, {Hartmann}  \& {Pinto}}{{Woosley}
  et~al.}{1989}]{Woosley89}
{Woosley} S.~E.,  {Hartmann} D.,   {Pinto} P.~A.,  1989, \mn@doi [\apj]
  {10.1086/168019}, \href {http://adsabs.harvard.edu/abs/1989ApJ...346..395W}
  {346, 395}

\bibitem[\protect\citeauthoryear{{Yan} et~al.,}{{Yan} et~al.}{2017a}]{Yan17}
{Yan} L.,  et~al., 2017a, \mn@doi [\apj] {10.3847/1538-4357/aa6b02}, \href
  {https://ui.adsabs.harvard.edu/abs/2017ApJ...840...57Y} {840, 57}

\bibitem[\protect\citeauthoryear{{Yan} et~al.,}{{Yan} et~al.}{2017b}]{Yan17a}
{Yan} L.,  et~al., 2017b, \mn@doi [\apj] {10.3847/1538-4357/aa8993}, \href
  {https://ui.adsabs.harvard.edu/abs/2017ApJ...848....6Y} {848, 6}

\bibitem[\protect\citeauthoryear{{Yaron} \& {Gal-Yam}}{{Yaron} \&
  {Gal-Yam}}{2012}]{Yaron12}
{Yaron} O.,  {Gal-Yam} A.,  2012, \mn@doi [\pasp] {10.1086/666656}, \href
  {http://adsabs.harvard.edu/abs/2012PASP..124..668Y} {124, 668}

\bibitem[\protect\citeauthoryear{{Young} et~al.}{{Young}
  et~al.}{2010}]{Young10}
{Young} D.~R.,  et~al., 2010, \mn@doi [\aap] {10.1051/0004-6361/200913004},
  \href {http://adsabs.harvard.edu/abs/2010A%26A...512A..70Y} {512, A70+}

\makeatother
\end{thebibliography}

\section*{Supporting information}

Supplementary data are available at MNRAS online.\\

\noindent \textbf{Table A1.} Optical photometry of \sn.\\
\textbf{Table A2.} JHK Vega photometry of \sn\ obtained with NOTCam.\\
\textbf{Table A3.} UV photometry obtained with Swift in the AB system.\\
\textbf{Table A4.} ATLAS AB optical photometry.\\
\textbf{Table A5.} ZTF AB optical photometry.\\
\textbf{Table A6.} Spectroscopic observations of \sn.

\appendix
\label{ap1}

\section{Tables}

\renewcommand{\thetable}{A\arabic{table}}
\setcounter{table}{0}
\begin{table*}
\centering
\scriptsize
\caption{Optical photometry of \sn.}
\label{table_photo}
\begin{tabular}[t]{cccccccccccccc}
\hline
\hline
UT date  &   MJD    & Phase           &     $u$        &      $B$	    &      $V$	     &      $g$       &	      $r$	   &    	$i$	    &   $z$         & Telescope         \\
         &          & (days)$^{\ast}$ &     (mag)      &     (mag)      &     (mag)      &     (mag)      &     (mag)      &    (mag)       &   (mag)       &                   \\
\hline
\hline                                              
20201121 & 59174.83 &     38.21       & $18.43\pm0.14$ & $17.91\pm0.05$ & $17.28\pm0.03$ & $17.57\pm0.03$ & $17.11\pm0.03$ & $16.94\pm0.06$ & \nodata        & Schmidt Telescope\\
20201121 & 59174.99 &     38.36       & $18.40\pm0.03$ & $17.93\pm0.02$ & $17.25\pm0.01$ & $17.50\pm0.01$ & $17.07\pm0.01$ & $16.92\pm0.01$ & $16.87\pm0.02$ & Copernico 1.82-m \\
20201124 & 59177.13 &     40.44       & $18.38\pm0.10$ & $17.84\pm0.04$ & $17.16\pm0.03$ & $17.44\pm0.03$ & $17.01\pm0.03$ & $16.84\pm0.04$ & \nodata        & Schmidt Telescope\\
20201126 & 59179.84 &     43.06       & \nodata        & $17.67\pm0.06$ & $17.05\pm0.06$ & $17.37\pm0.07$ & $16.86\pm0.07$ & $16.73\pm0.07$ & \nodata        & Schmidt Telescope\\
20201129 & 59182.83 &     45.96       & \nodata        & $17.63\pm0.07$ & $16.94\pm0.06$ & $17.28\pm0.11$ & $16.81\pm0.05$ & $16.69\pm0.06$ & \nodata        & Schmidt Telescope\\
20201203 & 59186.97 &     49.97       & $18.01\pm0.02$ & $17.44\pm0.02$ & $16.83\pm0.01$ & $17.08\pm0.01$ & $16.66\pm0.01$ & $16.49\pm0.01$ & $16.47\pm0.01$ & LT               \\
20201205 & 59188.01 &     50.98       & $17.98\pm0.02$ & \nodata        & \nodata        & $17.03\pm0.01$ & $16.64\pm0.01$ & $16.46\pm0.01$ & $16.50\pm0.02$ & LT               \\
20201207 & 59190.97 &     53.85       & $17.89\pm0.02$ & $17.34\pm0.01$ & $16.72\pm0.01$ & $16.93\pm0.01$ & $16.54\pm0.01$ & $16.37\pm0.01$ & $16.35\pm0.01$ & LT               \\
20201210 & 59193.97 &     56.75       & $17.84\pm0.02$ & $17.28\pm0.01$ & $16.66\pm0.01$ & $16.87\pm0.01$ & $16.49\pm0.01$ & $16.28\pm0.01$ & $16.28\pm0.01$ & LT               \\
20201212 & 59195.97 &     58.69       & $17.80\pm0.17$ & $17.23\pm0.03$ & $16.68\pm0.16$ & $16.94\pm0.05$ & $16.46\pm0.03$ & $16.33\pm0.03$ & $16.22\pm0.04$ & LT               \\
20201213 & 59196.09 &     58.81       & \nodata        & $17.25\pm0.14$ & $16.67\pm0.03$ & $16.91\pm0.03$ & $16.47\pm0.07$ & $16.39\pm0.07$ & \nodata        & Schmidt Telescope\\
20201213 & 59196.97 &     59.66       & $17.82\pm0.02$ & $17.24\pm0.01$ & $16.60\pm0.01$ & $16.87\pm0.01$ & $16.44\pm0.01$ & $16.28\pm0.03$ & $16.24\pm0.01$ & LT               \\
20201214 & 59197.98 &     60.64       & \nodata        & $17.26\pm0.04$ & $16.66\pm0.03$ & $16.90\pm0.03$ & $16.47\pm0.04$ & $16.35\pm0.04$ & \nodata        & Schmidt Telescope\\
20201216 & 59199.88 &     62.48       & \nodata        & \nodata        & \nodata        & $16.86\pm0.01$ & $16.47\pm0.01$ & $16.33\pm0.02$ & $16.23\pm0.02$ & Copernico 1.82-m \\
20201216 & 59199.95 &     62.55       & \nodata        & $17.24\pm0.01$ & $16.60\pm0.01$ & \nodata        & \nodata        & \nodata        & \nodata        & LT               \\
20201217 & 59200.95 &     63.52       & $17.82\pm0.02$ & $17.25\pm0.01$ & $16.58\pm0.02$ & $16.84\pm0.01$ & $16.41\pm0.01$ & $16.20\pm0.01$ & $16.20\pm0.01$ & LT               \\
20201218 & 59201.92 &     64.46       & \nodata        & $17.27\pm0.05$ & $16.65\pm0.03$ & $16.91\pm0.02$ & $16.46\pm0.03$ & $16.23\pm0.05$ & \nodata        & Schmidt Telescope\\
20201218 & 59201.94 &     64.48       & $17.83\pm0.02$ & $17.25\pm0.02$ & $16.60\pm0.03$ & $16.84\pm0.01$ & $16.43\pm0.00$ & $16.20\pm0.01$ & $16.18\pm0.01$ & LT               \\
20201220 & 59203.75 &     66.23       & $17.87\pm0.03$ & $17.24\pm0.03$ & $16.62\pm0.02$ & $16.88\pm0.02$ & $16.48\pm0.03$ & $16.24\pm0.01$ & $16.22\pm0.02$ & Copernico 1.82-m \\
20201227 & 59210.04 &     72.33       & $18.04\pm0.03$ & \nodata        & \nodata        & $16.87\pm0.01$ & $16.39\pm0.01$ & $16.18\pm0.01$ & $16.13\pm0.01$ & NOT              \\
20201229 & 59212.68 &     74.88       & \nodata        & \nodata        & $16.69\pm0.22$ & \nodata        & $16.35\pm0.06$ & $16.03\pm0.18$ & \nodata        & TNT              \\
20201230 & 59213.66 &     75.83       & \nodata        & \nodata        & $16.62\pm0.14$ & $16.85\pm0.06$ & $16.35\pm0.07$ & $16.15\pm0.12$ & \nodata        & TNT              \\
20201231 & 59214.69 &     76.83       & \nodata        & $17.31\pm0.03$ & $16.59\pm0.10$ & $16.83\pm0.04$ & $16.35\pm0.05$ & $16.18\pm0.10$ & \nodata        & TNT              \\
20210101 & 59215.02 &     77.15       & \nodata        & $17.27\pm0.15$ & $16.62\pm0.07$ & $16.88\pm0.06$ & $16.43\pm0.04$ & $16.18\pm0.03$ & $16.19\pm0.05$ & LT               \\
20210101 & 59215.95 &     78.05       & \nodata        & \nodata        & $16.64\pm0.23$ & \nodata        & $16.44\pm0.16$ & $16.18\pm0.19$ & \nodata        & LT               \\
20210105 & 59219.47 &     81.46       & \nodata        & $17.42\pm0.02$ & $16.59\pm0.01$ & $16.94\pm0.01$ & $16.39\pm0.05$ & $16.20\pm0.01$ & \nodata        & TNT              \\
20210109 & 59223.45 &     85.32       & \nodata        & $17.49\pm0.03$ & $16.60\pm0.02$ & $16.98\pm0.04$ & $16.40\pm0.01$ & $16.18\pm0.01$ & \nodata        & TNT              \\
20210110 & 59224.96 &     86.78       & \nodata        & $17.51\pm0.02$ & $16.70\pm0.02$ & $17.09\pm0.02$ & $16.51\pm0.02$ & $16.18\pm0.01$ & $16.14\pm0.01$ & Copernico 1.82-m \\
20210112 & 59226.65 &     88.42       & \nodata        & $17.66\pm0.02$ & $16.67\pm0.03$ & $17.12\pm0.03$ & $16.47\pm0.02$ & $16.25\pm0.02$ & \nodata        & TNT              \\
20210113 & 59227.55 &     89.29       & \nodata        & $17.66\pm0.02$ & $16.70\pm0.02$ & $17.19\pm0.02$ & $16.43\pm0.02$ & $16.23\pm0.02$ & \nodata        & TNT              \\
20210114 & 59228.05 &     89.78       & \nodata        & $17.67\pm0.02$ & $16.72\pm0.02$ & $17.15\pm0.02$ & $16.47\pm0.02$ & $16.20\pm0.01$ & \nodata        & Schmidt Telescope\\
20210115 & 59229.59 &     91.27       & \nodata        & $17.74\pm0.02$ & $16.75\pm0.02$ & $17.22\pm0.03$ & $16.50\pm0.02$ & $16.24\pm0.02$ & \nodata        & TNT              \\
20210116 & 59230.63 &     92.28       & \nodata        & $17.84\pm0.02$ & $16.78\pm0.02$ & $17.25\pm0.03$ & $16.51\pm0.02$ & $16.26\pm0.01$ & \nodata        & TNT              \\
20210117 & 59231.59 &     93.21       & \nodata        & $17.89\pm0.02$ & $16.82\pm0.02$ & $17.31\pm0.02$ & $16.54\pm0.02$ & $16.30\pm0.02$ & \nodata        & TNT              \\
20210117 & 59231.86 &     93.47       & \nodata        & $17.83\pm0.02$ & $16.83\pm0.02$ & $17.30\pm0.02$ & $16.54\pm0.03$ & $16.23\pm0.02$ & \nodata        & Schmidt Telescope\\
20210118 & 59232.81 &     94.39       & $18.87\pm0.04$ & $17.84\pm0.02$ & $16.85\pm0.02$ & $17.35\pm0.01$ & $16.56\pm0.01$ & $16.28\pm0.01$ & $16.18\pm0.01$ & Copernico 1.82-m \\
20210120 & 59234.86 &     96.38       & $19.06\pm0.05$ & $17.98\pm0.02$ & $16.85\pm0.03$ & $17.41\pm0.02$ & $16.52\pm0.02$ & $16.26\pm0.04$ & $16.11\pm0.02$ & NOT              \\
20210125 & 59239.95 &     101.31      & $19.41\pm0.07$ & $18.21\pm0.02$ & $16.95\pm0.01$ & $17.57\pm0.01$ & $16.57\pm0.01$ & $16.25\pm0.01$ & $16.15\pm0.01$ & LT               \\
20210128 & 59242.75 &     104.02      & \nodata        & $18.31\pm0.06$ & $17.01\pm0.02$ & $17.67\pm0.03$ & $16.66\pm0.02$ & $16.29\pm0.02$ & \nodata        & Schmidt Telescope\\
20210130 & 59244.57 &     105.78      & \nodata        & $18.40\pm0.02$ & $17.15\pm0.02$ & $17.78\pm0.03$ & $16.68\pm0.02$ & $16.40\pm0.02$ & \nodata        & TNT              \\
20210130 & 59244.94 &     106.14      & $19.77\pm0.10$ & $18.40\pm0.03$ & $17.10\pm0.01$ & $17.76\pm0.02$ & $16.68\pm0.01$ & $16.31\pm0.01$ & $16.23\pm0.01$ & LT               \\
20210131 & 59245.56 &     106.74      & \nodata        & $18.33\pm0.02$ & $17.18\pm0.02$ & $17.81\pm0.03$ & $16.75\pm0.02$ & $16.44\pm0.02$ & \nodata        & TNT              \\
20210201 & 59246.54 &     107.69      & \nodata        & $18.49\pm0.02$ & $17.16\pm0.02$ & $17.86\pm0.03$ & $16.76\pm0.02$ & $16.46\pm0.02$ & \nodata        & TNT              \\
20210202 & 59247.58 &     108.70      & \nodata        & $18.56\pm0.02$ & $17.21\pm0.02$ & $17.92\pm0.03$ & $16.82\pm0.02$ & $16.50\pm0.02$ & \nodata        & TNT              \\
20210204 & 59249.78 &     110.83      & \nodata        & $18.67\pm0.01$ & $17.39\pm0.08$ & $18.13\pm0.13$ & $16.90\pm0.05$ & $16.56\pm0.03$ & \nodata        & Schmidt Telescope\\
20210208 & 59253.86 &     114.79      & \nodata        & \nodata        & $17.52\pm0.02$ & $18.31\pm0.03$ & $17.13\pm0.03$ & $16.76\pm0.16$ & $16.55\pm0.03$ & Copernico 1.82-m \\
20210216 & 59261.87 &     122.55      & $21.19\pm0.09$ & $19.52\pm0.02$ & $17.89\pm0.01$ & $18.78\pm0.01$ & $17.44\pm0.01$ & $17.05\pm0.01$ & $16.76\pm0.01$ & NOT              \\
20210219 & 59264.48 &     125.08      & \nodata        & $19.49\pm0.02$ & $18.07\pm0.02$ & $18.75\pm0.03$ & $17.51\pm0.02$ & $17.09\pm0.09$ & \nodata        & TNT              \\
20210219 & 59264.75 &     125.34      & \nodata        & \nodata        & $18.15\pm0.06$ & \nodata        & $17.63\pm0.06$ & $17.14\pm0.03$ & \nodata        & Schmidt Telescope\\
20210220 & 59265.81 &     126.37      & \nodata        & $19.75\pm0.15$ & \nodata        & $18.89\pm0.11$ & \nodata        & \nodata        & \nodata        & Schmidt Telescope\\
20210221 & 59266.48 &     127.02      & \nodata        & $19.50\pm0.02$ & $18.17\pm0.02$ & $18.89\pm0.03$ & $17.55\pm0.02$ & $17.18\pm0.02$ & \nodata        & TNT              \\
20210222 & 59267.46 &     127.97      & \nodata        & $19.67\pm0.02$ & $18.20\pm0.02$ & $19.08\pm0.03$ & $17.58\pm0.02$ & $17.25\pm0.02$ & \nodata        & TNT              \\
20210226 & 59271.77 &     132.14      & \nodata        & \nodata        & $18.30\pm0.05$ & \nodata        & $17.77\pm0.04$ & $17.33\pm0.07$ & \nodata        & Schmidt Telescope\\
20210228 & 59273.77 &     134.08      & \nodata        & \nodata        & \nodata        & $19.17\pm0.13$ & \nodata        & \nodata        & \nodata        & Schmidt Telescope\\
20210304 & 59277.53 &     137.72      & \nodata        & $19.74\pm0.02$ & $18.33\pm0.07$ & $19.13\pm0.22$ & $17.70\pm0.02$ & $17.26\pm0.02$ & \nodata        & TNT              \\
20210306 & 59279.80 &     139.92      & \nodata        & \nodata        & $18.37\pm0.05$ & \nodata        & $17.80\pm0.04$ & $17.39\pm0.03$ & \nodata        & Schmidt Telescope\\
20210313 & 59286.79 &     146.70      & \nodata        & \nodata        & \nodata        & \nodata        & $17.90\pm0.05$ & $17.41\pm0.02$ & $17.06\pm0.03$ & Copernico 1.82-m \\
20210318 & 59291.86 &     151.61      & $21.65\pm0.25$ & $20.00\pm0.02$ & $18.36\pm0.01$ & $19.19\pm0.04$ & $17.87\pm0.01$ & $17.41\pm0.01$ & $17.01\pm0.01$ & NOT              \\
20210323 & 59296.79 &     156.39      & \nodata        & \nodata        & \nodata        & \nodata        & $17.96\pm0.05$ & $17.55\pm0.04$ & \nodata        & Schmidt Telescope\\
20210407 & 59311.49 &     170.63      & \nodata        & \nodata        & \nodata        & \nodata        & $18.06\pm0.25$ & $17.74\pm0.07$ & \nodata        & TNT	            \\
20210416 & 59320.81 &     179.66      & \nodata        & \nodata        & \nodata        & \nodata        & $18.18\pm0.08$ & $17.82\pm0.05$ & \nodata        & Schmidt Telescope\\
20210628 & 59393.07 &     249.68      & \nodata        & \nodata        & \nodata        & \nodata        & \nodata        & $18.46\pm0.10$ & \nodata        & Schmidt Telescope\\
20210630 & 59395.08 &     251.63      & \nodata        & \nodata        & \nodata        & $19.47\pm0.27$ & \nodata        & \nodata        & \nodata        & Schmidt Telescope\\
20210701 & 59396.06 &     252.58      & \nodata        & \nodata        & \nodata        & $19.44\pm0.14$ & \nodata        & \nodata        & \nodata        & Schmidt Telescope\\
20210706 & 59401.06 &     257.83      & \nodata        & $20.29\pm0.19$ & \nodata        & \nodata        & \nodata        & \nodata        & \nodata        & Schmidt Telescope\\
20210707 & 59402.07 &     257.42      & \nodata        & \nodata        & $19.13\pm0.10$ & \nodata        & \nodata        & \nodata        & \nodata        & Schmidt Telescope\\
20210708 & 59403.07 &     259.37      & \nodata        & \nodata        & \nodata        & $19.81\pm0.18$ & \nodata        & $18.55\pm0.08$ & \nodata        & Schmidt Telescope\\
20210717 & 59412.22 &     268.24      & $>20.87$       & $20.37\pm0.04$ & $19.05\pm0.05$ & $19.50\pm0.08$ & $18.91\pm0.03$ & $18.53\pm0.07$ & $17.84\pm0.05$ & LT               \\
20210722 & 59417.22 &     273.08      & $>21.29$       & $20.38\pm0.04$ & $19.22\pm0.05$ & $19.66\pm0.06$ & $18.97\pm0.02$ & $18.57\pm0.11$ & $17.97\pm0.17$ & LT               \\
20210729 & 59424.05 &     279.70      & \nodata        & \nodata        & \nodata        & \nodata        & $19.38\pm0.11$ & $19.23\pm0.10$ & \nodata        & Schmidt Telescope\\
20210729 & 59424.18 &     279.83      & $>20.97$       & $20.88\pm0.08$ & $19.53\pm0.08$ & $19.84\pm0.08$ & $19.36\pm0.09$ & $19.05\pm0.09$ & $18.32\pm0.07$ & LT               \\
20210801 & 59427.21 &     282.76      & $>21.66$       & $21.15\pm0.16$ & $19.62\pm0.04$ & $20.16\pm0.06$ & $19.51\pm0.06$ & $19.18\pm0.12$ & $18.54\pm0.13$ & LT               \\
20210804 & 59430.18 &     285.64      & \nodata        & $21.00\pm0.24$ & $19.75\pm0.06$ & $20.31\pm0.06$ & $19.57\pm0.09$ & $19.22\pm0.19$ & $18.58\pm0.13$ & LT               \\
20210808 & 59434.15 &     289.49      & \nodata        & $21.18\pm0.10$ & $19.95\pm0.10$ & $20.48\pm0.09$ & $19.71\pm0.11$ & $19.43\pm0.25$ & $18.80\pm0.17$ & LT               \\
20210811 & 59437.16 &     292.40      & \nodata        & $21.19\pm0.34$ & $20.10\pm0.11$ & $20.63\pm0.11$ & $19.82\pm0.07$ & $19.62\pm0.14$ & $18.96\pm0.18$ & LT               \\
20210814 & 59440.17 &     295.32      & \nodata        & $21.56\pm0.32$ & $20.32\pm0.38$ & $20.68\pm0.42$ & $20.05\pm0.59$ & $19.71\pm0.81$ & $18.98\pm0.41$ & LT               \\
20210818 & 59444.18 &     299.21      & \nodata        & \nodata        & $20.15\pm0.13$ & $20.67\pm0.13$ & $19.97\pm0.09$ & $19.73\pm0.11$ & $19.33\pm0.10$ & LT               \\
20210821 & 59447.19 &     302.12      & \nodata        & \nodata        & $20.61\pm0.15$ & $21.07\pm0.15$ & $20.28\pm0.15$ & $19.87\pm0.34$ & $19.50\pm0.21$ & LT               \\
20210828 & 59454.10 &     308.82      & \nodata        & \nodata        & $20.67\pm0.14$ & $21.10\pm0.24$ & $20.57\pm0.11$ & $20.08\pm0.25$ & $19.77\pm0.17$ & LT               \\
\hline                
\end{tabular}
\end{table*}

\begin{table*}
\centering
\scriptsize
\renewcommand{\thetable}{A\arabic{table}}
\setcounter{table}{0}
\caption{-- continued}
\label{table_photo}
\begin{tabular}[t]{cccccccccccccc}
\hline
\hline
UT date  &   MJD    & Phase           &     $u$        &      $B$	    &      $V$	     &      $g$       &	      $r$	   &    	$i$	    &   $z$         & Telescope         \\
         &          & (days)$^{\ast}$ &     (mag)      &     (mag)      &     (mag)      &     (mag)      &     (mag)      &    (mag)       &   (mag)       &                   \\
\hline
\hline 
20210831 & 59457.19 &     311.81      & \nodata        & \nodata        & $20.82\pm0.10$ & $21.33\pm0.12$ & $20.45\pm0.28$ & $20.07\pm0.44$ & $19.59\pm0.30$ & LT               \\
20210903 & 59460.04 &     314.57      & \nodata        & \nodata        & $21.10\pm0.12$ & \nodata        & $20.77\pm0.15$ & $20.48\pm0.18$ & $19.87\pm0.16$ & Copernico 1.82-m \\
20210909 & 59466.07 &     320.42      & \nodata        & $22.05\pm0.25$ & $21.24\pm0.25$ & $21.62\pm0.14$ & $20.56\pm0.20$ & $20.55\pm0.24$ & $20.11\pm0.27$ & TNG              \\
20211027 & 59514.94 &     367.77      & \nodata        & \nodata        & \nodata        & $21.30\pm0.06$ & $20.55\pm0.02$ & $20.60\pm0.07$ & $19.95\pm0.11$ & NOT              \\
20220227 & 59637.92 &     486.94      & \nodata        & $22.33\pm0.35$ & $21.56\pm0.25$ & $21.85\pm0.18$ & $21.38\pm0.18$ & $21.18\pm0.22$ & $20.78\pm0.15$ & NOT	            \\
\hline
\end{tabular}
\begin{list}{}{}
\item \textbf{Notes:} \\
$^{\ast}$ Rest-frame phase in days from explosion, MJD=$59135.42\pm0.98$. \\
$BV$ photometry is in the Vega system, while $ugriz$ photometry is in the AB system.\\
\textbf{Copernico 1.82-m:} Copernico 1.82 m telescope at the Asiago Observatory (Italy).
\textbf{LT:}  2-m Liverpool Telescope  at the Observatorio del Roque de Los Muchachos (Spain).
\textbf{NOT:} 2.56-m Nordic Optical Telescope at the Roque de Los Muchachos Observatory (Spain).
\textbf{Schmidt Telescope:} 67/91 Schmidt Telescope at the Asiago Observatory (Italy).
\textbf{TNG} 3.6-m Telescopio Nazionale Galileo at the Roque de Los Muchachos Observatory (Spain).
\textbf{TNT:}  0.8-m Tsinghua University–NAOC (National Astronomical Observatories of China) Telescope at Xinglong Observatory of NAOC. 
\end{list}
\end{table*}

\renewcommand{\thetable}{A\arabic{table}}
\setcounter{table}{1}
\begin{table*}
\centering
\caption{$JHK$ Vega photometry of \sn\ obtained with NOTCam.}
\label{table_photoNOTCam}
\begin{tabular}[t]{cccccccccccc}
\hline
\hline
UT date    &    MJD	  &    Phase           &       J        &        H       &       K        \\
           &          &   (days)$^{\ast}$  &      (mag)     &       (mag)    &      (mag)     \\
\hline                                
\hline   
20201216   & 59199.97 &       62.57        & $15.54\pm0.01$ & $15.57\pm0.01$ & \nodata         \\
20210126   & 59240.36 &       101.71       & $15.37\pm0.01$ & $15.08\pm0.01$ & $14.965\pm0.01$ \\
20210219   & 59264.43 &       125.03       & $16.23\pm0.03$ & $15.71\pm0.03$ & $15.577\pm0.02$ \\
20210317   & 59290.39 &       150.18       & $16.26\pm0.03$ & $15.80\pm0.03$ & $15.747\pm0.12$ \\
20210403   & 59307.40 &       166.67       & $16.64\pm0.04$ & $15.95\pm0.11$ & $15.960\pm0.12$ \\
20210716   & 59411.40 &       267.44       & $17.49\pm0.04$ & $16.76\pm0.05$ & $16.340\pm0.05$ \\
20210810   & 59436.31 &       291.58       & $18.39\pm0.07$ & $17.77\pm0.06$ & $17.275\pm0.11$ \\
20211012   & 59499.15 &       352.47       & $19.16\pm0.12$ & $18.38\pm0.15$ & $17.808\pm0.19$ \\
20211208   & 59556.30 &       407.85       & $19.56\pm0.20$ & $18.68\pm0.20$ & $18.083\pm0.15$ \\
20220204   & 59614.40 &       464.15       & $19.77\pm0.19$ & $19.00\pm0.18$ & $17.991\pm0.13$ \\
\hline
\end{tabular}
\begin{list}{}{}
\item \textbf{Notes:} \\
$^{\ast}$ Rest-frame phase in days from explosion, MJD=$59135.42\pm0.98$
\end{list}
\end{table*}

\renewcommand{\thetable}{A\arabic{table}}
\setcounter{table}{2}
\begin{table*}
\centering
\caption{UV photometry obtained with Swift in the AB system.}
\label{table_photoswift}
\begin{tabular}[t]{cccccccccc}
\hline
\hline
UT date    &  MJD     & Phase           &     UVW2       &    UVM2   &      UVW1     &      U         &      B         &      V         \\
           &          & (days)$^{\ast}$ &     (mag)      &    (mag)  &     (mag)     &    (mag)       &    (mag)       &    (mag)       \\
\hline                                
\hline        
20201123   & 59176.18 &    39.51        & $22.12\pm0.52$ & $22.22\pm0.56$ & $19.95\pm0.18$ & $18.53\pm0.13$ & $17.82\pm0.07$ & $17.14\pm0.09$ \\
20210114   & 59228.42 &    90.14        & $22.07\pm0.50$ &   $>21.64$     & $21.14\pm0.43$ & $18.62\pm0.13$ & $17.53\pm0.08$ & $16.59\pm0.08$ \\
20210203   & 59248.08 &    109.19       & $23.21\pm1.51$ &   $>21.43$     & \nodata        & $20.58\pm0.52$ & $18.55\pm0.09$ & $17.16\pm0.08$ \\
20210223   & 59268.20 &    128.68       & $22.47\pm2.31$ &   $>19.89$     & $21.99\pm3.16$ & $20.78\pm1.61$ & $19.41\pm0.30$ & $18.07\pm0.26$ \\
20210227   & 59272.37 &    132.73       & $22.94\pm0.96$ &   $>21.75$     & \nodata        & \nodata        & \nodata        & \nodata        \\
20210303   & 59276.82 &    137.04       & \nodata        &   $>21.39$     & $21.69\pm0.76$ & $21.07\pm0.80$ & $19.51\pm0.22$ & $18.25\pm0.24$ \\
20210315   & 59288.78 &    148.62       & \nodata        &   $>21.45$     & \nodata        & $22.37\pm2.56$ & $20.32\pm0.63$ & $18.06\pm0.26$ \\
20210318   & 59291.03 &    150.80       & $22.11\pm1.16$ &   $>20.50$     & $20.48\pm0.57$ & $20.65\pm1.16$ & $19.89\pm0.23$ & $18.29\pm0.19$ \\ 
\hline
\end{tabular}
\begin{list}{}{}
\item \textbf{Notes:} 
\item $^{\ast}$ Rest-frame phase in days from explosion, MJD=$59135.42\pm0.98$.
\end{list}
\end{table*}

\renewcommand{\thetable}{A\arabic{table}}
\setcounter{table}{3}
\begin{table}
\centering
\caption{ATLAS AB optical photometry.}
\label{table_photoatlas}
\begin{tabular}[t]{ccccccc}
\hline
\hline
UT date    &	MJD	  & Phase           &  Band  &   Magnitude    \\
           &          & (days)$^{\ast}$ &        &    (mag)       \\
\hline                                                           
\hline                                                           
20200926   & 59118.49 &     -16.386     &  $o$   & $>22.71$       \\
20201004   & 59126.48 &      -8.643     &  $o$   & $>22.00$       \\
20201014   & 59136.50 &       1.066     &  $c$   & $19.70\pm0.11$ \\
20201018   & 59140.48 &       4.922     &  $c$   & $20.05\pm0.22$ \\
20201020   & 59142.44 &       6.822     &  $o$   & $19.61\pm0.10$ \\
20201022   & 59144.41 &       8.731     &  $c$   & $19.61\pm0.09$ \\
20201024   & 59146.55 &      10.804     &  $o$   & $19.02\pm0.06$ \\
20201026   & 59148.50 &      12.694     &  $o$   & $18.70\pm0.11$ \\
20201026   & 59148.57 &      12.762     &  $c$   & $19.70\pm0.59$ \\
20201028   & 59150.59 &      14.719     &  $o$   & $18.65\pm0.04$ \\
20201031   & 59153.44 &      17.481     &  $o$   & $18.52\pm0.10$ \\
\hline                
\end{tabular}
\end{table}

\begin{table}
\centering
\renewcommand{\thetable}{A\arabic{table}}
\setcounter{table}{3}
\caption{-- continued}
\label{table_photoatlas}
\begin{tabular}[t]{cccccccccccccc}
\hline
\hline   
20201109   & 59162.46 &      26.221     &  $o$   & $17.81\pm0.02$ \\ 
20201117   & 59170.48 &      33.992     &  $o$   & $17.20\pm0.02$ \\
20201123   & 59176.41 &      39.738     &  $o$   & $16.92\pm0.04$ \\
20201125   & 59178.48 &      41.744     &  $o$   & $16.90\pm0.01$ \\
20201127   & 59180.42 &      43.624     &  $o$   & $16.86\pm0.02$ \\
20201202   & 59185.47 &      48.517     &  $o$   & $16.66\pm0.02$ \\
20201203   & 59186.47 &      49.486     &  $o$   & $16.66\pm0.02$ \\
20201207   & 59190.35 &      53.246     &  $o$   & $16.53\pm0.01$ \\
20201209   & 59192.34 &      55.174     &  $c$   & $16.79\pm0.01$ \\
20201211   & 59194.41 &      57.180     &  $c$   & $16.73\pm0.01$ \\
20201217   & 59200.38 &      62.965     &  $c$   & $16.67\pm0.01$ \\
20201219   & 59202.54 &      65.058     &  $c$   & $16.67\pm0.01$ \\
20201223   & 59206.38 &      68.779     &  $o$   & $16.33\pm0.01$ \\ 
\hline                
\end{tabular}
\end{table}

\begin{table}
\centering
\renewcommand{\thetable}{A\arabic{table}}
\setcounter{table}{3}
\caption{-- continued}
\label{table_photoatlas}
\begin{tabular}[t]{cccccccccccccc}
\hline
\hline    
20201224   & 59207.30 &      69.671     &  $o$   & $16.30\pm0.01$ \\
20201225   & 59208.36 &      70.698     &  $o$   & $16.28\pm0.01$ \\
20201231   & 59214.35 &      76.502     &  $o$   & $16.27\pm0.01$ \\
20210102   & 59216.33 &      78.421     &  $o$   & $16.28\pm0.01$ \\
20210104   & 59218.39 &      80.417     &  $o$   & $16.28\pm0.01$ \\
20210105   & 59219.39 &      81.386     &  $c$   & $16.64\pm0.01$ \\
20210106   & 59220.32 &      82.287     &  $c$   & $16.63\pm0.01$ \\
20210108   & 59222.31 &      84.215     &  $c$   & $16.72\pm0.01$ \\
20210112   & 59226.35 &      88.130     &  $c$   & $16.78\pm0.01$ \\
20210114   & 59228.39 &      90.107     &  $c$   & $16.83\pm0.01$ \\
20210116   & 59230.30 &      91.957     &  $c$   & $16.86\pm0.01$ \\
20210128   & 59242.37 &     103.653     &  $o$   & $16.54\pm0.03$ \\
20210130   & 59244.34 &     105.562     &  $o$   & $16.55\pm0.01$ \\
20210131   & 59245.26 &     106.453     &  $o$   & $16.55\pm0.01$ \\
20210201   & 59246.29 &     107.452     &  $o$   & $16.58\pm0.01$ \\
20210205   & 59250.30 &     111.337     &  $c$   & $17.42\pm0.01$ \\
20210207   & 59252.32 &     113.295     &  $c$   & $17.56\pm0.02$ \\
20210209   & 59254.29 &     115.203     &  $c$   & $17.65\pm0.01$ \\
20210211   & 59256.27 &     117.122     &  $c$   & $17.83\pm0.03$ \\
20210213   & 59258.30 &     119.089     &  $c$   & $17.88\pm0.02$ \\
20210217   & 59262.26 &     122.926     &  $o$   & $17.31\pm0.02$ \\
20210223   & 59268.30 &     128.779     &  $o$   & $17.54\pm0.04$ \\
20210227   & 59272.24 &     132.597     &  $o$   & $17.60\pm0.03$ \\
20210303   & 59276.24 &     136.473     &  $o$   & $17.60\pm0.03$ \\
20210311   & 59284.26 &     144.244     &  $c$   & $18.41\pm0.04$ \\
20210325   & 59298.23 &     157.781     &  $o$   & $17.78\pm0.05$ \\
20210327   & 59300.25 &     159.738     &  $o$   & $17.81\pm0.06$ \\
20210711   & 59406.60 &     262.791     &  $o$   & $18.73\pm0.09$ \\
20210719   & 59414.61 &     270.552     &  $o$   & $18.72\pm0.07$ \\
20210721   & 59416.60 &     272.481     &  $o$   & $19.12\pm0.28$ \\
20210727   & 59422.61 &     278.304     &  $o$   & $19.69\pm0.23$ \\
20210806   & 59432.61 &     287.994     &  $o$   & $20.07\pm0.22$ \\
20210808   & 59434.62 &     289.942     &  $c$   & $20.46\pm0.19$ \\
20210812   & 59438.62 &     293.818     &  $c$   & $20.82\pm0.31$ \\
20210820   & 59446.61 &     301.560     &  $o$   & $20.76\pm0.30$ \\
20210905   & 59462.56 &     317.016     &  $o$   & $21.09\pm0.35$ \\
20210911   & 59468.58 &     322.849     &  $c$   & $21.05\pm0.38$ \\
20210915   & 59472.56 &     326.705     &  $c$   & $21.71\pm0.71$ \\
20211011   & 59498.50 &     351.841     &  $o$   & $20.42\pm0.30$ \\
20211015   & 59502.47 &     355.688     &  $o$   & $20.18\pm0.34$ \\
20211017   & 59504.43 &     357.587     &  $o$   & $19.99\pm0.29$ \\
20211104   & 59522.52 &     375.116     &  $o$   & $20.17\pm0.28$ \\
20211106   & 59524.47 &     377.006     &  $c$   & $21.64\pm0.65$ \\
20220204   & 59614.28 &     464.031     &  $c$   & $22.00\pm0.97$ \\
20220205   & 59615.28 &     465.000     &  $c$   & $21.96\pm1.07$ \\ 
\hline
\end{tabular}
\begin{list}{}{}
\item \textbf{Notes:} 
\item $^{\ast}$ Rest-frame phase in days from explosion, MJD=$59135.42\pm0.98$.
\end{list}
\end{table}

\renewcommand{\thetable}{A\arabic{table}}
\setcounter{table}{4}
\begin{table}
\centering
\caption{ZTF AB optical photometry.}
\label{table_photoZTF}
\begin{tabular}[t]{ccccccc}
\hline
\hline
UT date    &   MJD    & Phase           &  Band  &    Magnitude   \\
           &          & (days)$^{\ast}$ &        &      (mag)     \\
\hline                                
\hline   
20201014   & 59136.40 &     0.95        &  $g$   & $19.63\pm0.06$ \\
20201014   & 59136.47 &     1.02        &  $r$   & $19.57\pm0.05$ \\
20201017   & 59139.37 &     3.83        &  $r$   & $19.96\pm0.09$ \\
20201017   & 59139.41 &     3.87        &  $g$   & $20.24\pm0.13$ \\
20201019   & 59141.34 &     5.74        &  $r$   & $19.90\pm0.08$ \\
20201019   & 59141.42 &     5.81        &  $g$   & $20.29\pm0.09$ \\
20201021   & 59143.32 &     7.66        &  $r$   & $19.62\pm0.06$ \\
20201021   & 59143.39 &     7.72        &  $g$   & $19.90\pm0.06$ \\
20201023   & 59145.34 &     9.61        &  $r$   & $19.21\pm0.03$ \\
20201023   & 59145.42 &     9.69        &  $g$   & $19.65\pm0.05$ \\
\hline                
\end{tabular}
\end{table}

\begin{table}
\centering
\renewcommand{\thetable}{A\arabic{table}}
\setcounter{table}{4}
\caption{-- continued}
\label{table_photoZTF}
\begin{tabular}[t]{cccccccccccccc}
\hline
\hline    
UT date    &   MJD    & Phase           &  Band  &    Magnitude   \\
           &          & (days)$^{\ast}$ &        &      (mag)     \\
\hline
\hline  
20201027   & 59149.37 &     13.52       &  $r$   & $18.83\pm0.05$ \\
20201027   & 59149.46 &     13.60       &  $g$   & $19.36\pm0.06$ \\
20201029   & 59151.33 &     15.42       &  $g$   & $19.02\pm0.08$ \\
20201029   & 59151.36 &     15.45       &  $r$   & $18.67\pm0.05$ \\
20201031   & 59153.34 &     17.36       &  $r$   & $18.58\pm0.05$ \\
20201031   & 59153.40 &     17.42       &  $g$   & $19.03\pm0.11$ \\
20201102   & 59155.29 &     19.25       &  $r$   & $18.39\pm0.04$ \\
20201102   & 59155.39 &     19.35       &  $g$   & $19.00\pm0.09$ \\
20201104   & 59157.39 &     21.29       &  $g$   & $18.92\pm0.08$ \\
20201104   & 59157.44 &     21.34       &  $r$   & $18.29\pm0.04$ \\
20201111   & 59164.32 &     28.00       &  $r$   & $17.70\pm0.01$ \\
20201111   & 59164.43 &     28.11       &  $g$   & $18.17\pm0.02$ \\
20201113   & 59166.30 &     29.92       &  $r$   & $17.77\pm0.09$ \\
20201113   & 59166.41 &     30.03       &  $g$   & $17.99\pm0.02$ \\
20201115   & 59168.26 &     31.82       &  $g$   & $17.89\pm0.02$ \\
20201115   & 59168.30 &     31.86       &  $r$   & $17.42\pm0.01$ \\
20201117   & 59170.23 &     33.73       &  $g$   & $17.73\pm0.01$ \\
20201117   & 59170.28 &     33.78       &  $r$   & $17.30\pm0.01$ \\
20201119   & 59172.23 &     35.67       &  $g$   & $17.63\pm0.01$ \\
20201122   & 59175.30 &     38.64       &  $g$   & $17.42\pm0.02$ \\
20201122   & 59175.37 &     38.71       &  $r$   & $17.00\pm0.01$ \\
20201124   & 59177.39 &     40.67       &  $g$   & $17.38\pm0.01$ \\
20201126   & 59179.25 &     42.47       &  $g$   & $17.33\pm0.02$ \\
20201126   & 59179.31 &     42.53       &  $r$   & $16.95\pm0.01$ \\
20201128   & 59181.22 &     44.38       &  $g$   & $17.35\pm0.03$ \\
20201128   & 59181.28 &     44.44       &  $r$   & $16.88\pm0.02$ \\
20201130   & 59183.21 &     46.31       &  $g$   & $17.12\pm0.03$ \\
20201202   & 59185.19 &     48.23       &  $g$   & $17.08\pm0.02$ \\
20201202   & 59185.26 &     48.29       &  $r$   & $16.70\pm0.01$ \\
20201204   & 59187.20 &     50.17       &  $g$   & $17.10\pm0.01$ \\
20201204   & 59187.24 &     50.21       &  $r$   & $16.72\pm0.01$ \\
20201206   & 59189.22 &     52.13       &  $r$   & $16.60\pm0.01$ \\
20201209   & 59192.16 &     54.98       &  $r$   & $16.64\pm0.02$ \\
20201209   & 59192.22 &     55.04       &  $g$   & $17.06\pm0.02$ \\
20201211   & 59194.22 &     56.98       &  $g$   & $16.98\pm0.01$ \\
20201211   & 59194.26 &     57.02       &  $r$   & $16.53\pm0.01$ \\
20201214   & 59197.24 &     59.90       &  $r$   & $16.44\pm0.02$ \\
20201217   & 59200.18 &     62.75       &  $g$   & $16.79\pm0.01$ \\
20201217   & 59200.24 &     62.81       &  $r$   & $16.40\pm0.01$ \\
20201220   & 59203.22 &     65.70       &  $r$   & $16.45\pm0.01$ \\
20201220   & 59203.31 &     65.78       &  $g$   & $16.88\pm0.01$ \\
20201222   & 59205.15 &     67.57       &  $r$   & $16.38\pm0.01$ \\
20201222   & 59205.26 &     67.67       &  $g$   & $16.79\pm0.01$ \\
20201224   & 59207.18 &     69.53       &  $g$   & $16.95\pm0.16$ \\
20201224   & 59207.20 &     69.55       &  $r$   & $16.55\pm0.13$ \\
20201228   & 59211.16 &     73.39       &  $r$   & $16.38\pm0.01$ \\
20201228   & 59211.22 &     73.45       &  $g$   & $16.81\pm0.02$ \\
20210102   & 59216.18 &     78.26       &  $g$   & $16.69\pm0.01$ \\
20210102   & 59216.28 &     78.35       &  $r$   & $16.30\pm0.01$ \\
20210104   & 59218.23 &     80.24       &  $g$   & $16.81\pm0.01$ \\
20210104   & 59218.35 &     80.36       &  $r$   & $16.32\pm0.01$ \\
20210106   & 59220.20 &     82.15       &  $g$   & $16.82\pm0.01$ \\
20210106   & 59220.24 &     82.19       &  $r$   & $16.41\pm0.01$ \\
20210108   & 59222.18 &     84.07       &  $g$   & $16.89\pm0.02$ \\
20210108   & 59222.20 &     84.09       &  $r$   & $16.35\pm0.01$ \\
20210110   & 59224.20 &     86.03       &  $g$   & $17.05\pm0.01$ \\
20210110   & 59224.25 &     86.08       &  $r$   & $16.44\pm0.01$ \\
20210112   & 59226.21 &     87.97       &  $g$   & $17.07\pm0.01$ \\
20210112   & 59226.24 &     88.00       &  $r$   & $16.48\pm0.01$ \\
20210114   & 59228.20 &     89.90       &  $g$   & $17.10\pm0.01$ \\
20210114   & 59228.22 &     89.92       &  $r$   & $16.49\pm0.01$ \\
20210116   & 59230.20 &     91.84       &  $g$   & $17.17\pm0.01$ \\
20210116   & 59230.22 &     91.86       &  $r$   & $16.50\pm0.01$ \\ 
\hline                
\end{tabular}
\end{table}

\begin{table}
\centering
\renewcommand{\thetable}{A\arabic{table}}
\setcounter{table}{4}
\caption{-- continued}
\label{table_photoZTF}
\begin{tabular}[t]{cccccccccccccc}
\hline
\hline    
UT date    &   MJD    & Phase           &  Band  &    Magnitude   \\
           &          & (days)$^{\ast}$ &        &      (mag)     \\
\hline
\hline    
20210118   & 59232.18 &     93.76       &  $r$   & $16.48\pm0.01$ \\
20210118   & 59232.22 &     93.80       &  $g$   & $17.26\pm0.01$ \\ 
20210129   & 59243.24 &     104.48      &  $g$   & $17.53\pm0.18$ \\
20210204   & 59249.14 &     110.19      &  $g$   & $18.06\pm0.07$ \\
20210204   & 59249.19 &     110.24      &  $r$   & $16.90\pm0.01$ \\
20210206   & 59251.14 &     112.13      &  $g$   & $18.10\pm0.02$ \\
20210206   & 59251.18 &     112.17      &  $r$   & $16.95\pm0.01$ \\
20210208   & 59253.16 &     114.09      &  $r$   & $17.04\pm0.01$ \\
20210208   & 59253.18 &     114.11      &  $g$   & $18.22\pm0.02$ \\
20210210   & 59255.14 &     116.01      &  $r$   & $17.14\pm0.01$ \\
20210210   & 59255.16 &     116.03      &  $g$   & $18.40\pm0.02$ \\
20210212   & 59257.14 &     117.95      &  $r$   & $17.28\pm0.01$ \\
20210212   & 59257.16 &     117.97      &  $g$   & $18.52\pm0.02$ \\
20210215   & 59260.18 &     120.89      &  $g$   & $18.68\pm0.04$ \\
20210215   & 59260.22 &     120.93      &  $r$   & $17.35\pm0.02$ \\
20210218   & 59263.16 &     123.78      &  $g$   & $18.81\pm0.07$ \\
20210218   & 59263.17 &     123.79      &  $r$   & $17.53\pm0.02$ \\
20210220   & 59265.14 &     125.70      &  $r$   & $17.57\pm0.01$ \\
20210220   & 59265.18 &     125.74      &  $g$   & $18.92\pm0.05$ \\
20210222   & 59267.14 &     127.64      &  $r$   & $17.64\pm0.02$ \\
20210224   & 59269.14 &     129.57      &  $r$   & $17.62\pm0.02$ \\
20210224   & 59269.25 &     129.68      &  $g$   & $19.02\pm0.10$ \\
20210226   & 59271.14 &     131.51      &  $r$   & $17.72\pm0.02$ \\
20210228   & 59273.16 &     133.47      &  $g$   & $19.24\pm0.12$ \\
20210228   & 59273.21 &     133.52      &  $r$   & $17.80\pm0.04$ \\
20210302   & 59275.16 &     135.41      &  $g$   & $19.35\pm0.09$ \\
20210302   & 59275.26 &     135.50      &  $r$   & $17.78\pm0.04$ \\
20210305   & 59278.14 &     138.29      &  $r$   & $17.86\pm0.03$ \\
20210305   & 59278.21 &     138.36      &  $g$   & $19.34\pm0.11$ \\
20210318   & 59291.19 &     150.94      &  $r$   & $17.91\pm0.02$ \\
20210320   & 59293.18 &     152.87      &  $r$   & $17.91\pm0.03$ \\
20210323   & 59296.20 &     155.79      &  $g$   & $19.37\pm0.20$ \\
20210325   & 59298.16 &     157.69      &  $r$   & $17.96\pm0.08$ \\
20210625   & 59390.48 &     247.15      &  $r$   & $18.89\pm0.14$ \\
20210626   & 59391.48 &     248.12      &  $r$   & $19.04\pm0.17$ \\
20210627   & 59392.48 &     249.09      &  $r$   & $18.92\pm0.16$ \\
\hline                
\end{tabular}
\end{table}

\begin{table}
\centering
\renewcommand{\thetable}{A\arabic{table}}
\setcounter{table}{4}
\caption{-- continued}
\label{table_photoZTF}
\begin{tabular}[t]{cccccccccccccc}
\hline
\hline    
UT date    &   MJD    & Phase           &  Band  &    Magnitude   \\
           &          & (days)$^{\ast}$ &        &      (mag)     \\
\hline
\hline   
20210628   & 59393.48 &     250.06      &  $r$   & $18.91\pm0.10$ \\
20210629   & 59394.48 &     251.03      &  $r$   & $18.85\pm0.09$ \\
20210630   & 59395.48 &     252.00      &  $r$   & $18.94\pm0.10$ \\
20210701   & 59396.48 &     252.97      &  $r$   & $18.89\pm0.09$ \\
20210702   & 59397.48 &     253.93      &  $r$   & $18.94\pm0.09$ \\
20210704   & 59399.48 &     255.87      &  $r$   & $19.02\pm0.10$ \\
20210705   & 59400.48 &     256.84      &  $r$   & $19.02\pm0.12$ \\
20210706   & 59401.48 &     257.81      &  $r$   & $18.92\pm0.09$ \\
20210709   & 59404.47 &     260.71      &  $r$   & $18.90\pm0.08$ \\
20210710   & 59405.47 &     261.68      &  $r$   & $19.02\pm0.09$ \\
20210711   & 59406.47 &     262.65      &  $r$   & $18.90\pm0.06$ \\
20210731   & 59426.45 &     282.01      &  $r$   & $19.64\pm0.10$ \\
20210808   & 59434.47 &     289.78      &  $g$   & $20.66\pm0.14$ \\
20210808   & 59434.47 &     289.78      &  $r$   & $20.27\pm0.12$ \\
20210808   & 59434.47 &     289.78      &  $r$   & $20.27\pm0.12$ \\
20210810   & 59436.47 &     291.72      &  $r$   & $20.40\pm0.10$ \\
20210810   & 59436.47 &     291.72      &  $r$   & $20.40\pm0.10$ \\
20210817   & 59443.48 &     298.51      &  $r$   & $20.79\pm0.19$ \\
20210817   & 59443.48 &     298.51      &  $r$   & $20.79\pm0.19$ \\
20210930   & 59487.39 &     341.06      &  $r$   & $20.67\pm0.20$ \\
20210930   & 59487.39 &     341.06      &  $r$   & $20.67\pm0.20$ \\
20211009   & 59496.41 &     349.80      &  $r$   & $20.43\pm0.17$ \\
20211009   & 59496.41 &     349.80      &  $r$   & $20.43\pm0.17$ \\
20211011   & 59498.37 &     351.70      &  $r$   & $20.68\pm0.13$ \\
20211011   & 59498.37 &     351.70      &  $r$   & $20.68\pm0.13$ \\
20211016   & 59503.40 &     356.57      &  $r$   & $20.65\pm0.20$ \\
20211016   & 59503.40 &     356.57      &  $r$   & $20.65\pm0.20$ \\
20211030   & 59517.34 &     370.08      &  $r$   & $21.21\pm0.18$ \\
20211030   & 59517.34 &     370.08      &  $r$   & $21.21\pm0.18$ \\
20211103   & 59521.29 &     373.91      &  $r$   & $20.92\pm0.15$ \\
20211103   & 59521.29 &     373.91      &  $r$   & $20.92\pm0.15$ \\
20211107   & 59525.31 &     377.80      &  $r$   & $21.04\pm0.17$ \\
20211107   & 59525.31 &     377.80      &  $r$   & $21.04\pm0.17$ \\
\hline
\end{tabular}
\begin{list}{}{}
\item \textbf{Notes:} 
\item $^{\ast}$ Rest-frame phase in days from explosion, MJD=$59135.42\pm0.98$.
\end{list}
\end{table}

\renewcommand{\thetable}{A\arabic{table}}
\setcounter{table}{5}
\begin{table*}
\centering
\caption{Spectroscopic observations of SN~2020wnt}
\label{tspectra}
\begin{tabular}[t]{cccccccccc}
\hline
\hline
UT date   &	MJD   & Rest-frame       	&  Range       &  Telescope   & Grism/Grating   \\
          &           & phase (days)$^{\ast}$ &  (\AA)       & +Instrument  &             \\
\hline
\hline
20201115  & 59168.00  &    31.6         & $3205-9775$  & LICK+KAST    &  \nodata    \\  
20201121  & 59174.92  &    38.3         & $3390-7915$  & Ekar+AFOSC   &  gm4        \\ 
20201124  & 59177.13  &    40.4         & $3550-9005$  & Ekar+AFOSC   &  VPH7       \\  
20201129  & 59182.79  &    45.9         & $3680-7945$  & Ekar+AFOSC   &  gm4        \\  
20201202  & 59185.56  &    48.6         & $6700-24620$ & IRTF+Spex    &  Prism      \\  
20201203  & 59186.10  &    49.1         & $3430-10075$ & TNG+LRS      &  LR-B+LR-R  \\ 
20201212  & 59195.10  &    57.8         & $3295-9335$  & NOT+ALFOSC   &  Grism\#4   \\  
20201215  & 59198.05  &    60.7         & $3440-9350$  & NOT+ALFOSC   &  Grism\#4   \\  
20201216  & 59199.85  &    62.4         & $4845-9010$  & Ekar+AFOSC   &  VPH6       \\  
20201218  & 59201.10  &    63.6         & $2975-10070$ & TNG+LRS      &  LR-B+LR-R  \\  
20201220  & 59203.78  &    66.2         & $4845-9010$  & Ekar+AFOSC   &  VPH6       \\  
20210101  & 59215.70  &    77.8         & $3390-8495$  & LJT+YFOSC    &  G3         \\
20210102  & 59216.44  &    78.5         & $3740-8550$  & XLT+BFOSC    &  G4         \\  
20210110  & 59224.92  &    86.7         & $3250-9010$  & Ekar+AFOSC   &  VPH6       \\    
20210114  & 59228.53  &    90.2         & $3390-8490$  & LJT+YFOSC    &  G3         \\  
20210115  & 59229.48  &    91.1         & $3740-8540$  & XLT+BFOSC    &  G4         \\  
20210118  & 59232.76  &    94.3         & $3295-9010$  & Ekar+AFOSC   &  VPH6       \\  
20210122  & 59236.53  &    98.0         & $3390-8495$  & LJT+YFOSC    &  G3         \\  
20210127  & 59241.89  &    103.2        & $3295-9375$  & NOT+ALFOSC   &  Grism\#4   \\  
20210204  & 59249.69  &    110.7        & $3390-8490$  & LJT+YFOSC    &  G3         \\  
20210211  & 59256.52  &    117.3        & $3390-8495$  & LJT+YFOSC    &  G3         \\  
20210216  & 59261.84  &    122.5        & $3295-9350$  & NOT+ALFOSC   &  Grism\#4   \\  
20210311  & 59284.86  &    144.8        & $3295-9385$  & NOT+ALFOSC   &  Grism\#4   \\  
20210317  & 59290.86  &    150.6        & $3000-10050$ & TNG+LRS      &  LR-B+LR-R  \\
20210407  & 59311.80  &    170.9        & $4845-9010$  & Ekar+AFOSC   &  VPH6       \\
20210728  & 59423.20  &    278.9        & $3540-7630$  & GTC + OSIRIS &  R1000B     \\
20210811  & 59438.16  &    293.3        & $4945-10100$ & GTC + OSIRIS &  R1000R     \\
\hline 
\end{tabular}
\begin{list}{}{}
\item \textbf{NOTES:}
\item $^{\ast}$ Rest-frame phase in days from explosion, MJD=$59135.42\pm0.98$.
\item \textbf{Telescope code:}
EKAR: Copernico 1.82-m Telescope, INAF (Mount Ekar); AFOSC: Asiago Faint Object Spectrograph and Camera; IRTF: NASA Infrared Telescope Facility; SpeX: 0.7-5.3 Micron Medium-Resolution Spectrograph and Imager; LICK: 3-m Shane telescope at Lick Observatory; KAST: Kast double spectrograph; LJT: Lijiang 2.4-m Telescope; YFOSC: Yunnan Faint Object Spectrograph and Camera; XLT: Xinglong 2.16-m Telescope; BFOSC: Beijing Faint Object Spectrograph and Camera; NOT: Nordic Optical Telescope; ALFOSC: Alhambra Faint Object Spectrograph and Camera; TNG: 3.6-m Telescopio Nazionale Galileo, INAF; LRS: Low Resolution Spectrograph; GTC: Gran Telescopio Canarias; OSIRIS: Optical System for Imaging and low-Intermediate-Resolution Integrated Spectroscopy.
\end{list}
\end{table*}

\section*{Affiliations}

$^{1}$ Finnish Centre for Astronomy with ESO (FINCA), FI-20014 University of Turku, Finland \\
$^{2}$ Tuorla Observatory, Department of Physics and Astronomy, FI-20014 University of Turku, Finland \\
$^{3}$ INAF - Osservatorio Astronomico di Padova, Vicolo dell'Osservatorio 5, I-35122 Padova, Italy\\
$^{4}$ Facultad de Ciencias Astron\'omicas y Geof\'isicas, Universidad Nacional de La Plata, Paseo del Bosque S/N, B1900FWA, \\ La Plata, Argentina\\
$^{5}$ Instituto de Astrof\'isica de La Plata (IALP), CCT-CONICET-UNLP. Paseo del Bosque S/N, B1900FWA, La Plata, Argentina\\
$^{6}$ Kavli Institute for the Physics and Mathematics of the Universe (WPI), The University of Tokyo, 5-1-5 Kashiwanoha, \\  Kashiwa, Chiba 277-8583, Japan\\
$^{7}$ Universidad Nacional de Río Negro. Sede Andina, Mitre 630 (8400), Bariloche, Argentina.\\
$^{8}$ Consejo Nacional de Investigaciones Científicas y Tećnicas (CONICET), Argentina.\\
$^{9}$ European Centre for Theoretical Studies in Nuclear Physics and Related Areas (ECT$^{\ast}$), Fondazione Bruno Kessler, Trento, Italy\\
$^{10}$ INFN-TIFPA, Trento Institute for Fundamental Physics and Applications, Via Sommarive 14, I-38123 Trento, Italy\\
$^{11}$ Department of Physics and Astronomy, Aarhus University, Ny Munkegade 120, DK-8000 Aarhus C, Denmark\\ 
$^{12}$ Departamento de Ciencias F\'{i}sicas – Universidad Andrés Bello, Avda. Rep\'{u}blica 252, Santiago, Chile\\
$^{13}$ Millennium Institute of Astrophysics, Nuncio Monsenor S\'{o}tero Sanz 100, Providencia, Santiago, Chile\\
$^{14}$ Cosmic Dawn Center (DAWN), Denmark\\
$^{15}$ Niels Bohr Institute, University of Copenhagen, Jagtvej 128, 2200 Copenhagen N, Denmark\\
$^{16}$ Department of Physics and Astronomy G. Galilei, University of Padova, Vicolo dell’Osservatorio 3, 35122, Padova, Italy\\
$^{17}$ Astrophysics Research Institute, Liverpool John Moores University, IC2, Liverpool Science Park, 146 Brownlow Hill, Liverpool L3 5RF, UK\\
$^{18}$ Max-Planck-Institut f\"ur Astrophysik, Karl-Schwarzschild Str. 1, D-85748 Garching, Germany\\
$^{19}$ School of Physics and Astronomy, University of Southampton,  Southampton, SO17 1BJ, UK\\
$^{20}$ Physics Department and Tsinghua Center for Astrophysics (THCA), Tsinghua University, Beijing 100084, China\\
$^{21}$ Institute of Space Sciences (ICE, CSIC), Campus UAB, Carrer de Can Magrans s/n, 08193 Barcelona, Spain\\
$^{22}$ School of Physics, O'Brien Centre for Science North, University College Dublin, Dublin, Ireland.\\
$^{23}$ Department of Physics, Florida State University, 77 Chieftan Way, Tallahassee, FL 32306, USA\\
$^{24}$ Turku Collegium for Science, Medicine and Technology, University of Turku, FI-20014 Turku, Finland\\
$^{25}$ National Astronomical Observatories, Chinese Academy of Sciences, Beijing 100101, China\\
$^{26}$ School of Astronomy and Space Science, University of Chinese Academy of Sciences, Beijing 101408, China\\
$^{27}$ Beijing Planetarium, Beijing Academy of Science and Technology, Beijing 100044, China\\
$^{28}$ Yunnan Observatories, Chinese Academy of Sciences, Kunming 650216, China\\
$^{29}$ Key Laboratory for the Structure and Evolution of Celestial Objects, Chinese Academy of Sciences, Kunming 650216, China\\

\bsp	
\label{lastpage}
\end{document}